%% file: main.tex
\documentclass[10pt,twocolumn,letterpaper]{article}

\usepackage[accsupp]{axessibility}  

\usepackage{iccv}              

\input{preamble}

\definecolor{iccvblue}{rgb}{0.21,0.49,0.74}
\usepackage[pagebackref,breaklinks,colorlinks,allcolors=iccvblue]{hyperref}

\def\paperID{} 
\def\confName{ICCV}
\def\confYear{2025}

\title{Global Motion Corresponder for 3D Point-Based \\ Scene Interpolation under Large Motion}

\author{Junru Lin$^{* 1, 2}$ \quad Chirag Vashist$^{* 3}$ \quad Mikaela Angelina Uy$^{2, 4}$ \quad Colton Stearns$^{2}$ 
\\ 
Xuan Luo$^{5}$ \quad Leonidas Guibas$^{2}$ \quad Ke Li$^{3}$
\\
$^{1}$University of Toronto \quad $^{2}$ Stanford University \quad  $^{3}$Simon Fraser University \quad $^{4}$ Nvidia \quad $^{5}$ Google
\\
$^*$ equal contribution
\\
\hyperlink{https://junrul.github.io/gmc/}{https://junrul.github.io/gmc/}
}

\begin{document}

\twocolumn[{
    \renewcommand\twocolumn[1][]{#1}
    \maketitle
    \centering
    \footnotesize
    \begin{tabularx}{\linewidth}{*{3}{Y}}
        \multicolumn{1}{c}{\cellcolor[HTML]{cfdcef}{{Local motion}}} & 
        \multicolumn{1}{c}{\cellcolor[HTML]{FFCCC9}{{Global motion}}} & 
        \multicolumn{1}{c}{\cellcolor[HTML]{A8D5BA}{{Ideal behaviour}}} \\
        \\
        \includegraphics[width=0.85\hsize,valign=m]{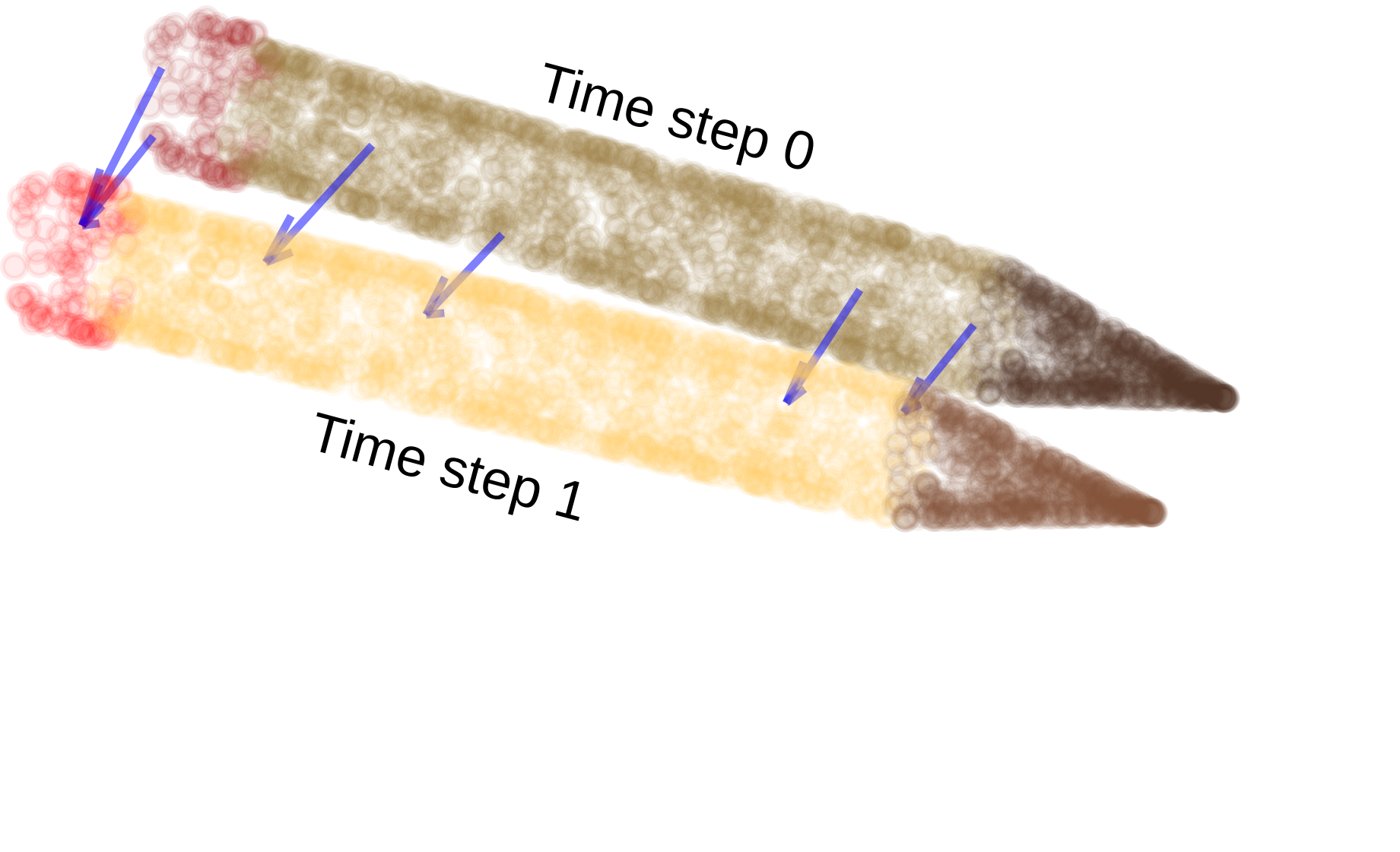} &
        \includegraphics[width=0.85\hsize,valign=m]{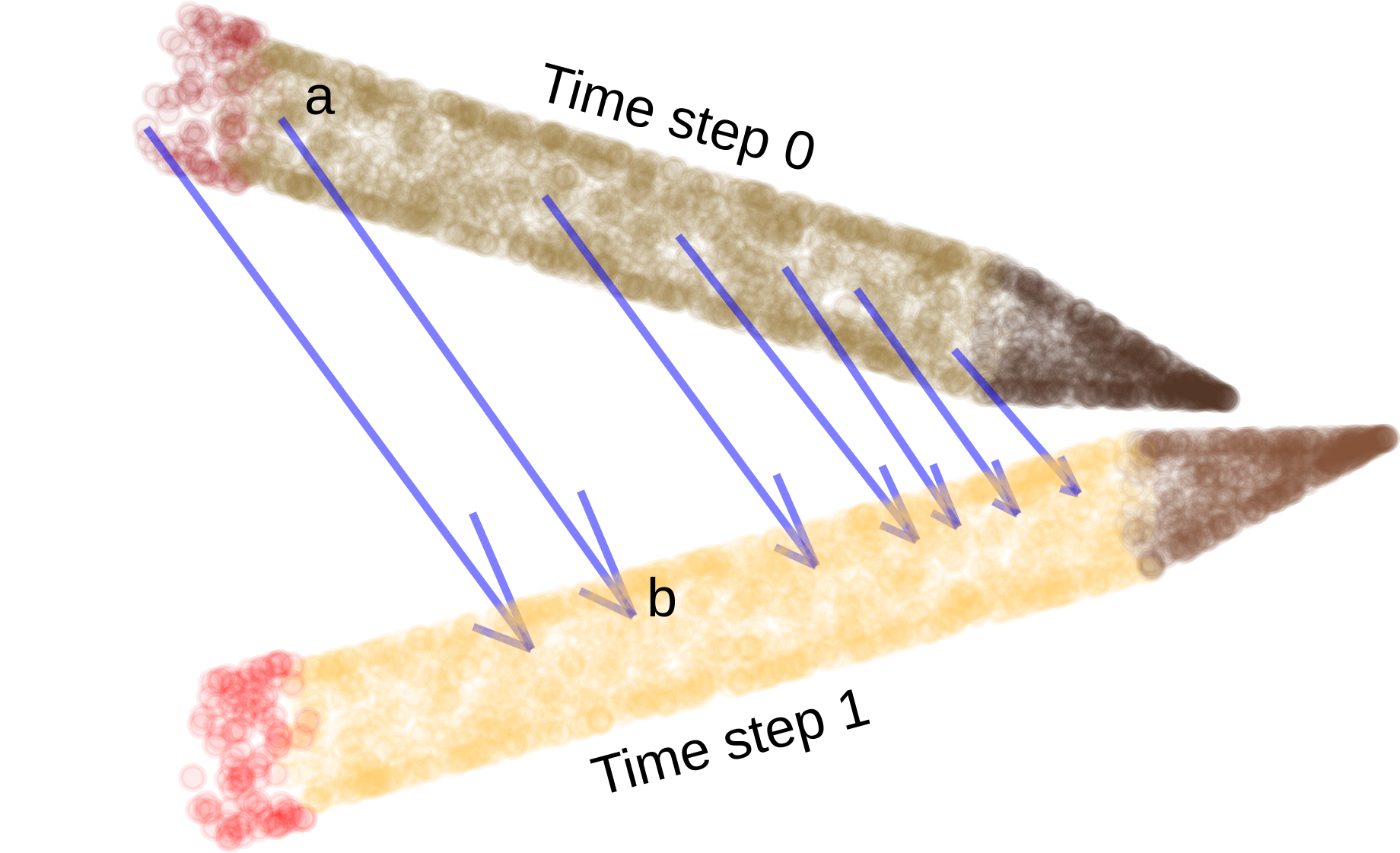} &
        \includegraphics[width=0.85\hsize,valign=m]{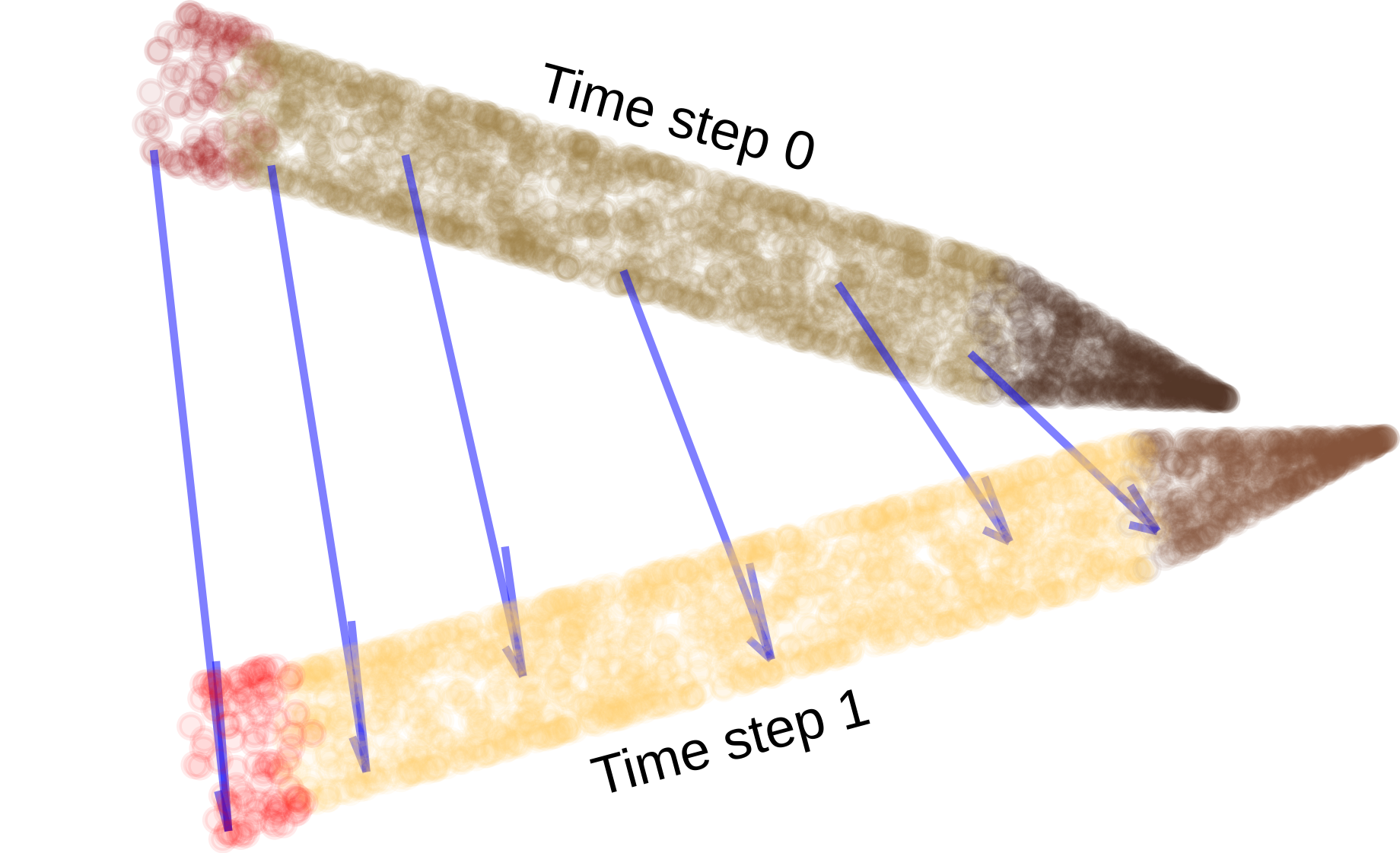} \\
    
    \end{tabularx}
    \captionof{figure}{
        \textbf{Global Motion Challenge.} 
        (1) Left: For small inter-frame motion, determining a point’s motion is effectively equivalent to matching it with a corresponding point within a small local neighborhood. In this case, local neighborhood searches yield correct correspondence and motion prediction. 
        (2) Middle: With large global motion, local searches lead to a wrong correspondence between adjacent timesteps.
        (3) Right: An ideal method would be able to predict correct correspondence and achieve global motion.
    }
    \label{fig:teaser}
    \vspace{2em}
}]

\input{sec/0_abstract}

\input{sec/1_intro}
\input{sec/2_related_work}

\input{sec/3_method}

\input{sec/4_experiments}
\input{sec/5_conclusion}
{
    \small
    \bibliographystyle{ieeenat_fullname}
    \bibliography{main}
}
\input{sec/suppl/0_overview}
\input{sec/suppl/1_method}

\input{sec/suppl/2_results}

\end{document}

%% file: preamble.tex
\newcommand{\Method}{\text{Global Motion Corresponder }}

\usepackage{multirow}
\usepackage{array}
\usepackage{booktabs}
\usepackage{graphicx}
\usepackage{subcaption}

\usepackage{makecell, tabularx}
\usepackage[table,xcdraw, dvipsnames]{xcolor}
\newcolumntype{Y}{>{\centering\arraybackslash}X}
\usepackage{rotating}
\usepackage[export]{adjustbox}

%% file: sec/0_abstract.tex
\begin{abstract}

Existing dynamic scene interpolation methods typically assume that the motion between consecutive timesteps is small enough so that displacements can be locally approximated by linear models. In practice, even slight deviations from this small-motion assumption can cause conventional techniques to fail. In this paper, we introduce \textbf{Global Motion Corresponder} (GMC), a novel approach that robustly handles large motion and achieves smooth transitions. GMC learns unary potential fields that predict SE(3) mappings into a shared canonical space, balancing correspondence, spatial and semantic smoothness, and local rigidity. We demonstrate that our method significantly outperforms existing baselines on 3D scene interpolation when the two states undergo large global motions. Furthermore, our method enables extrapolation capabilities where other baseline methods cannot.

\end{abstract}

%% file: sec/1_intro.tex
\input{figures/interpolation_compare_baseline}

\section{Introduction}
\label{sec:intro}

Dynamic scene interpolation reconstructs continuous motion from two sets of discrete multi-view frames, which is a fundamental challenge in computer vision driven by the increasing demand for photorealistic animation. Recent methods built on point-based representations~\cite{kerbl20233dgaussiansplattingrealtime, papr} are especially appealing as they enable efficient training and real-time rendering. 

The core challenge in scene interpolation with point-based representations lies in establishing reliable correspondences: each point must predict its motion to a corresponding location in another frame. Most existing methods rely on a critical assumption that the motion between adjacent timesteps is small enough that point positions do not change significantly. Under this assumption, determining correspondence reduces to matching points within small local neighborhoods, which has been successfully explored by existing works~\cite{4DGS, Deformable-3DGS, luiten2023dynamic, peng2024papr}. However, this small-motion assumption does not always hold in practice. Many real-world scenarios, such as recording athletes in motion, vehicles on a highway, or any dynamic scene where temporally dense capture is expensive or infeasible, routinely violate this fundamental assumption. 

When motion becomes sufficiently large, dynamic scene interpolation faces a fundamental breakdown. As shown in Figure~\ref{fig:interpolation_compare_baseline}, existing works that rely on the motion locality assumption would fail. This failure stems from the ill-posed nature of point correspondence under a large motion: 1) a point's local neighborhood becomes unreliable as objects move far from their original positions, and 2) globally, a point may have multiple plausible matches, making the point-to-point correspondence ambiguous. As illustrated in Figure~\ref{fig:teaser}, naïvely matching nearest neighbors in these cases results in criss-cross matches, making it unusable for scene interpolation.

We introduce GMC to address this fundamental limitation by learning smooth global correspondences through transformation to a shared canonical space for each timestep. Our key insight is to replace direct point-to-point matching with unary potential fields that predict SE(3) transformations for each timestep. These fields take universal features such as DINO~\cite{caron2021emerging} as input and leverage the inductive bias of MLPs to address the ill-posed nature of global matching. By construction, this approach ensures that semantically similar points move coherently by predicting coherent transformation.

Our approach seamlessly integrates point-based representations, such as 3D Gaussian Splatting~\cite{kerbl20233dgaussiansplattingrealtime} (3DGS) with continuous and queryable SE(3) fields, enabling robust interpolation and extrapolation under large motion. Specifically, we train two unary potential fields alongside two sets of 3D Gaussians, where each field learns SE(3) transformations that map its respective Gaussians into a shared canonical space. By defining the GMC over both Gaussian 3D coordinates and semantic features, our SE(3) fields exhibit smoothness and continuity across both space and semantics. Forward and backward mapping through the canonical space yields dense per-point trajectories for both interpolation and extrapolation.  

We evaluate GMC on \textit{interpolation} using eight scenes from the PAPR In Motion dataset~\cite{peng2024papr} and twelve additional challenging scenes with multi-object interactions and large motions. Results demonstrate that GMC significantly outperforms baseline methods under large motion conditions. Additionally, GMC enables realistic motion \textit{extrapolation}, a capability lacking in existing baselines. We also provide preliminary results showing that our method can not only better reason in the setting of a sparse \textit{temporal} capture but can also improve reconstruction in sparse \textit{spatial} capture.

In summary, our contributions are:
\begin{itemize}
    \item \textbf{Global Motion Corresponder} (GMC), a novel method for 3D scene interpolation under large motion.
    \item Comprehensive experimental results demonstrating GMC's superior performance over prior methods for interpolation on challenging large-motion scenarios.
    \item Further demonstration of GMC's capabilities in enabling extrapolation and improving reconstruction quality under sparse temporal and spatial capture conditions.
\end{itemize}

%% file: figures/interpolation_compare_baseline.tex
\begin{figure*}[h]
\setlength\tabcolsep{1pt}
\footnotesize
\begin{tabularx}{\linewidth}{l@{\hskip 2em}YYYYY@{\hskip 1em}|@{\hskip 1em}YYYYY}
& \cellcolor[HTML]{FFCCC9}{Start} & \multicolumn{3}{c}{\cellcolor[HTML]{ebdbe5}{Intermediate}} & \cellcolor[HTML]{DAE8FC}{End} & \cellcolor[HTML]{FFCCC9}{Start} & \multicolumn{3}{c}{\cellcolor[HTML]{ebdbe5}{Intermediate}} & \cellcolor[HTML]{DAE8FC}{End} \\ 
\rotatebox[origin=c]{90}{\parbox[l]{0.05\textwidth}{\centering \text{PAPR in}\\\text{\;Motion}}}          &   
    \includegraphics[width=\hsize,valign=m]{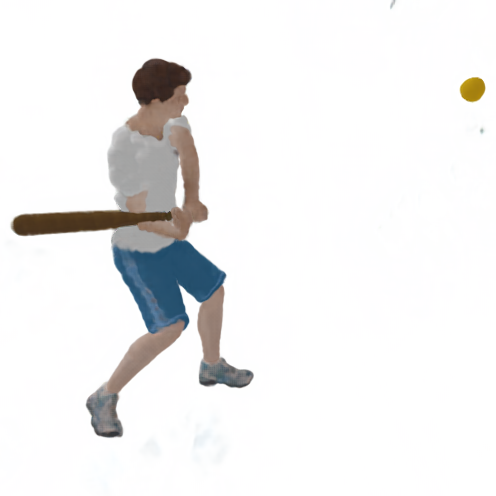}
    &   \includegraphics[width=\hsize,valign=m]{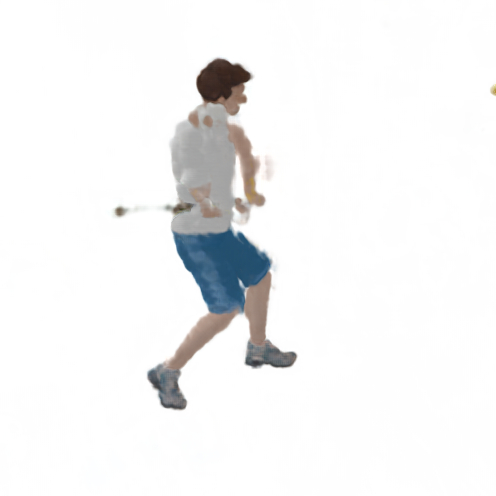}
    &   \includegraphics[width=\hsize,valign=m]{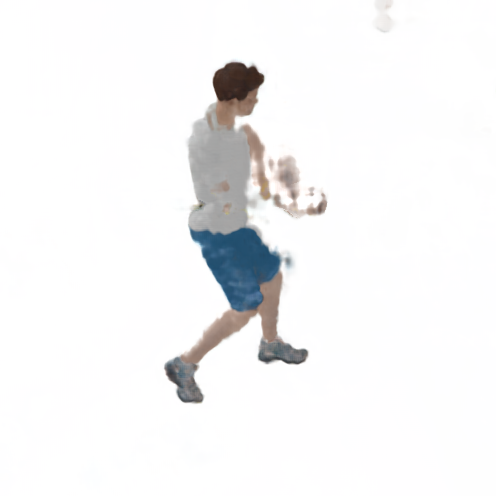}
    &   \includegraphics[width=\hsize,valign=m]{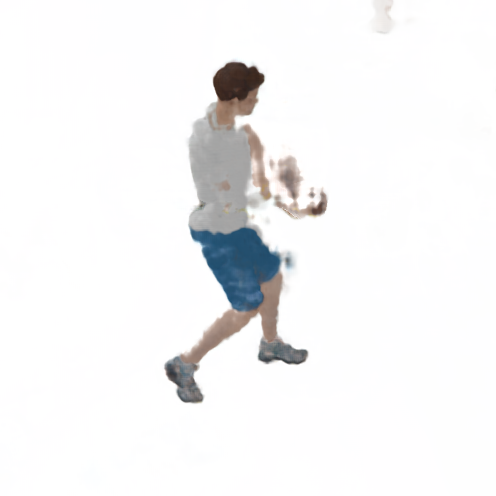}
    &   \includegraphics[width=\hsize,valign=m]{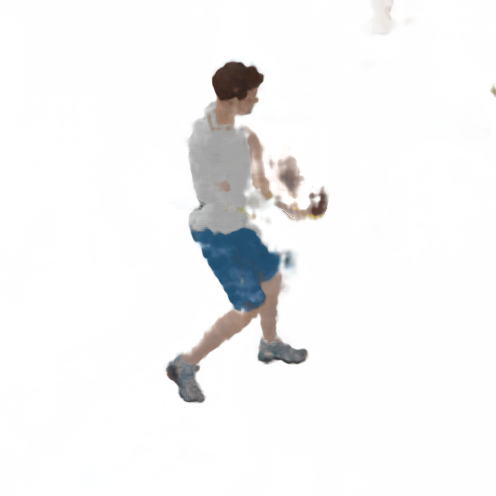}
    & \includegraphics[width=\hsize,valign=m]{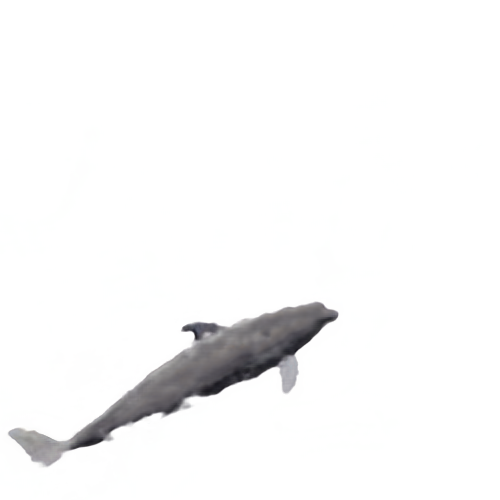}
    & \includegraphics[width=\hsize,valign=m]{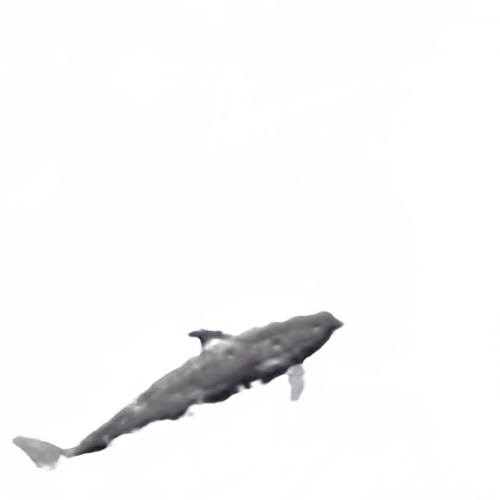}
    & \includegraphics[width=\hsize,valign=m]{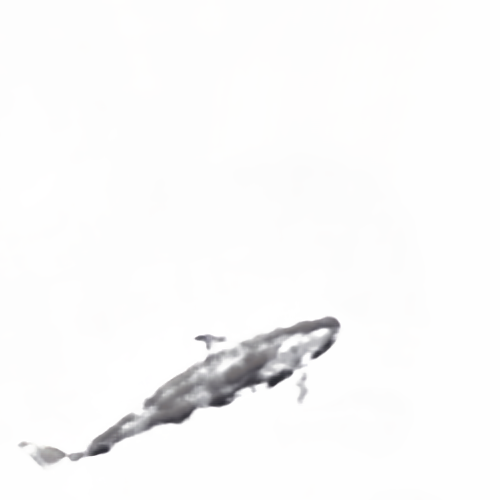}
    & \includegraphics[width=\hsize,valign=m]{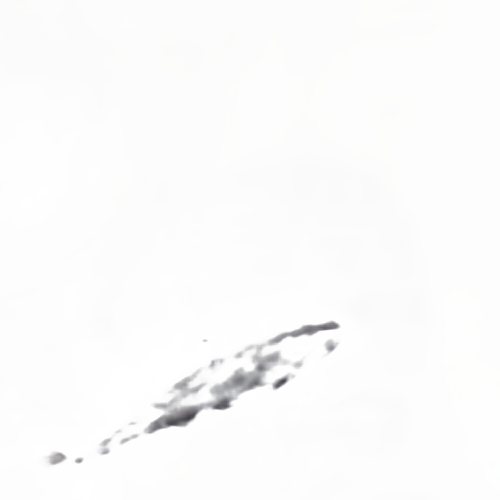}
    & \includegraphics[width=\hsize,valign=m]{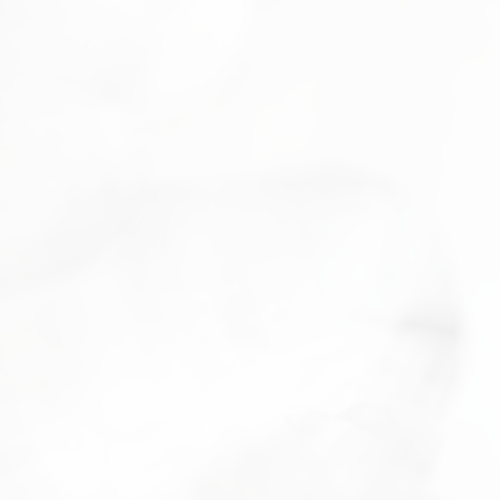}
    
    \\  
    
\rotatebox[origin=c]{90}{\parbox[l]{0.05\textwidth}{\text{\;Dynamic}\\\text{Gaussians}}}          &  
    
    \includegraphics[width=\hsize,valign=m]{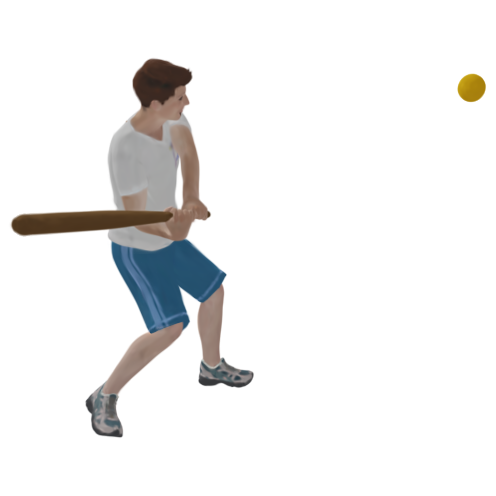}
    &   \includegraphics[width=\hsize,valign=m]{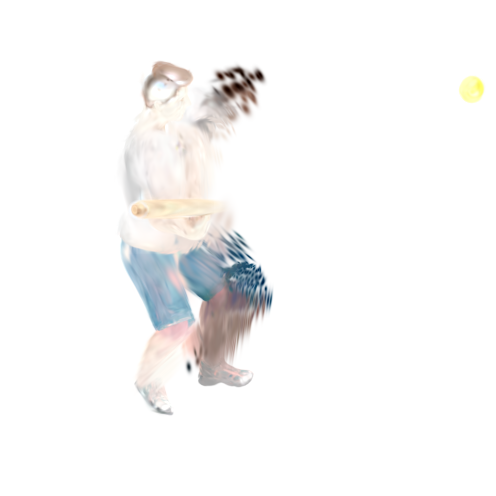}
    &   \includegraphics[width=\hsize,valign=m]{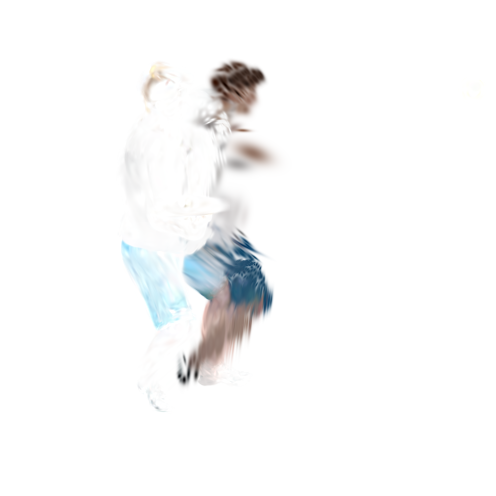}
    &   \includegraphics[width=\hsize,valign=m]{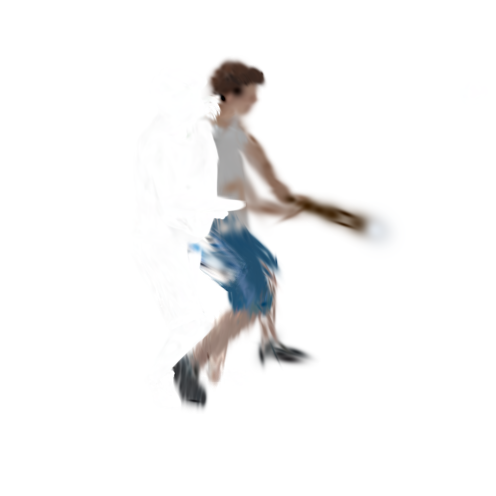}
    &   \includegraphics[width=\hsize,valign=m]{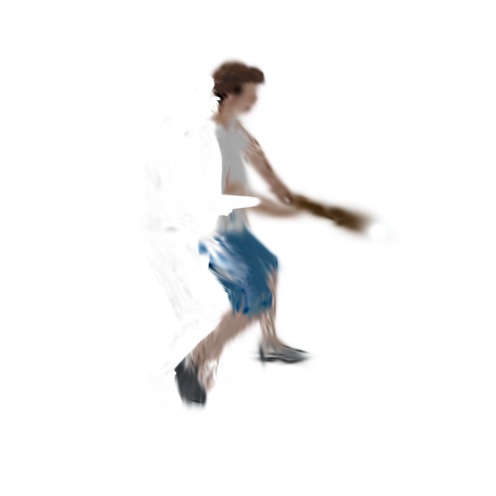}
    &   \includegraphics[width=\hsize,valign=m]{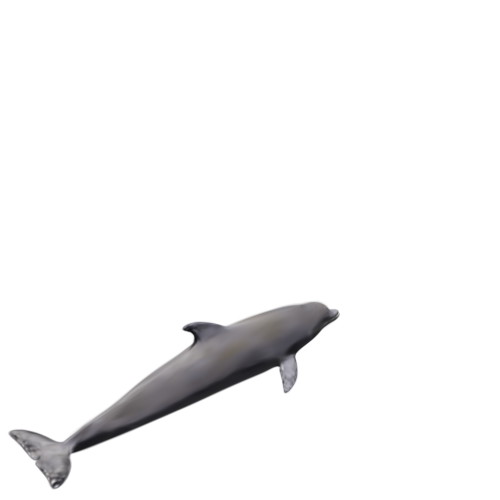}
    &   \includegraphics[width=\hsize,valign=m]{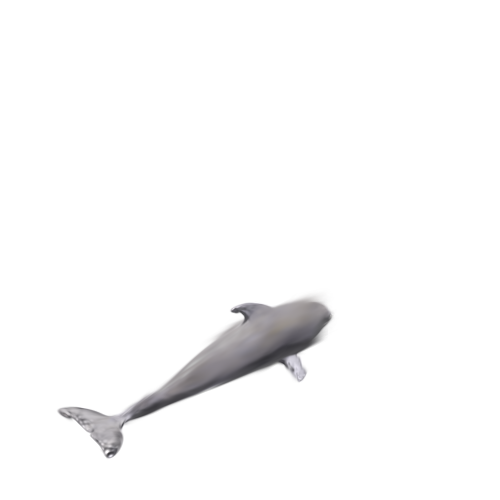}
    &   \includegraphics[width=\hsize,valign=m]{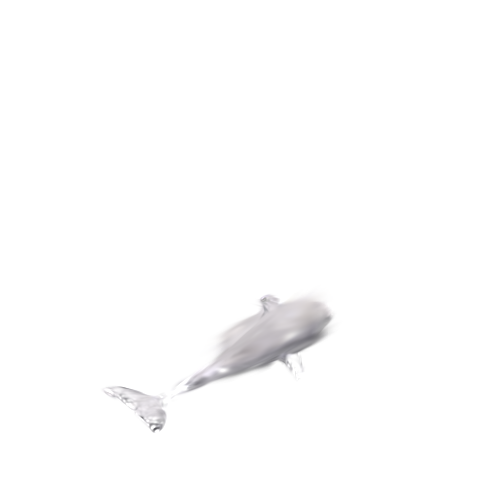}
    &   \includegraphics[width=\hsize,valign=m]{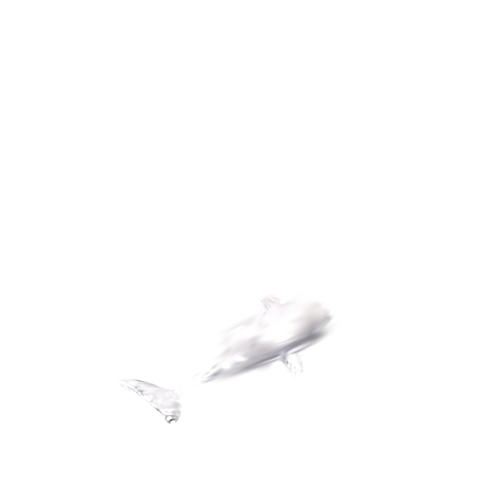}
    &   \includegraphics[width=\hsize,valign=m]{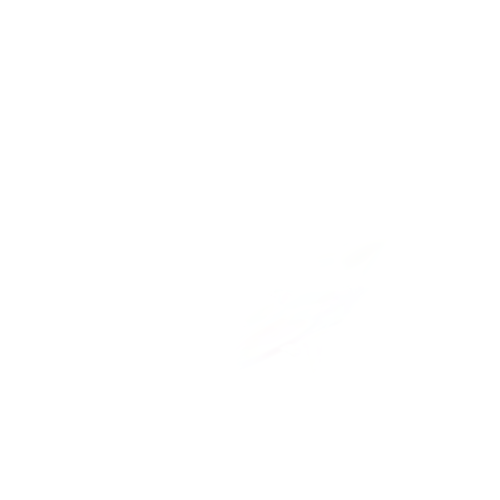}
    \\  

\rotatebox[origin=c]{90}{\parbox[l]{0.05\textwidth}{\text{\;\;\;\;\textbf{Our} }\\\text{\;\textbf{Method}}}}          &   
    \includegraphics[width=\hsize,valign=m]{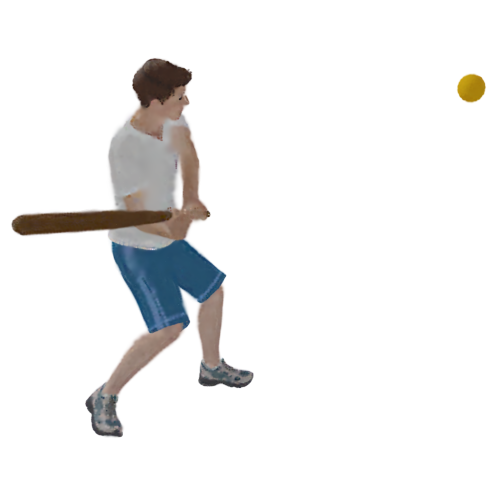}
    &   \includegraphics[width=\hsize,valign=m]{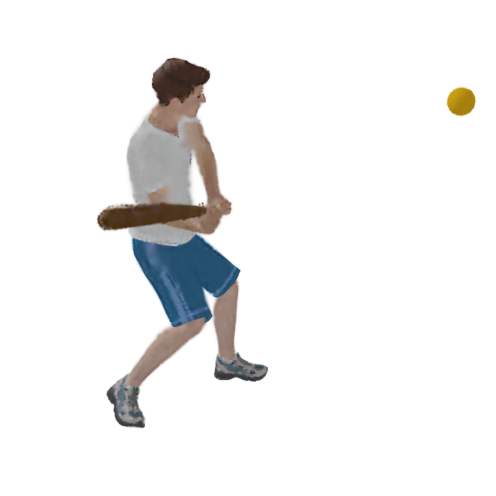}
    &   \includegraphics[width=\hsize,valign=m]{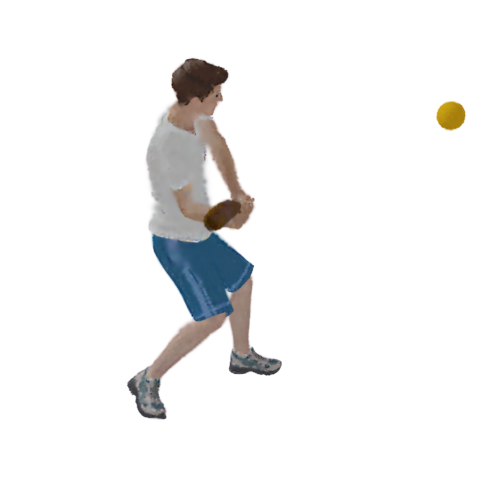}
    &   \includegraphics[width=\hsize,valign=m]{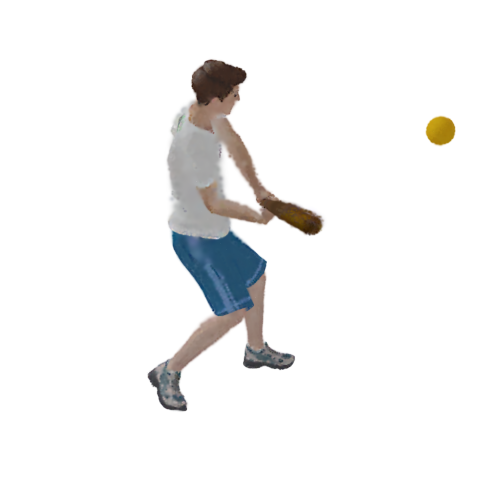}
    &   \includegraphics[width=\hsize,valign=m]{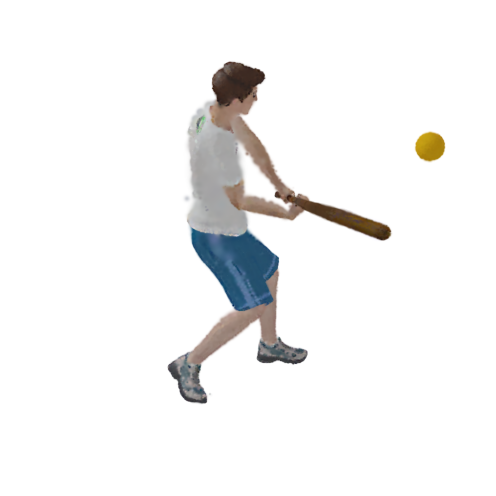}
    & \includegraphics[width=\hsize,valign=m]{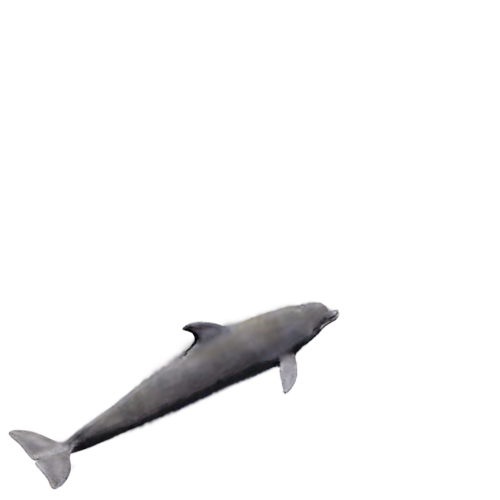}
    & \includegraphics[width=\hsize,valign=m]{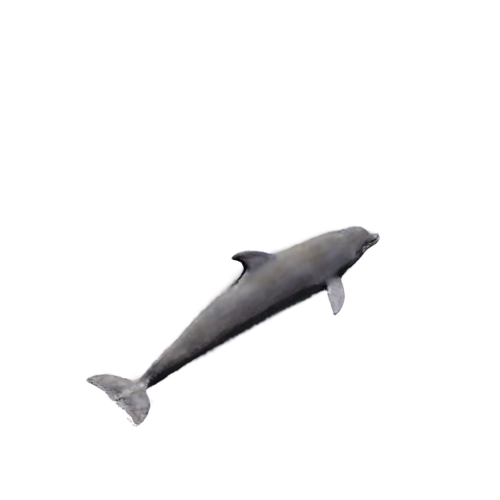}
    & \includegraphics[width=\hsize,valign=m]{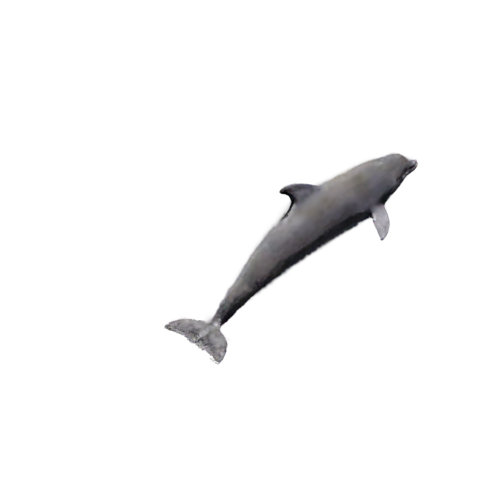}
    & \includegraphics[width=\hsize,valign=m]{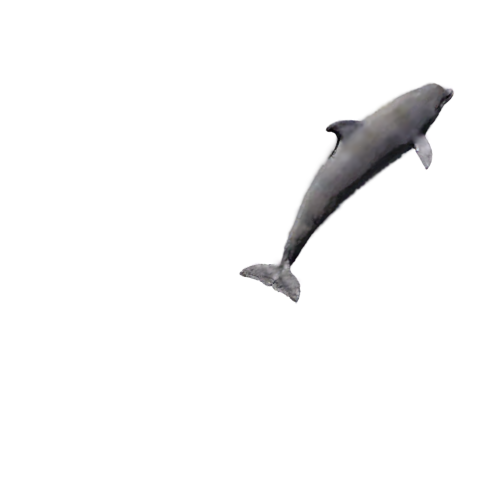}
    & \includegraphics[width=\hsize,valign=m]{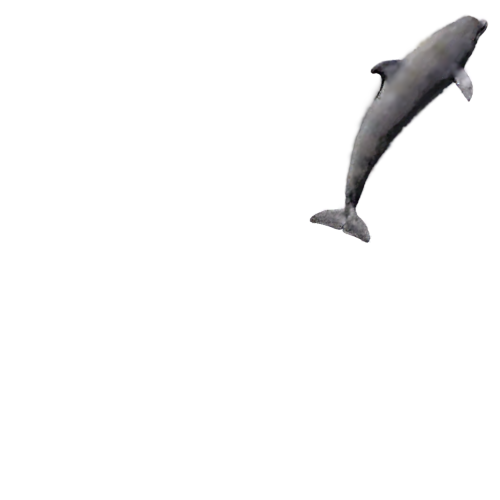}
    \\  
\end{tabularx}
\caption{\textbf{Novel View Synthesis of Scene Interpolation.} 
\texttt{Ball} scene (left): A person swings a bat with dynamic body poses as a ball flies toward them.
\texttt{Dolphin} scene (right): A dolphin jumps, undergoing large and non-rigid motion. 
Only two baseline methods are presented, since the other two (4DGS~\cite{4DGS} and Deformable 3DGS~\cite{Deformable-3DGS}) fail to produce any reasonable rendering on these two scenes.
}
\label{fig:interpolation_compare_baseline}
\end{figure*}

%% file: sec/2_related_work.tex
\section{Related Work}

\input{figures/interpolation_dg_scenes}

3DGS~\cite{kerbl20233dgaussiansplattingrealtime} is a point-based representation that rasterizes small differentiable Gaussians for novel views synthesis of a 3D scene. Unlike volumetric representations like Neural Radiance Fields~\cite{nerf}, which require querying multiple points along each ray, 3DGS represents pixels using only a few Gaussians, achieving faster training and real-time rendering. 

The challenge of dynamic scene interpolation becomes particularly acute when scenes involve complex and large motions across multiple objects. Traditional point cloud interpolation methods~\cite{zheng2023neuralpci, zeng2022ideanet, zhang2024fastpci, Lu2020_PointINet} fail under such conditions. Numerous 3DGS-based approaches~\cite{Lin_2024_CVPR, das2024neuralparametricgaussiansmonocular, duisterhof2024deformgssceneflowhighly, bae2024pergaussianembeddingbaseddeformationdeformable, luiten2023dynamic, DrivingGaussian, Spacetime-Gaussian, Street-Gaussians, 4DGS, Deformable-3DGS} have emerged to tackle dynamic scene interpolation under the small-motion assumption. However, our problem requires both high-quality scene reconstruction and smooth interpolation, demanding robust performance when this assumption breaks down. Current 3DGS-based approaches fall into two categories: deformation fields and iterative refinement.

\paragraph{Scene Interpolation with Deformation Fields.}
These methods use neural deformation fields to predict pre-Gaussian displacement at given timesteps, jointly optimizing deformation parameters alongside 3D Gaussian models. Representative works include 4DGS~\cite{4DGS} and Deformable 3DGS~\cite{Deformable-3DGS}. However, under large motion, correspondence estimation across timesteps becomes fundamentally ill-posed. Hence, the training of the deformation network is very unstable, resulting in poor reconstruction quality and implausible motion estimates.

\paragraph{Scene Interpolation by Iterative Refinement.}
These approaches achieve interpolation by iteratively refining point primitives from previous timesteps. Under small-motion assumptions, point geometry and semantics do not change drastically between timesteps. These methods pre-train 3D scene representations at initial states, then update the point positions and semantics via rendering loss gradients at subsequent timesteps, storing model weights at each timestep. Interpolation proceeds by blending adjacent timestep weights. Notable examples include Dynamic Gaussian~\cite{luiten2023dynamic} and PAPR In Motion~\cite{peng2024papr}. However, under large motions, simple position fine-tuning from previous states proves insufficient, causing these methods to fail.

\paragraph{Visual Features for Correspondence.}
Features extracted from self-supervised vision transformers~\cite{dosovitskiy2021imageworth16x16words, caron2021emerging} serve as powerful visual descriptors that preserve local semantic information. These features excel in downstream tasks such as zero-shot segmentation and semantic correspondence~\cite{amir2022deepvitfeaturesdense}. Zhang et al. ~\cite{zhang2023talefeaturesstablediffusion} demonstrate semantic correspondence using features from large vision models such as DINO~\cite{caron2021emerging} and Stable Diffusion~\cite{stablediff, NEURIPS2021_4b6538a4}. However, while achieving impressive 2D correspondence results, these methods fail to extend to 3D representations or dynamic scenes due to their lack of 3D awareness under complex object motions and occlusion. To address this issue, Zero-Shot 3D Shape Correspondence~\cite{abdelreheem2023zeroshot3dshapecorrespondence} proposes a method for establishing correspondence between 3D shapes without explicit supervision. Although innovative, the method is not specifically designed to capture temporal dynamics. In contrast, our method leverages DINO~\cite{caron2021emerging} features to learn motion-aware point-based correspondences, uniquely enabling both 3D scene interpolation and extrapolation under large motion.

%% file: figures/interpolation_dg_scenes.tex
\begin{figure*}[]
\setlength\tabcolsep{1pt}
\footnotesize
\begin{tabularx}{\linewidth}{l@{\hskip 2em}YYYYY}
& \cellcolor[HTML]{FFCCC9}{Start} & \multicolumn{3}{c}{\cellcolor[HTML]{ebdbe5}{Intermediate}} & \cellcolor[HTML]{DAE8FC}{End} \\
\\
    
\rotatebox[origin=c]{90}{\parbox[l]{0.05\textwidth}{\text{\;{Dynamic} }\\\text{{Gaussians}}}}          &   
    \includegraphics[width=\hsize,valign=m]{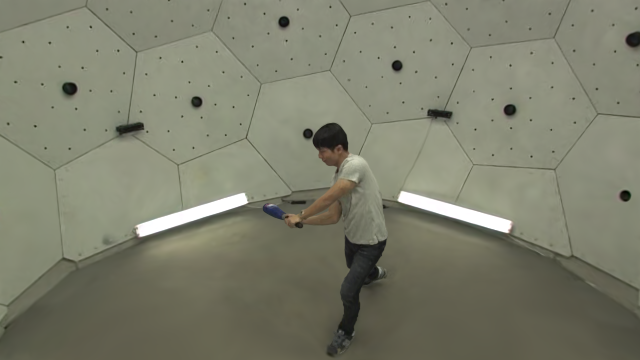}
    &   \includegraphics[width=\hsize,valign=m]{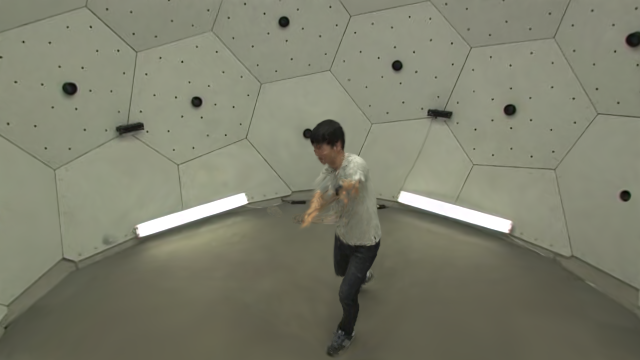}
    &   \includegraphics[width=\hsize,valign=m]{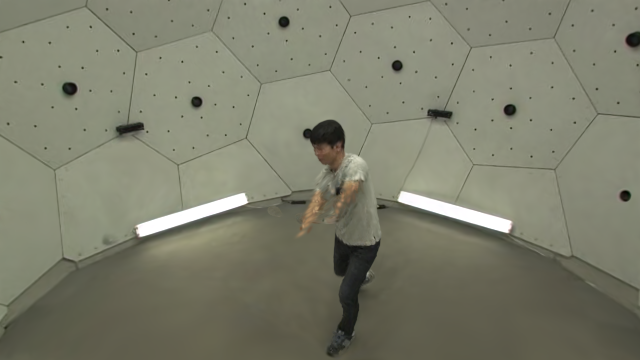}
    &   \includegraphics[width=\hsize,valign=m]{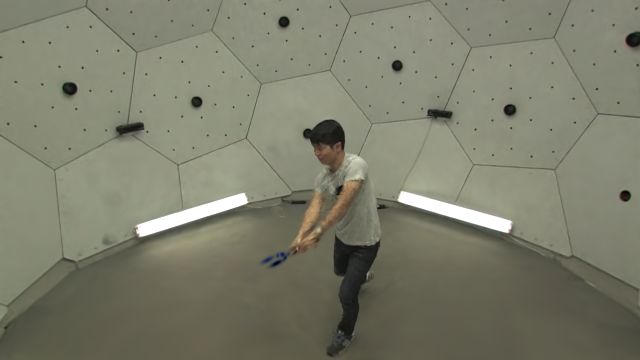}
    &   \includegraphics[width=\hsize,valign=m]{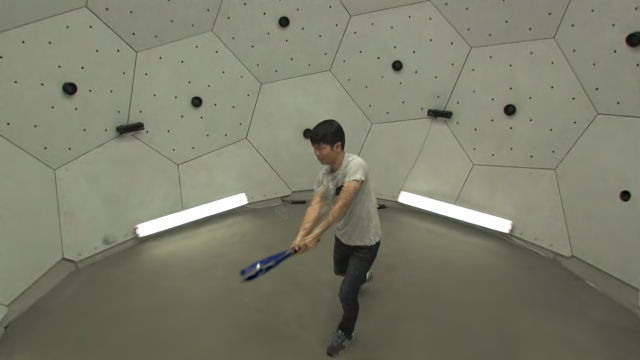}
    \\  

\rotatebox[origin=c]{90}{\parbox[l]{0.05\textwidth}{\text{\;\;\;\;\textbf{Our} }\\\text{\;\textbf{Method}}}}          &   
    \includegraphics[width=\hsize,valign=m]{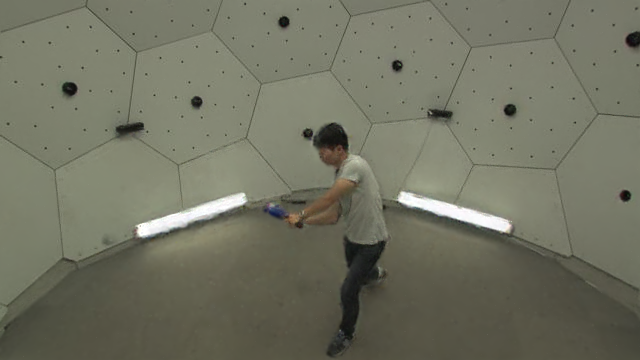}
    &   \includegraphics[width=\hsize,valign=m]{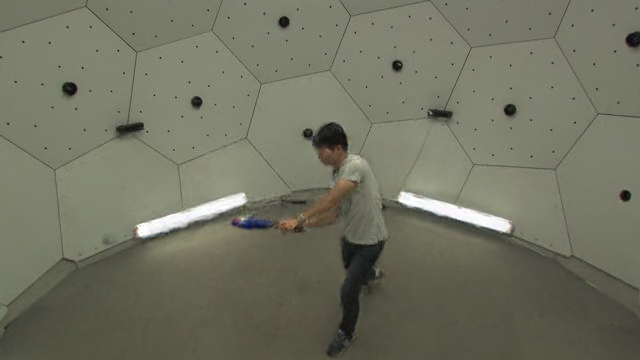}
    &   \includegraphics[width=\hsize,valign=m]{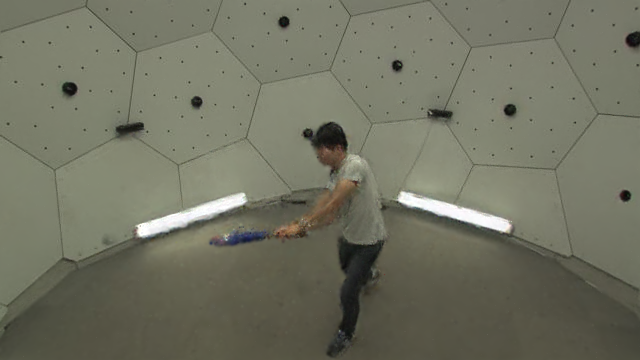}
    &   \includegraphics[width=\hsize,valign=m]{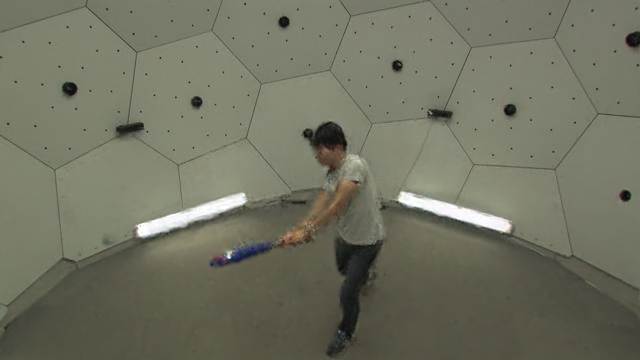}
    &   \includegraphics[width=\hsize,valign=m]{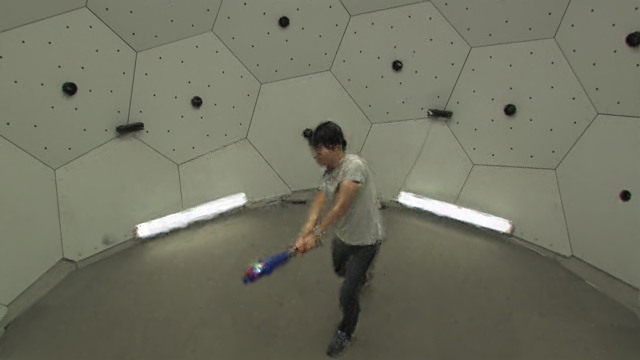}
    \\  
    
\end{tabularx}
\caption{\textbf{Real-World Interpolation.} In the \texttt{Softball} scene~\cite{luiten2023dynamic}, Dynamic Gaussian~\cite{luiten2023dynamic} fails on large inter-frame motion (note the missing baseball bat). The five columns correspond to five timesteps: {0.00, 0.25, 0.50, 0.75, 1.00}.}
\label{fig:interpolation_dg_scenes}
\end{figure*}

%% file: sec/3_method.tex
\section{Method}

In the previous sections, we explained that traditional methods assume spatial positions of 3D points change minimally between adjacent timesteps, which holds only under small, local motion. Since each Gaussian's position remains nearly constant, it is easy to establish correspondences between points from adjacent timesteps by simply identifying the Euclidean nearest neighbors. Using these correspondences, existing methods either refine point positions from previous timesteps or train neural deformation fields to predict per-point displacements.

However, nearest-neighbor correspondence fails when scenes undergo substantial motion. Under this premise, finding correspondences between adjacent timesteps is not as straightforward. When correspondences rely solely on spatial proximity, severe mismatches can occur (Figure~\ref{fig:teaser}), causing existing methods to produce implausible trajectories that violate spatial rigidity and physical constraints (Figure~\ref{fig:interpolation_compare_baseline}). Hence, the core challenge under large motion is \textit{actually} analogous to the challenge of establishing smooth global correspondences between primitives across time. 

\input{figures/3D_illustration_butterfly}
Although establishing global correspondences remains challenging for current works in literature, it is intuitive for human perception. Humans can effortlessly perceive the geometry and physical properties of dynamic 3D scenes, intuitively predicting intermediate or future motions from just two timesteps. This is because humans have inherent \textit{semantic} knowledge about the physical world and understanding of \textit{local spatial rigidity}. The latter property is important, which means that even though the overall motion may be non-rigid, the movement of nearby object parts move similarly.

\subsection{Challenges in Learning Correspondence}
Since a naive nearest-neighbor search in Euclidean space using point coordinates does not work under the large motion setup, we need to find a way to conduct a global search to obtain point-to-point correspondences. For instance, we can try to define a function that allows us to measure the distance between points of two arbitrary Gaussian $g_i$ and $g_j$. One option is to use a weighted sum of color ($\boldsymbol{c}$), semantic ($\boldsymbol{f}$), and spatial ($\boldsymbol{\mu}$) features to define the following distance function:
\begin{align}
     {D}_{i, j}^{\text{ toy}} &= w_c \|\boldsymbol{c}_i - \boldsymbol{c}_j \|_2^2 + w_f \|\boldsymbol{f}_i - \boldsymbol{f}_j\|_2^2 \nonumber \\
     &+ w_{\mu}\|{\boldsymbol{\mu}}_i - {\boldsymbol{\mu}}_j\|_2^2.
     \label{eqn-toy-energy}
\end{align}
It can be reasoned that even this distance function is not enough to establish reliable correspondences. This is because there may be many Gaussian elements within a scene that can have the same color and semantic features. For instance, consider point ``a" in column 2 of Figure~\ref{fig:teaser}: by the distance function defined in Eq.~\ref{eqn-toy-energy}, its closest match in timestep 1 is point ``b", because both of them have the same color and semantic features. In general, it is difficult to find analytical distance metrics that yield the correct correspondences.

Since it is hard to come up with a pre-defined analytical function, we propose to learn these correspondences through a differentiable optimization-based method. Before describing our approach, we first highlight key properties essential for tackling the challenges posed in establishing correspondences and consequently recovering the scene motion under the large-motion assumption:
\begin{enumerate}
    \item The semantics and color of corresponding points should be in agreement. 
    \item Semantically similar points in a neighborhood should move coherently. 
\end{enumerate}

\subsection{Global Motion Corresponder}\label{sec:interpolation-field}

We give a quick overview of our approach to tackle the challenge of finding smooth and accurate global correspondence (Figure~\ref{fig:3D_illustration_butterfly} and~\ref{fig:GMF}). We propose transforming both sets of Gaussians into a learnable shared canonical space where corresponding Gaussians occupy identical spatial locations: 
\begin{equation}
    \underset{\hat{\boldsymbol{\mu}}^{(0)}_i}{\underbrace{\boldsymbol{R}^{(0)}_i \boldsymbol{\mu}_i^{(0)}+ \boldsymbol{t}^{(0)}_i}}
    = \underset{\hat{\boldsymbol{\mu}}^{(1)}_j}{\underbrace{\boldsymbol{R}^{(1)}_j \boldsymbol{\mu}_j^{(1)} + \boldsymbol{t}^{(1)}_{j}}},
    \label{eq:canonical_matching}
\end{equation}
where $\boldsymbol{R}$ and $\boldsymbol{t}$ represent \textit{learnable} point-wise transformations for each timestep. The transformations $\left(\boldsymbol{R}^{(0)}_i, \boldsymbol{t}^{(0)}_i\right)$ and $\left(\boldsymbol{R}^{(1)}_j, \boldsymbol{t}^{(1)}_j\right)$ are obtained from our Unary Potential Fields $\mathcal{F}_{0}$ and $\mathcal{F}_{1}$, which are parameterized as MLPs. The parameters of these MLPs are optimized using our proposed Energy-based loss. Now, we go into the technical details of the design of our Unary Potential Fields and Energy-based loss.

\subsubsection{Unary Potential Field}
We define a single Unary Potential Field $\mathcal{F}$, as function that maps 3D coordinates $\boldsymbol{\mu} \in \mathbb{R}^3$ and feature vectors 
$\tilde{\boldsymbol{f}}\in \mathbb{R}^4$ to SE(3) transformations. It outputs a quaternion representation $\boldsymbol{R} \in \mathbb{R}^{4}$ and translation $\boldsymbol{t} \in \mathbb{R}^{3}$:
\begin{equation}
\mathcal{F}(\tilde{\boldsymbol{f}}, \boldsymbol{\mu}) = (\boldsymbol{R}, \boldsymbol{t}).
\end{equation}
In practice, we use PCA-projected DINO features for $\tilde{\boldsymbol{f}}$ to capture semantic information efficiently, and $\boldsymbol{R}$ is converted to a $3\times3$ rotation matrix when applied in subsequent transformations.

It is a well-known fact that neural networks are able to learn smooth functions such that the output does not change abruptly. Since we wish to keep the predicted motion smooth with respect to both semantic features and spatial positions, we leverage the inductive bias of MLPs for our unary potential. Hence, $\mathcal{F}$ is parameterized as an MLP. Since $\mathcal{F}$ takes features $\tilde{\boldsymbol{f}}$ as input, semantically similar points will naturally predict similar transformations by the inductive bias of the MLP. 

We term these potential fields ``unary" because each timestep requires individual mapping to the canonical space. The same object at different timesteps needs distinct transformations to the shared canonical space. Hence, we need to have separate MLPs per timestep ($\mathcal{F}_0, \mathcal{F}_1$). 

\input{figures/GMF}

\subsubsection{Energy-based Loss}

We now define our energy-based loss used to optimize unary potential fields for global correspondence learning. Concretely, using the means of Gaussians mapped from $\mathcal{G}_0$ and $\mathcal{G}_1$ to the canonical space, we have:
\begin{align}
     E_{i, j} &= w_c \|\boldsymbol{c}_i - \boldsymbol{c}_j \|_2^2 + w_f \|\boldsymbol{f}_i - \boldsymbol{f}_j\|_2^2 \nonumber \\
     &+ w_{\mu}\|\hat{\boldsymbol{\mu}}_i - \hat{\boldsymbol{\mu}}_j\|_2^2,
     \label{eqn:final-energy}
\end{align}
where $\hat{\boldsymbol{\mu}}_i$ and $\hat{\boldsymbol{\mu}}_j$ represent the transformed means in the canonical space. Comparing this term with Eq.~\ref{eqn-toy-energy}, rather than using spatial distance in the original Euclidean space, we are now using spatial distance in the canonical space.

To train the parameters of $\mathcal{F}_0$ and $\mathcal{F}_1$, we define a bidirectional loss using the aforementioned energy term:
\begin{align}
    \label{eq:energy_loss_func}
    \mathcal{L}_{\text{E}} = \sum_{g_i \in \mathcal{G}_0} \min_{g_j \in \mathcal{G}_1} E_{i, j} + \sum_{g_j \in \mathcal{G}_1} \min_{g_i \in \mathcal{G}_0} E_{j, i}.
\end{align}
This loss encourages Gaussians that are similar in color and feature space, and that are close in the canonical space, to be matched, prompting smooth and realistic correspondences. It is important to note two crucial designs: 
\begin{enumerate}
    \item The \textit{bidirectional} loss that operates on both sets of Gaussians $\mathcal{G}_0, \mathcal{G}_1$ is essential. Optimizing the energy term for just one timestep can lead to a scenario similar to column 2 of Figure~\ref{fig:teaser}, where the back of the pencil at timestep 1 does not have any correspondence. The two-way nature of the loss ensures that all points in both timesteps find correspondences.
    \item Color $\boldsymbol{c}$ and features $\boldsymbol{f}$ terms in Eq.~\ref{eqn:final-energy} are crucial. Since unary potential fields are randomly initialized, the contribution of $w_{\mu}\|\hat{\boldsymbol{\mu}}_i - \hat{\boldsymbol{\mu}}_j\|_2^2$ at the beginning of the training is arbitrary. Hence, the color and feature terms help to establish initial ``soft" correspondences.
\end{enumerate}

\subsubsection{Global Motion}
Given learned unary potentials, we can then recover the global motion for Gaussian $g^0_i$ in $\mathcal{G}_0$ with position $\boldsymbol{\mu}^0_i$ by first mapping it to the shared canonical space using its learned unary potential field $\mathcal{F}_0$: $\hat{\boldsymbol{\mu}}^{(0)}_i = \boldsymbol{R}^{(0)}_i \boldsymbol{\mu}^{(0)}_i + \boldsymbol{t}^{(0)}_i$. Next, we can find its corresponding Gaussian $g_j$ in $\mathcal{G}_1$ using the energy term defined in Eq~\ref{eqn:final-energy} in the shared canonical space. We refer to the mean of $g_j$ as $\boldsymbol{\mu}^{(1)}_j$.
Finally, we can see the inverse transformation from its learned unary potential field $\mathcal{F}_1$ (refer Eq.~\ref{eq:canonical_matching}):
\begin{align}
    \boldsymbol{\mu}_{i}^{(0), t=1} &= \left(\boldsymbol{R}_{j}^{{(1)}}\right)^{-1} \left(\boldsymbol{R}^{(0)}_i \boldsymbol{\mu}_i^{(0)} + \boldsymbol{t}^{(0)}_i - \boldsymbol{t}^{(1)}_{j}\right) \nonumber \\
    &= \left(\boldsymbol{R}_{j}^{{(1)}}\right)^{\top} \left(\boldsymbol{R}^{(0)}_i \boldsymbol{\mu}_i^{(0)} + \boldsymbol{t}^{(0)}_i - \boldsymbol{t}^{(1)}_{j}\right).
    \label{eq:transform_1}
\end{align}
The overall global motion for $g^0_i$ is then the displacement vector $|\boldsymbol{\mu}_{i}^{(0), t=1} - \boldsymbol{\mu}_{i}^{(0)}|$.

\input{figures/inter_extra}


\subsection{Training Details}
\paragraph{Local Isometry Loss.}
To further enforce the smoothness and rigidity, we incorporate a local isometry loss applied to the transformations into the canonical frame, inspired by prior works~\cite{stearns2024dynamicgaussianmarblesnovel,luiten2023dynamic, peng2024papr}. For each Gaussian $g_i$, we find its $k$ nearest neighbors $\mathrm{NN}_i$ in the original space and enforce that distances are preserved after transformation with Local Isometry Loss $\mathcal{L}_{\text{iso}} =$
\begin{align}
     \frac{1}{kN} \sum_{g_i \in \mathcal{G}_k} \sum_{g_j \in \mathrm{NN}_i}\left| \| {\boldsymbol{\mu}}_i - {\boldsymbol{\mu}}_j \|_2^2 - \|\hat{\boldsymbol{\mu}}_i - \hat{\boldsymbol{\mu}}_j \|_2^2 \right|,
\end{align}
where $N$ is the total number of Gaussians. This loss penalizes changes in local geometric relationships, promoting locally rigid transformations. In practice, we apply this loss to the transformation into the canonical space, as well as the full transformation between frames $t=0$ and $t=1$ (see \cref{sec:inference-time-motion} for this full transformation).
The total loss for training each GMC combines the energy and the local isometry terms:
\begin{align}
    \mathcal{L} = \mathcal{L}_{\text{E}} + \alpha \mathcal{L}_{\text{iso}},
\end{align}
where $\alpha$ is a weight that balances the two terms. We gradually increase $\alpha$ during training, starting from zero, to allow the model to first establish correspondences before enforcing rigidity.

\paragraph{Joint Refinement.}
After GMC learning, we perform a joint refinement to update both GMC and the Gaussian sets $\mathcal{G}_0$ and $\mathcal{G}_1$ simultaneously. Without loss of generality, at each iteration, we render Gaussians from $\mathcal{G}_0$ at both $t=0$ and $t=1$ (where Gaussians are appropriately transformed into $t=1$ before rendering). Let  $\hat{\boldsymbol{I}}_0$ and $\hat{\boldsymbol{I}}_1$ represent the set of rendered images at timestep $t_0$ and $t_1$ respectively. Similarly, let $\boldsymbol{I}_0$ and $\boldsymbol{I}_1$ represent the set of ground truth images at timestep $t_0$ and $t_1$ respectively. We use the following RGB loss that combines L1 and LPIPS~\cite{lpips}: 
\begin{align}
\label{eq:joint_training_render_loss}
    \mathcal{L}_{\mathrm{render}} = \beta \mathcal{L}_{\mathrm{RGB}}(\boldsymbol{I}_0, \hat{\boldsymbol{I}}_0) +  \mathcal{L}_{\mathrm{RGB}}(\boldsymbol{I}_{1}, \hat{\boldsymbol{I}}_{1}),
\end{align}
where $\beta$ is a weight balancing the rendering contributions from each state. By backpropagating this loss, the parameters of both the Gaussian models and the unary potential fields are refined, thereby improving rendering quality and transformation coherence.

\paragraph{Dropout.}
Notice that a trivial solution to Eq~\ref{eq:canonical_matching} occurs when all Gaussians are mapped to the origin in the canonical space, collapsing to a single point to achieve minimal positional difference. This can happen when the potential fields simply make the outputs satisfy $\boldsymbol{R} \boldsymbol{\mu} = - \boldsymbol{t}$. To avoid this trivial solution, random dropout is applied to the position input $\boldsymbol{\mu}$ during the training.

\subsection{Interpolation and Extrapolation}\label{sec:inference-time-motion}

\input{tables/global-motion-interp}


Eq.~\ref{eq:transform_1} provides relative rotation $\boldsymbol{R}_{\text{r}}=\left(\boldsymbol{R}_{j}^{{(1)}}\right)^{\top}  \boldsymbol{R}_i^{(0)}$ and relative translation $\boldsymbol{t}_{\text{r}}=\left(\boldsymbol{R}_{j}^{{(1)}}\right)^{\top}\left(\boldsymbol{t}_i^{0} - \boldsymbol{t}_{j}^{1}\right)$. Interpolation between $t=0$ and $t=1$ proceeds by interpolating between the identity transformation and this relative transformation. Note that our method is flexible with the interpolation strategy, resulting in diverse motion speeds, and we will demonstrate simple linear interpolation. Specifically, for interpolation parameter $t \in [0,1]$, the interpolated rotation $\boldsymbol{R}_t$ is computed using spherical linear interpolation (SLERP)~\cite{slerp} between the identity matrix and $\boldsymbol{R}_{\text{r}}$, while the interpolated translation $\boldsymbol{t}_t$ employs linear interpolation between zero and $\boldsymbol{t}_{\text{r}}$. For extrapolation, we simply set parameter $t < 0$ or $t > 1$, extending motion beyond observed timesteps.

%% file: figures/3D_illustration_butterfly.tex
\begin{figure*}[t]
    \centering
    \includegraphics[width=\linewidth]{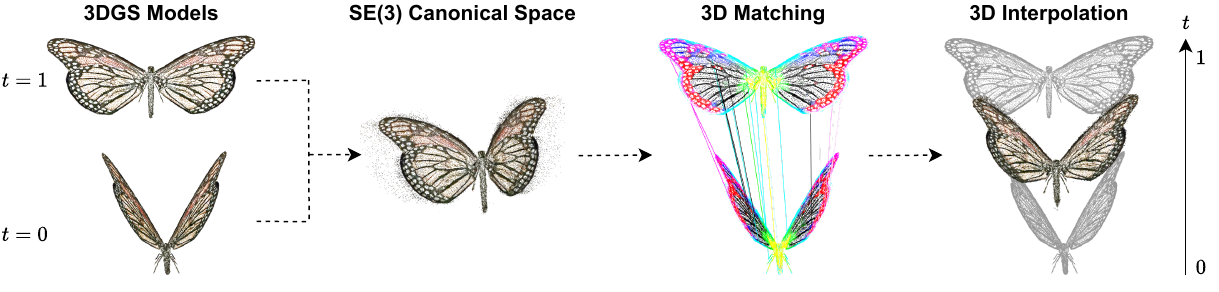}
    \caption{\textbf{Method Overview.} 
    (1) Left: 3DGS models at $t=0$ and $t=1$.
    (2) Middle Left: Alignment in a canonical space through SE(3) transformation.
    (3) Middle Right: 3D matching (colored by PCA-DINO features) is established based on the alignment.
    (4) Right: Continuous 3D interpolation is derived from the 3D transformation and 3D matching.
    }
    \label{fig:3D_illustration_butterfly}
\end{figure*}

%% file: figures/GMF.tex
\begin{figure}[t]
    \centering
    \includegraphics[width=\linewidth]{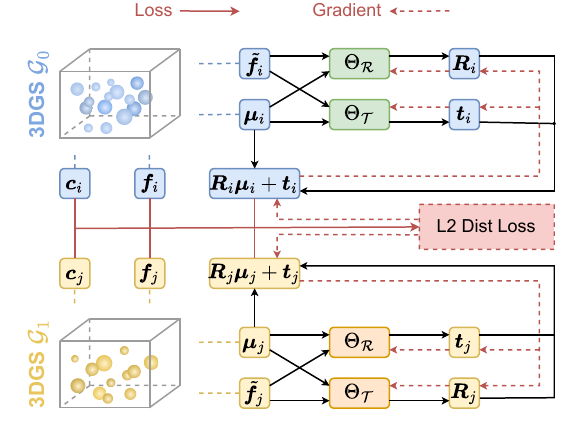}
    \caption{\textbf{GMC Learning Illustration.} 
    Each GMC uses two MLPs ($\Theta_{\mathcal{R}}$ and $\Theta_{\mathcal{T}}$), which input a Gaussian's mean ($\boldsymbol{\mu}$) and PCA-DINO feature ($\tilde{\boldsymbol{f}}$) and output rotation $\boldsymbol{R}$ and transformation $\boldsymbol{t}$, to calculate the new position $\hat{\boldsymbol{\mu}} = \boldsymbol{R} \boldsymbol{\mu} + \boldsymbol{t}$ in the shared canonical space. The energy loss is the L2 distance between nearest neighbors in the joint space of color, feature, and $\hat{\boldsymbol{\mu}}$, with gradient backpropagated to the MLPs. 
    }
    \label{fig:GMF}
\end{figure}

%% file: figures/inter_extra.tex
\begin{figure*}[h]
\centering
\footnotesize
\begin{tabularx}{\linewidth}{YYYYYYYYY}
    \multicolumn{2}{c}{\cellcolor[HTML]{e6b8b5}{Extrapolation}} & \multicolumn{1}{c}{\cellcolor[HTML]{FFCCC9}{Start}} & \multicolumn{3}{c}{\cellcolor[HTML]{ebdbe5}{Interpolation}} & \multicolumn{1}{c}{\cellcolor[HTML]{DAE8FC}{End}} & \multicolumn{2}{c}{\cellcolor[HTML]{cfdcef}{Extrapolation}} 
    
    \\
    \\
    
    \includegraphics[width=\hsize,valign=m]{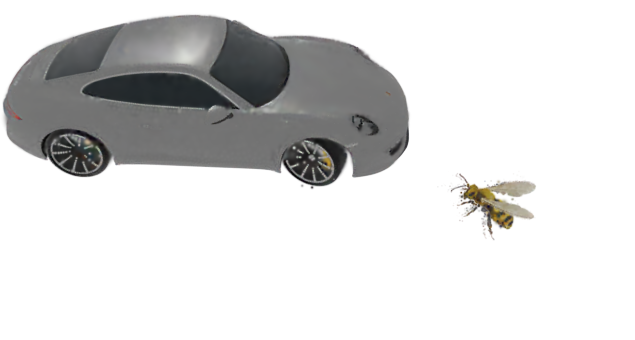}
    &   \includegraphics[width=\hsize,valign=m]{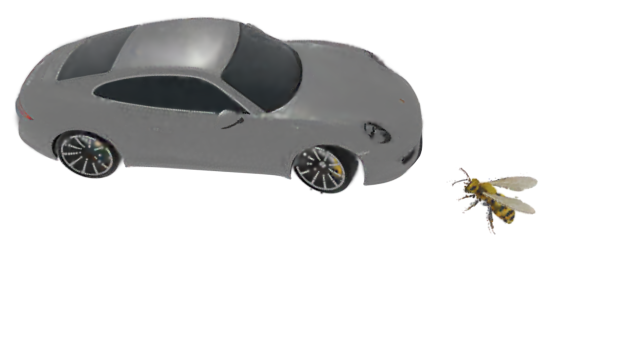}
    &   \includegraphics[width=\hsize,valign=m]{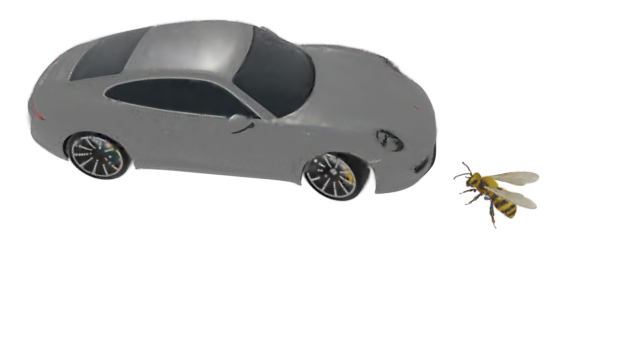}
    &   \includegraphics[width=\hsize,valign=m]{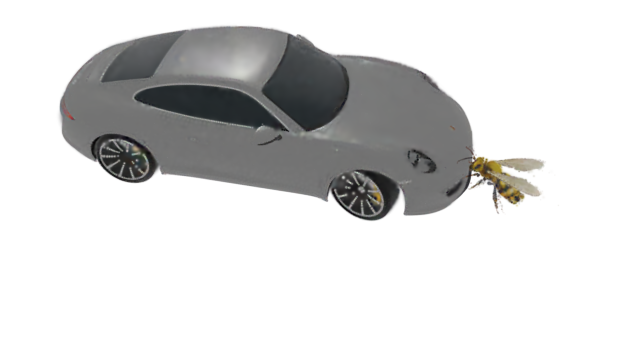}
     &   \includegraphics[width=\hsize,valign=m]{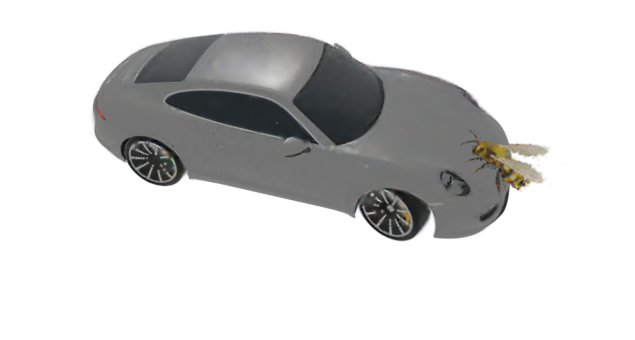}
    &   \includegraphics[width=\hsize,valign=m]{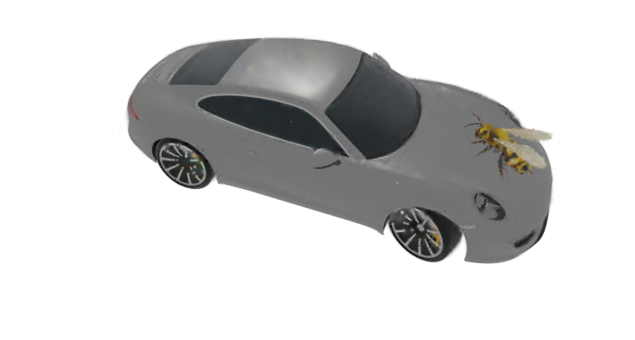}
    &   \includegraphics[width=\hsize,valign=m]{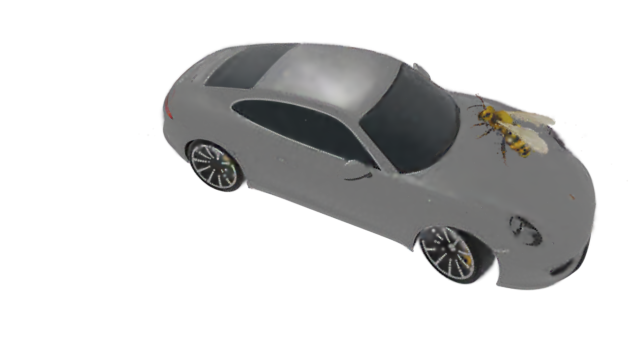}
    &   \includegraphics[width=\hsize,valign=m]{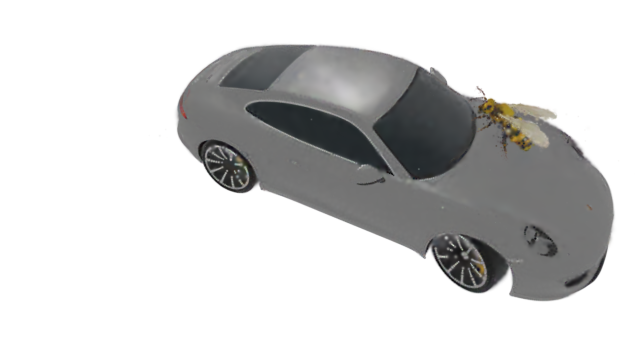}
    &   \includegraphics[width=\hsize,valign=m]{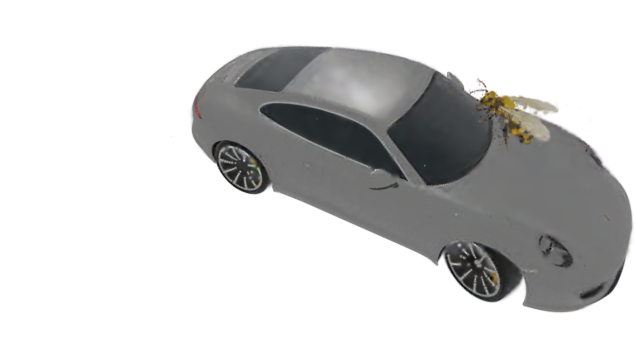}
    \\  
    
    \midrule

    \includegraphics[width=\hsize,valign=m]{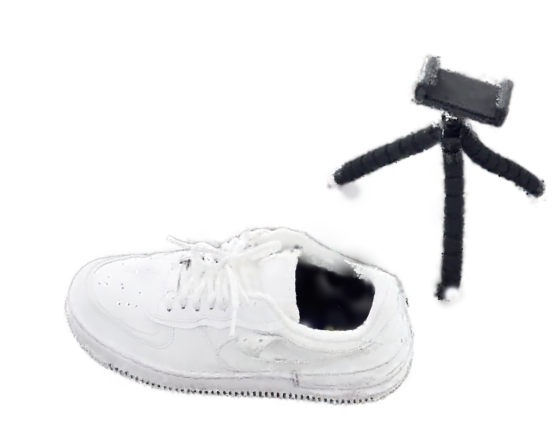}
    &   \includegraphics[width=\hsize,valign=m]{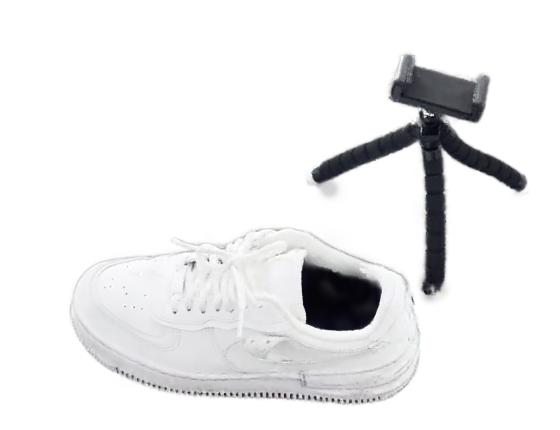}
    &   \includegraphics[width=\hsize,valign=m]{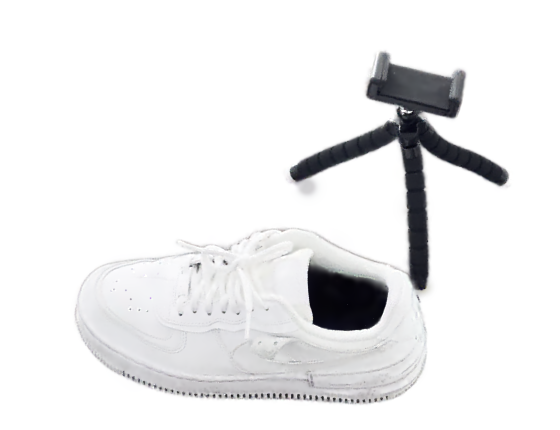}
    &   \includegraphics[width=\hsize,valign=m]{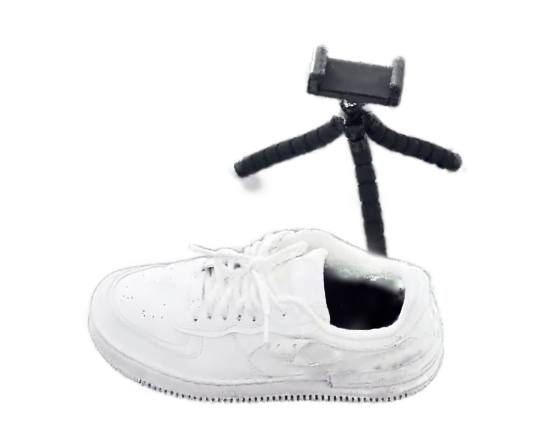}
     &   \includegraphics[width=\hsize,valign=m]{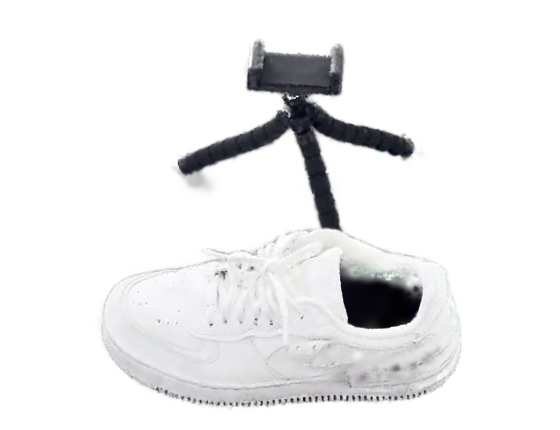}
    &   \includegraphics[width=\hsize,valign=m]{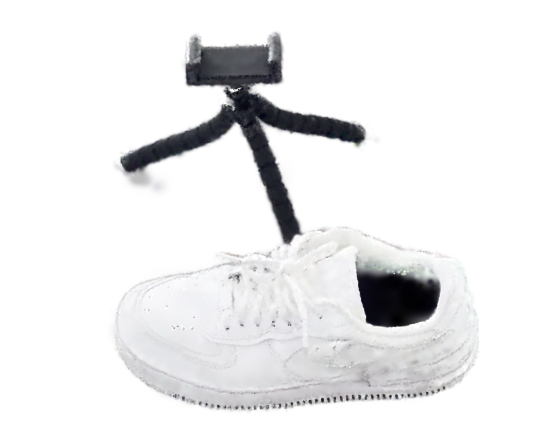}
    &   \includegraphics[width=\hsize,valign=m]{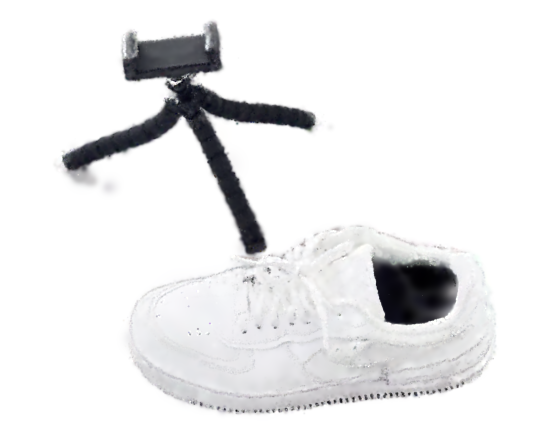}
    &   \includegraphics[width=\hsize,valign=m]{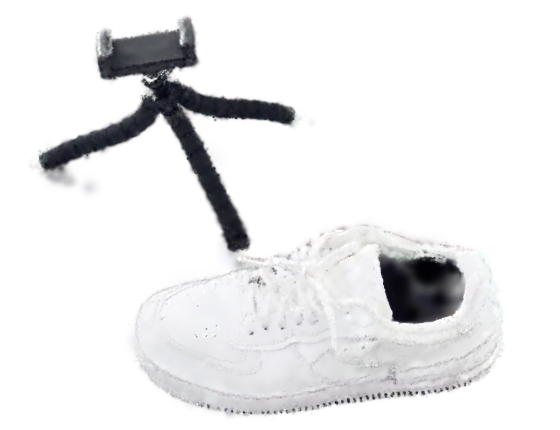}
    &   \includegraphics[width=\hsize,valign=m]{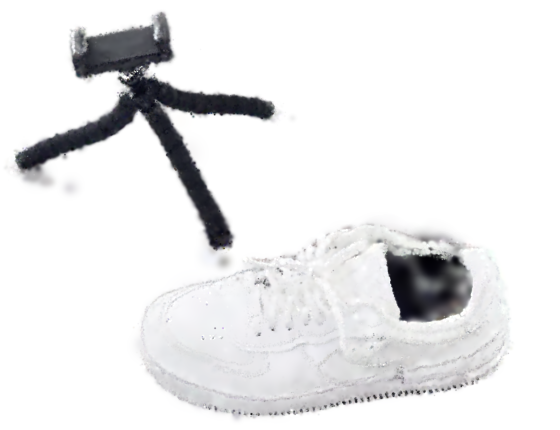}
    \\  

    \midrule

    \includegraphics[width=\hsize,valign=m]{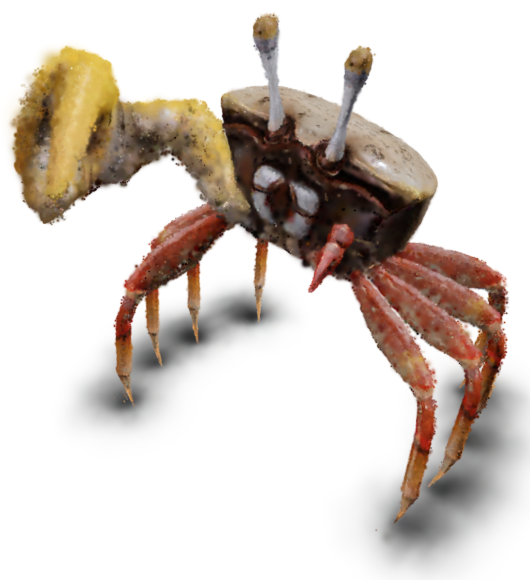}
    &   \includegraphics[width=\hsize,valign=m]{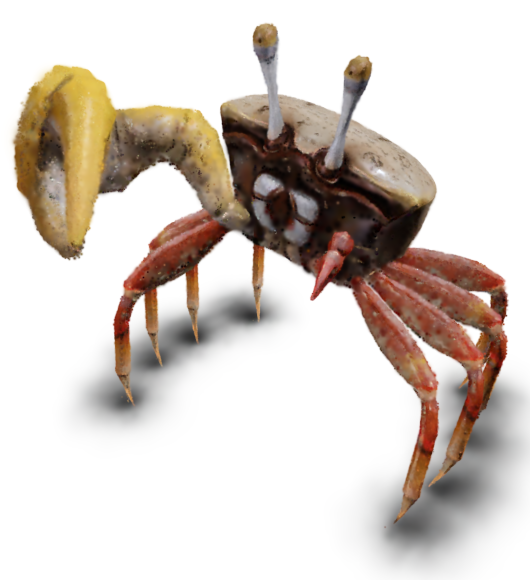}
    &   \includegraphics[width=\hsize,valign=m]{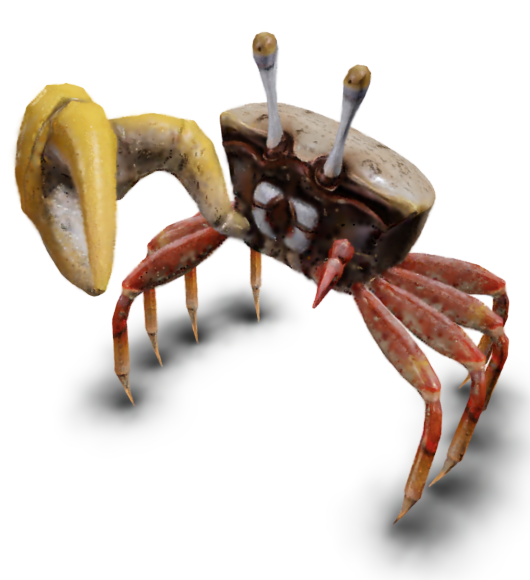}
    &   \includegraphics[width=\hsize,valign=m]{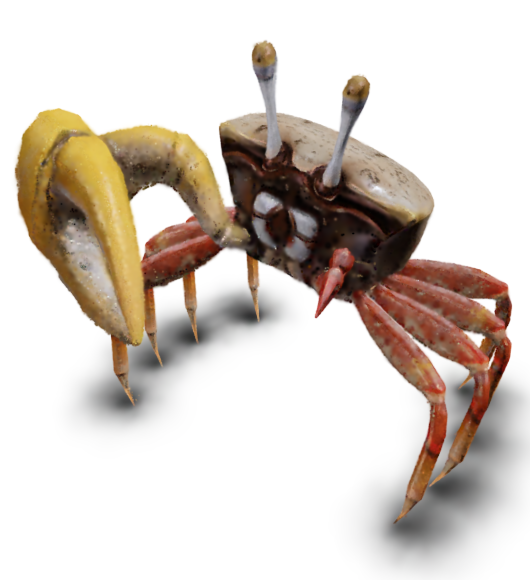}
     &   \includegraphics[width=\hsize,valign=m]{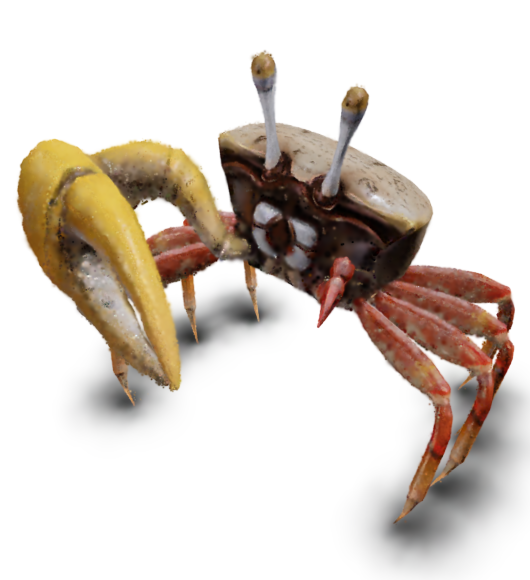}
    &   \includegraphics[width=\hsize,valign=m]{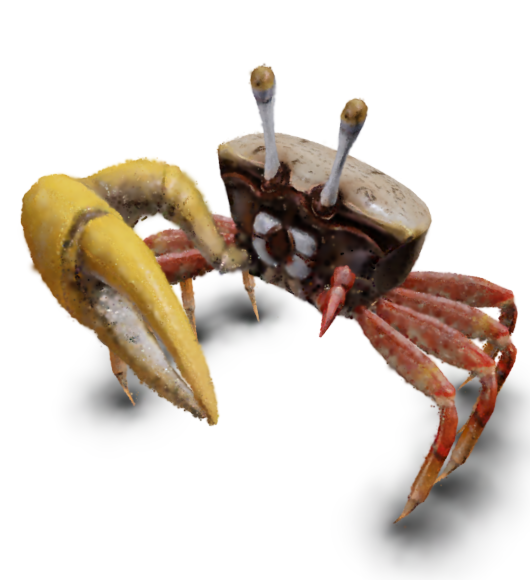}
    &   \includegraphics[width=\hsize,valign=m]{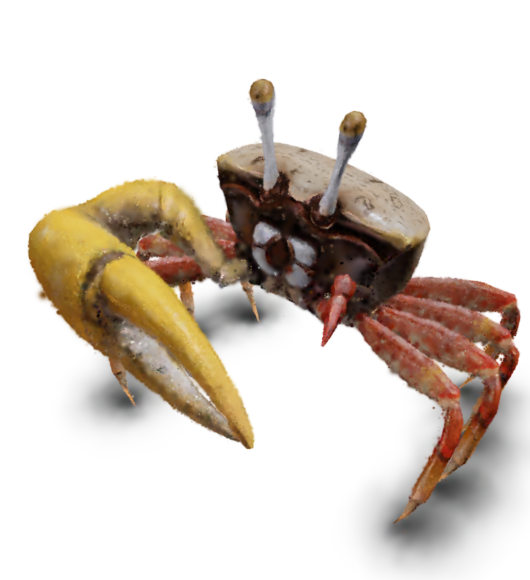}
    &   \includegraphics[width=\hsize,valign=m]{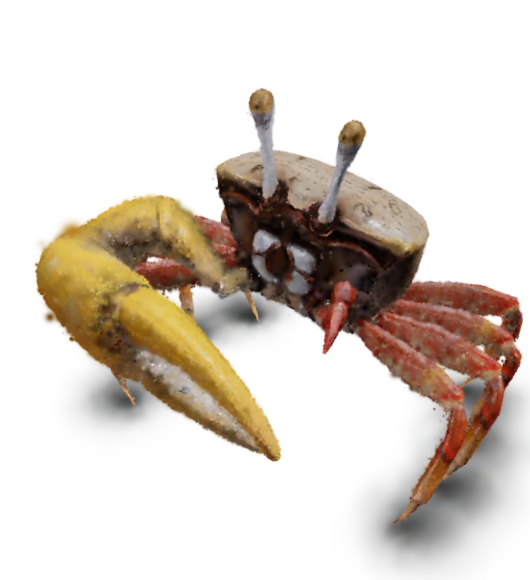}
    &   \includegraphics[width=\hsize,valign=m]{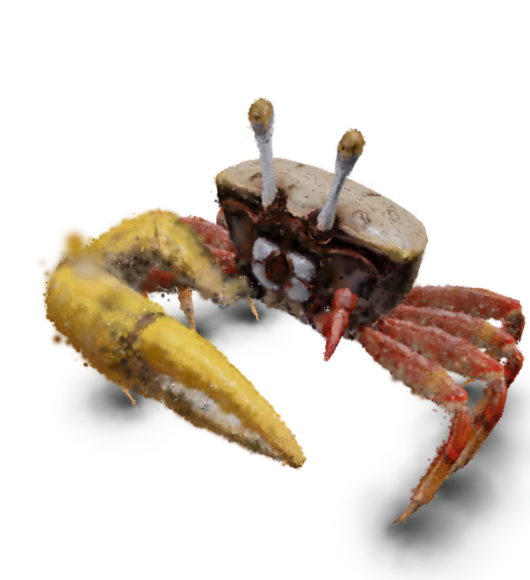}
    \\  

    \midrule

    \includegraphics[width=\hsize,valign=m]{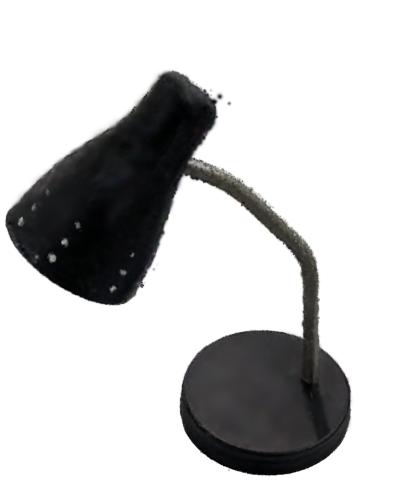}
    &   \includegraphics[width=\hsize,valign=m]{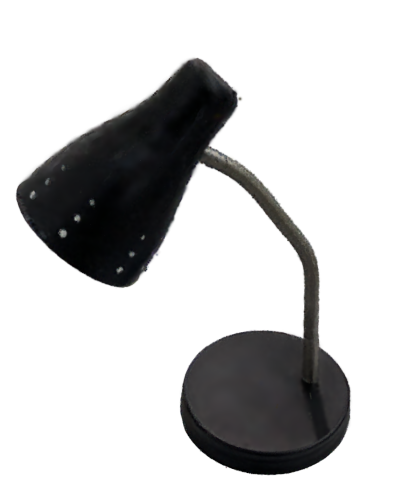}
    &   \includegraphics[width=\hsize,valign=m]{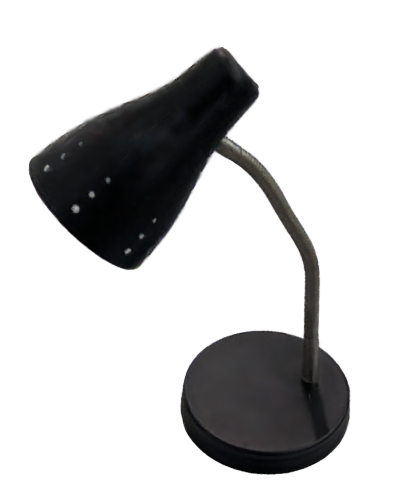}
    &   \includegraphics[width=\hsize,valign=m]{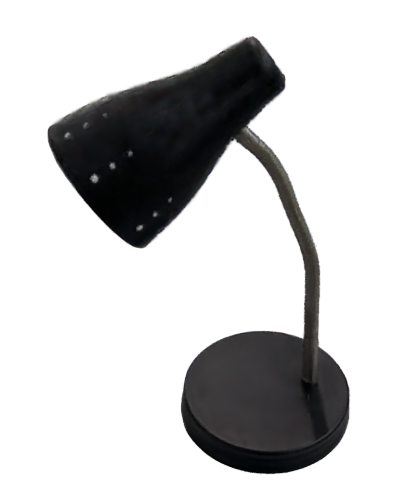}
     &   \includegraphics[width=\hsize,valign=m]{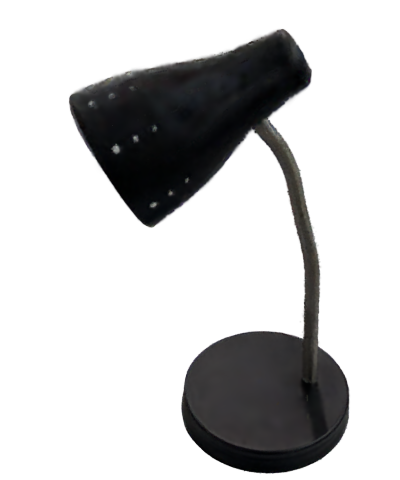}
    &   \includegraphics[width=\hsize,valign=m]{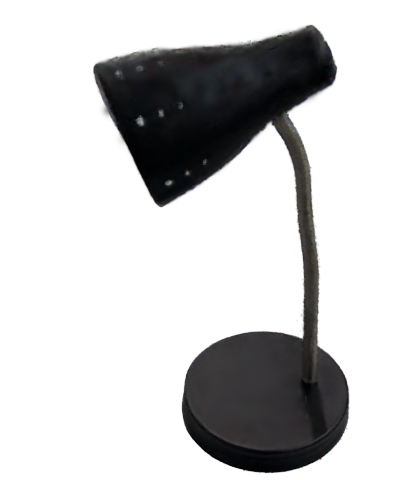}
    &   \includegraphics[width=\hsize,valign=m]{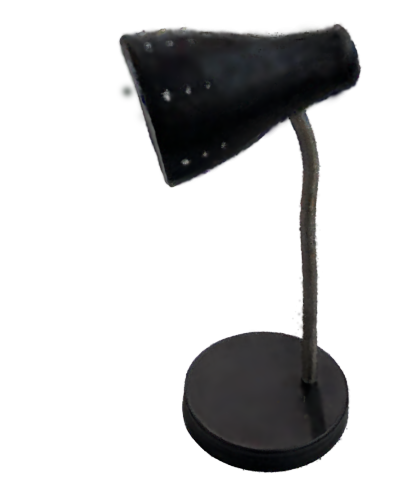}
    &   \includegraphics[width=\hsize,valign=m]{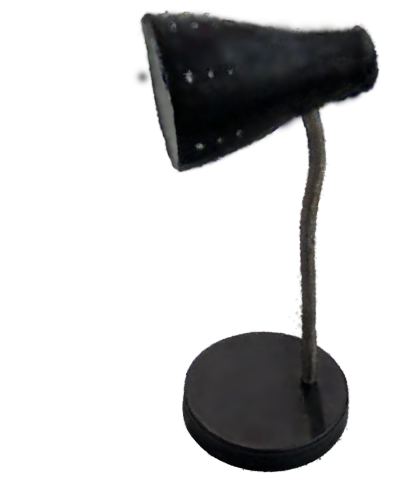}
    &   \includegraphics[width=\hsize,valign=m]{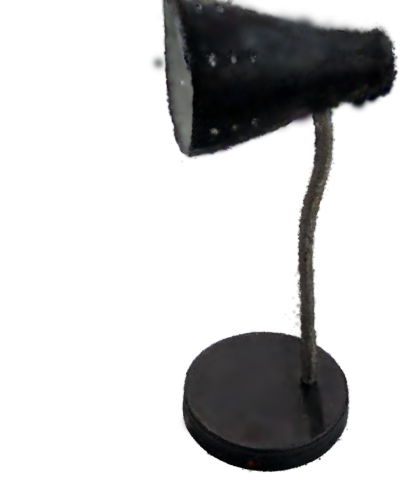}
    \\

\end{tabularx}
\caption{\textbf{Interpolation and Extrapolation Results.} 
(1) Rows 1-2: Multiple objects with global motion (synthetic \texttt{Car} and the real-world \texttt{Shoe}).
(2) Rows 3-4: Single objects with local motion~\cite{peng2024papr} (synthetic \texttt{Crab} and real-world \texttt{Lamp}).
Our method provides plausible interpolation and extrapolation given multi-view input of two states.
}
\label{fig:inter-extra}
\end{figure*}

%% file: tables/global-motion-interp.tex
\begin{table*}[t]
    \centering
    \begin{adjustbox}{max width=\textwidth}
    \begin{tabular}{lcccccccccc}
    \toprule

    Metric & Method & \texttt{Ball} & \texttt{Boat} & \texttt{Butterfly} & \texttt{Car} & \texttt{Dolphin} & \texttt{Knight} & \texttt{Microwave} & \texttt{Seagull} & Avg \\
    
    \midrule

    \multirow{5}{*}{SI-FID $\downarrow$} 
    & 4DGS~\cite{4DGS}  & -    & 328.84      & -    & 460.02      & -    & -    & \underline{258.16} & 294.02      & -      \\
    & Deformable 3DGS~\cite{Deformable-3DGS} & -    & 811.06      & -    & 800.08      & -    & -    & 709.55      & 633.55      & -      \\
    & Dynamic Gaussian~\cite{luiten2023dynamic} & 192.16 & \underline{267.84} & 303.63 & \underline{290.16} & \underline{281.37} & \underline{298.67} & 270.54 & \underline{278.00} & \underline{283.53} \\
    & PAPR in Motion~\cite{peng2024papr} & \underline{154.02} & 288.67      & \underline{269.90} & 339.79      & 297.43      & 360.81      & 289.02      & 291.50      & 315.75  \\
    & \textbf{Ours} & \textbf{109.37} & \textbf{171.32} & \textbf{100.75} & \textbf{170.87} & \textbf{201.01} & \textbf{262.10} & \textbf{210.15} & \textbf{166.94} & \textbf{224.42}     \\

    \midrule
    
    \multirow{5}{*}{SI-EMD $\downarrow$} 
    & 4DGS~\cite{4DGS}  & -          & \textbf{59.90} & -         & 1404.70       & -          & -          & \textbf{45.79} & \textbf{65.63}  & -          \\
    & Deformable 3DGS~\cite{Deformable-3DGS} & -          & 329.22      & -         & 392.66        & -          & -          & 195.4      & 324.78      & -          \\
    & Dynamic Gaussian~\cite{luiten2023dynamic} & 84.19      & 197.36      & 568.30    & \underline{264.01} & 1402.41    & \underline{55.89}  & 106.24     & 224.20      & 521.51     \\
    & PAPR in Motion~\cite{peng2024papr}  & \underline{78.18} & 130.42      & \underline{495.24} & 334.69        & \underline{458.82}  & 116.72     & 164.60     & 141.64      & \underline{246.71} \\
    & \textbf{Ours}  & \textbf{62.62}  & \underline{98.37}  & \textbf{435.32}  & \textbf{192.94}  & \textbf{321.59}  & \textbf{52.96}  & \underline{73.60}  & \underline{125.20} & \textbf{149.38}  \\

    \midrule

    \multirow{5}{*}{SI-MPED $\downarrow$} 
    & 4DGS~\cite{4DGS}  & - & 436.14   & - & 5202.3   & - & - & 173.04   & 489.79   & - \\
    & Deformable 3DGS~\cite{Deformable-3DGS} & - & 153.18   & - & 310.48   & - & - & 155.94   & 122.13   & - \\
    & Dynamic Gaussian~\cite{luiten2023dynamic} & 92.31   & 268.38   & 1036.32  & 223.32   & 2325.01  & \underline{54.96}  & 93.53   & 231.85   & 824.50 \\
    & PAPR in Motion~\cite{peng2024papr}  & \underline{42.83}  & \underline{67.34}   & \underline{285.55}  & \underline{88.42}   & \underline{242.89}  & 67.86   & \underline{32.60}   & \underline{57.50}   & \underline{114.45} \\
    & \textbf{Ours}  & \textbf{14.37}   & \textbf{18.96}   & \textbf{45.74}   & \textbf{35.10}   & \textbf{20.13}   & \textbf{15.31}   & \textbf{13.98}   & \textbf{16.95}   & \textbf{16.47} \\

    \bottomrule
  \end{tabular}
  \end{adjustbox}
  \vspace{2pt}
  \caption{\textbf{Scene Interpolation Evaluation.} This table compares methods on synthetic global-motion scenes (``-'' indicates failure). Our method achieves the lowest SI-FID and SI-MPED scores, indicating smoother interpolation of rendered images and geometry. In most scenes, our method also achieves lower SI-EMD scores, demonstrating better overall geometry fidelity.}
    \label{tab:global_motion_synthetic}
\end{table*}

%% file: sec/4_experiments.tex
\input{tables/local_motion_interp}


\section{Experiments}
\subsection{Dataset}
We evaluate our method on both synthetic and real-world scenes, including single-object and multi-object scenes with local or global motion. Our evaluation dataset comprises: (1) the local-motion dataset from PAPR~\cite{peng2024papr} containing six synthetic scenes and two real-world scenes, (2) real-world scenes from Dynamic Gaussian~\cite{luiten2023dynamic} with frame gaps of 5 timesteps to generate large-motion scenarios, (3) eight synthetic scenes created using the Objaverse objects~\cite{deitke2022objaverseuniverseannotated3d}, including three single-object scenes with large global motion and pose changes, and five multi-object scenes with global motion exhibiting rigid or non-rigid behavior, and (4) three manually captured real-world scenes, including one single-object and two multi-object scenes with global motion.

\subsection{Motion Interpolation}
\paragraph{Implementation Details.}
For the 3DGS pre-training, we build upon Feature 3DGS~\cite{fgs} codebase and incorporate gsplat~\cite{gsplat} for accelerated rasterization. For each state, we train for 30k iterations. Before the MLPs training, normalization scalars for the DINO, PCA-DINO, RGB, and Gaussian means are pre-calculated from the first 3DGS model. These normalizations are applied to both models during the query and database construction. The weights are set as
$w_{c}=1$,
$w_{f}=10$,
$w_{\mu}=10$,
$k=256$,
$\beta=1$,
with $\alpha$ starting at $0$ and linearly increasing to $10$ over 10k iterations. 
The four MLPs are trained for 20k iterations, followed by 20k joint refinement iterations with enabled pruning and densification for both 3DGS models. More details can be found in the supplementary material.\\

\paragraph{Scene Interpolation Metrics.}
When objects undergo large motions, determining the motion trajectories is an ill-posed problem, as there can be infinitely many paths for an object to transit from one state to another. Since we do not have access to a single ``ground truth" trajectory, we cannot define metrics that measure the \textit{correctness} of the predicted trajectory. One way to evaluate the trajectory is by the smoothness of interpolation.
PAPR in Motion~\cite{peng2024papr} proposed various such metrics to measure the \textit{smoothness} of the overall interpolation by using only start and end state information. Specifically, rendering quality is evaluated using Scene Interpolation Fréchet inception distance~\cite{fid} (SI-FID), while geometry quality is assessed using Scene Interpolation Earth Mover’s Distance (SI-EMD). Furthermore, we propose using Scene Interpolation Multiscale Potential Energy Discrepancy~\cite{mped} (SI-MPED) to measure the local geometry preservation (details in supplementary material). To achieve a low SI-MPED value, a method needs to predict the correct end-state while maintaining good local geometry. Therefore, a high SI-MPED indicates either poor motion learning or loss of local geometry, or both. Note that SI-EMD and SI-MPED results are reported in units of $10^{-3}$ in this paper, and for all three metrics, lower scores indicate better quality.\\

\paragraph{Note about Baselines.}
Methods that use deformation fields, namely 4DGS \cite{4DGS} and Deformable 3DGS \cite{Deformable-3DGS}, jointly optimize over the parameters of the deformation field and the Gaussian model. For large motion, this problem is ill-posed, and thus the training of these methods is very unstable. Hence, in tables and figures, we do not report results where the methods fail to train.\\

\paragraph{Quantitative Results.}
Evaluation results of interpolation on synthetic scenes are reported in Table~\ref{tab:global_motion_synthetic} and Table~\ref{tab:main_local_motion_interp}, and on real-world scenes in Table~\ref{tab:global_motion_real}. For both synthetic and real-world scenes, our method outperforms the baselines by a significant margin, indicating better geometry and rendering quality during interpolation. 
\input{figures/interpolation_real}
\paragraph{Qualitative Results.}
Qualitative results are shown in Figure~\ref{fig:interpolation_compare_baseline}, Figure~\ref{fig:interpolation_real}, Figure~\ref{fig:interpolation_dg_scenes} and Figure~\ref{fig:inter-extra}. In Figure~\ref{fig:interpolation_compare_baseline}, our method is compared with the baseline methods PAPR in Motion~\cite{peng2024papr} and Dynamic Gaussian~\cite{luiten2023dynamic} on novel-view synthesis during interpolation for synthetic scenes. 
Consider the \texttt{Dolphin} scene in Figure~\ref{fig:interpolation_compare_baseline}, where the baselines fail because they update the point positions locally and are not able to correspond to true positions that are far away. Figure~\ref{fig:interpolation_real} shows the results of novel-view synthesis during interpolation of the real-world scene \texttt{Box}. While the interpolation given by the baseline methods is suboptimal, our method correctly identifies such complex motion and produces feasible interpolation of the lid while moving the box body. Furthermore, Figure~\ref{fig:interpolation_dg_scenes} shows that our method is able to produce plausible interpolation in real-world scenes when the bat has large motion between start and end states, while the bat disappears during the interpolation for Dynamic Gaussian~\cite{luiten2023dynamic}. Figure~\ref{fig:inter-extra} presents qualitative results on both interpolation and extrapolation. Since neither baseline method can perform extrapolation, we focus on our method, demonstrating that it performs well on synthetic and real-world data, single and multiple objects, and both local and global motion.

\input{tables/real-world_global-interp}

\subsection{Ablation Study}

Each component is removed separately from the full model, and the average resulting metrics are reported in Table~\ref{tab:ablation}. Specifically, we study the importance of (1) the DINO input to the Motion Network, (2) the position (Gaussian mean) input to the Motion Network, (3) local distance preservation loss, and (4) the appearance refinement stage. Table~\ref{tab:ablation} shows that the full model achieves the best SI-FID, SI-EMD, and SI-MPED.

\input{tables/ablation}

%% file: tables/local_motion_interp.tex
\begin{table}[t]
    \centering
    \begin{adjustbox}{max width=\linewidth}
    \begin{tabular}{lccc}
    \toprule
    Method & SI-FID $\downarrow$ & SI-EMD $\downarrow$ & SI-MPED $\downarrow$ \\
    \midrule
    4DGS~\cite{4DGS}  & 150.42 & 15.35 & 156.64  \\
    Deformable 3DGS~\cite{Deformable-3DGS} & 404.46 & 27.35 & 40.38   \\
    Dynamic Gaussian~\cite{luiten2023dynamic} & 228.07 & 64.27 & 114.91  \\
    PAPR in Motion~\cite{peng2024papr}   & \underline{117.94} & \underline{12.99} & \underline{9.38} \\
    \textbf{Ours}  & \textbf{112.62} & \textbf{12.96} & \textbf{9.27}  \\

    \bottomrule
  \end{tabular}
  \end{adjustbox}
  \vspace{2pt}
  \caption{\textbf{Synthetic Local-Motion Evaluation.} Comparison of average metrics on synthetic local-motions scnees~\cite{peng2024papr} against baselines.}

    \label{tab:main_local_motion_interp}
\end{table}

%% file: figures/interpolation_real.tex
\begin{figure}[t]
\setlength\tabcolsep{1pt}
\footnotesize
\begin{tabularx}{\linewidth}{l@{\hskip 2em}YYYYY}
& \cellcolor[HTML]{FFCCC9}{Start} & \multicolumn{3}{c}{\cellcolor[HTML]{ebdbe5}{Intermediate}} & \cellcolor[HTML]{DAE8FC}{End} \\ \\

\rotatebox[origin=c]{90}{\parbox[l]{0.05\textwidth}{\text{} \\ \text{\;\;\;\;\;4DGS}}}
&  
    \includegraphics[width=\hsize,valign=m]{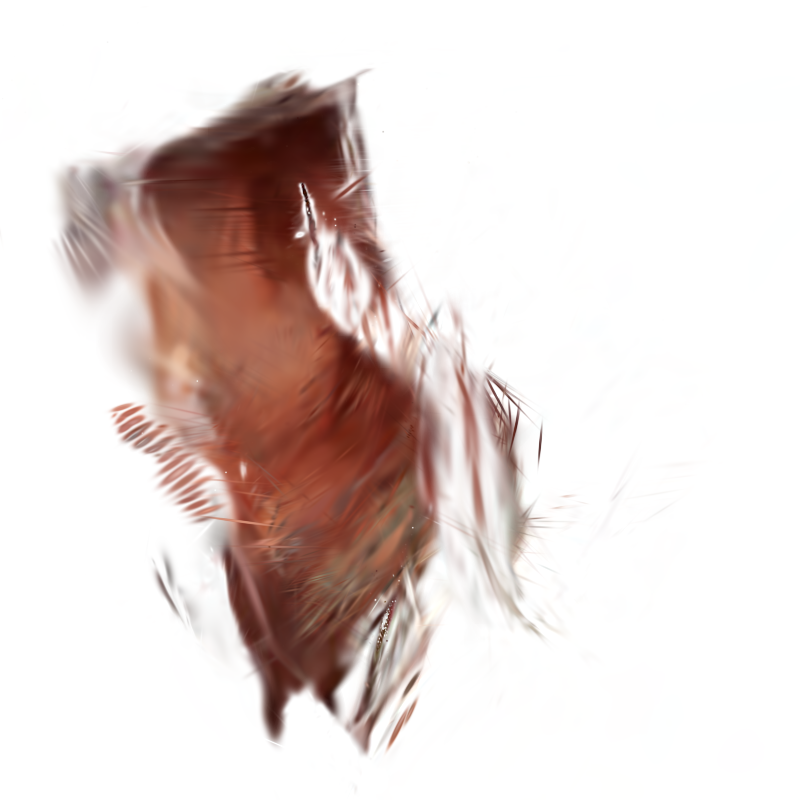}
    &   \includegraphics[width=\hsize,valign=m]{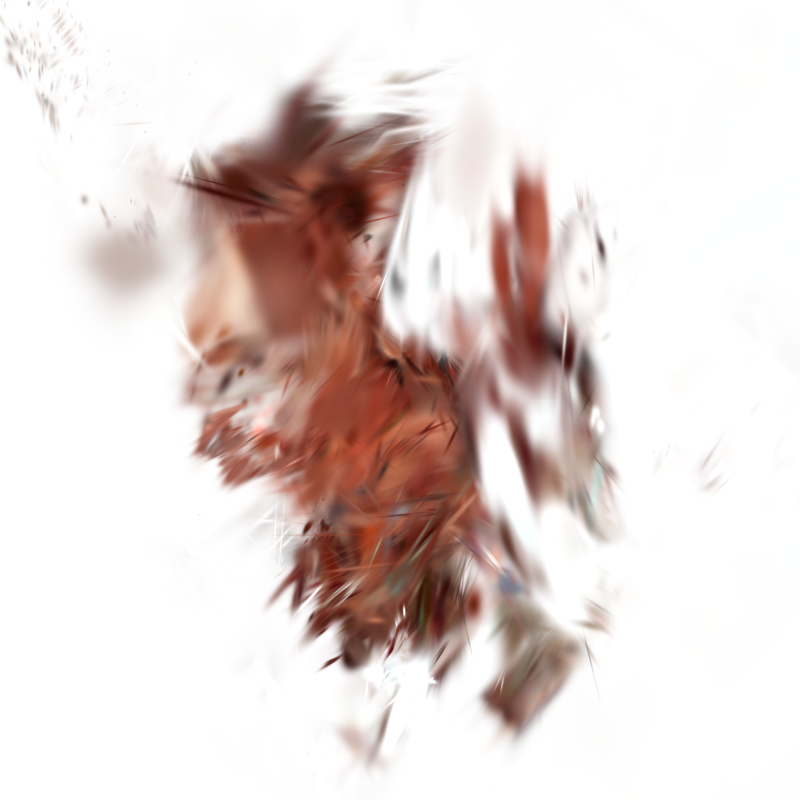}
    &   \includegraphics[width=\hsize,valign=m]{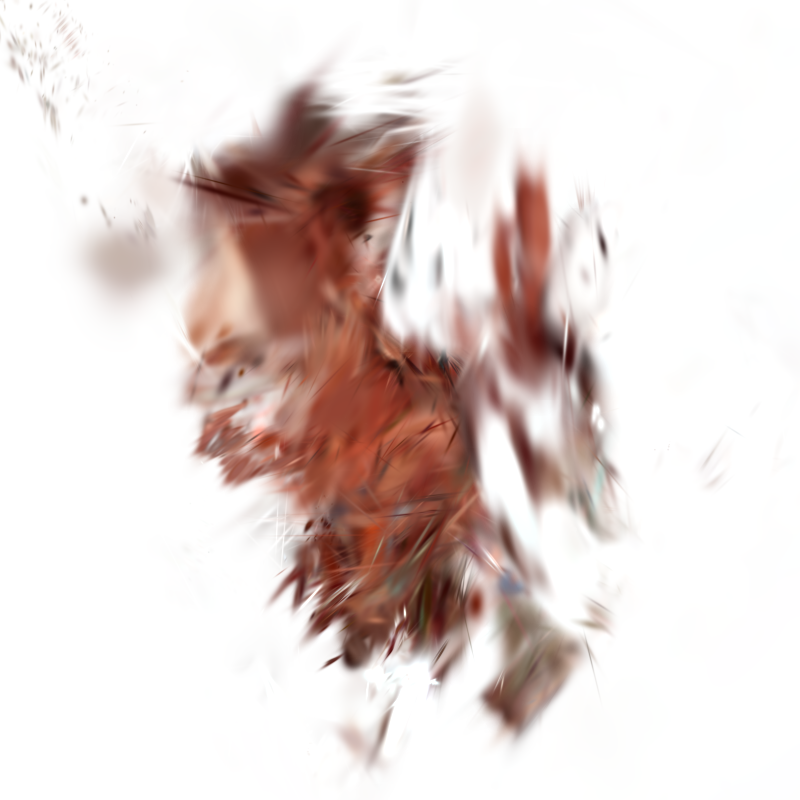}
    &   \includegraphics[width=\hsize,valign=m]{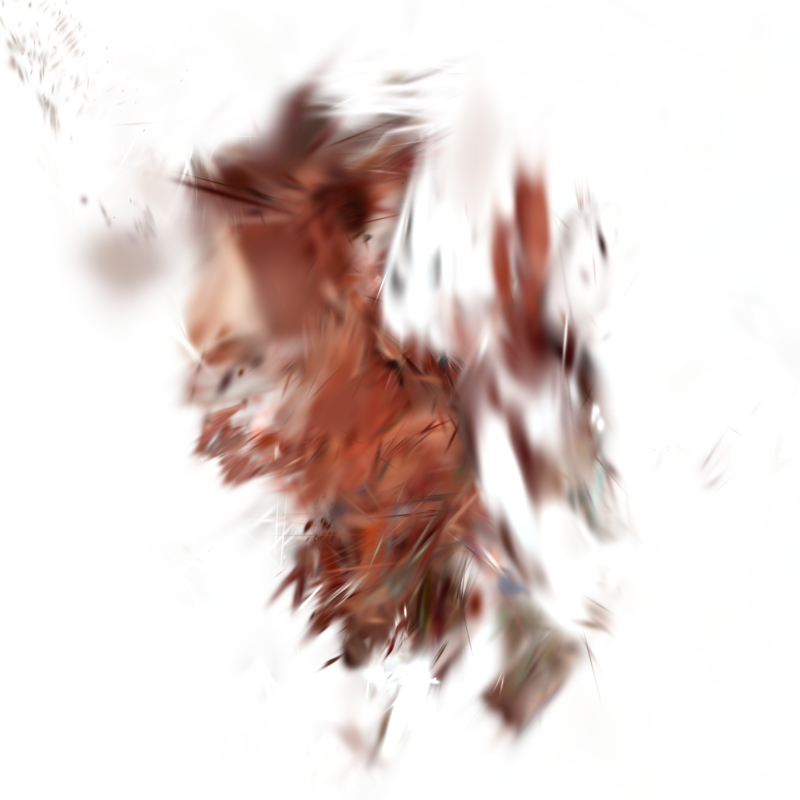}
    &   \includegraphics[width=\hsize,valign=m]{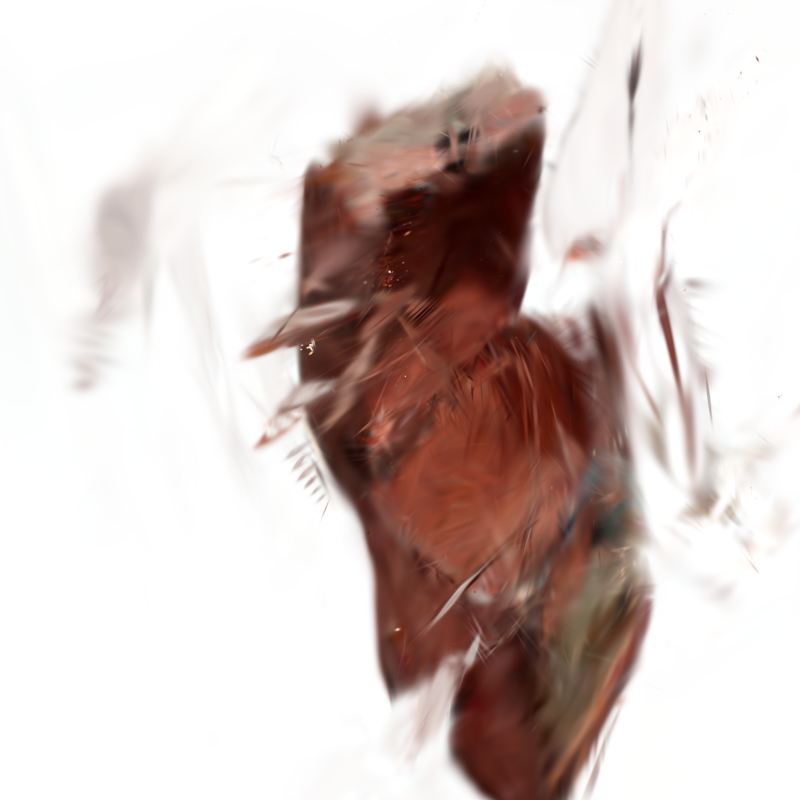}\\

    \\
        
\rotatebox[origin=c]{90}{\parbox[l]{0.05\textwidth}{\text{\;Dynamic}\\\text{Gaussians}}}          &  
    \includegraphics[width=\hsize,valign=m]{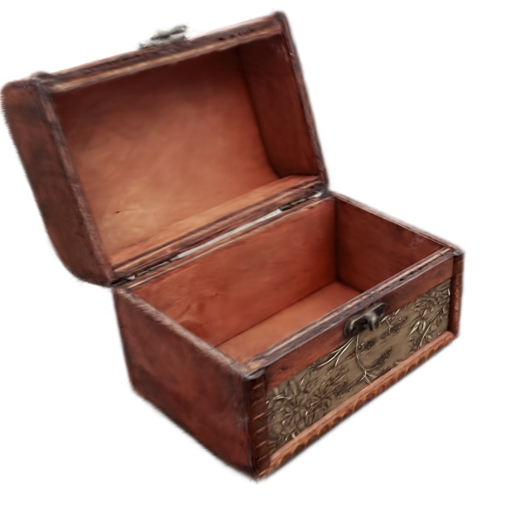}
    &   \includegraphics[width=\hsize,valign=m]{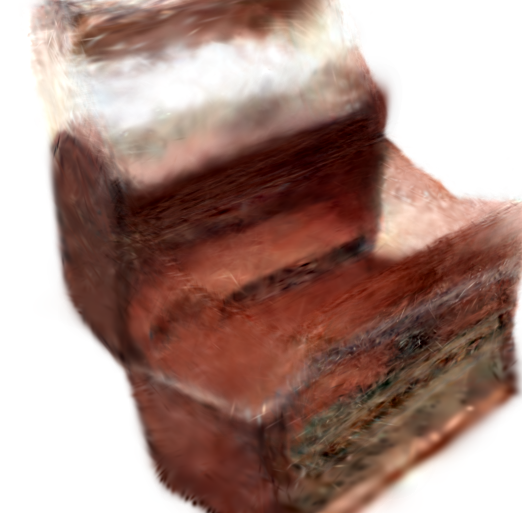}
    &   \includegraphics[width=\hsize,valign=m]{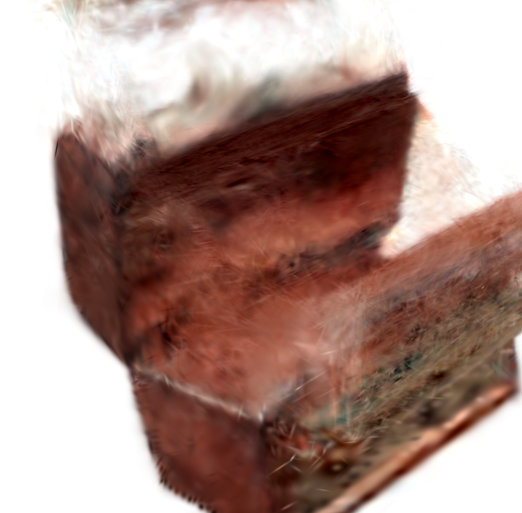}
    &   \includegraphics[width=\hsize,valign=m]{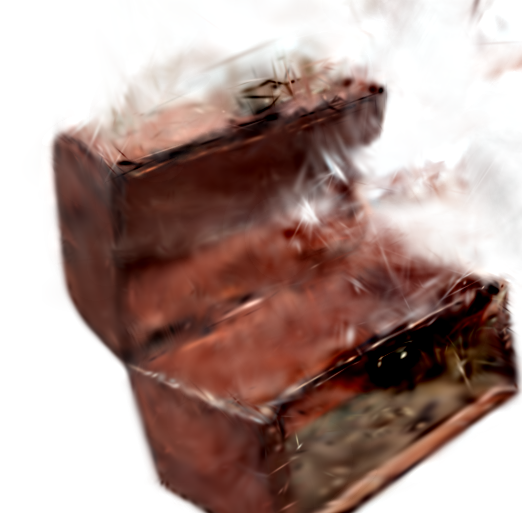}
    &   \includegraphics[width=\hsize,valign=m]{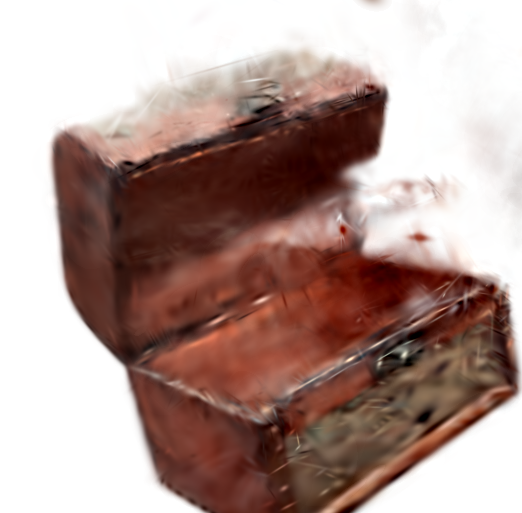}\\

    \\

\rotatebox[origin=c]{90}{\parbox[l]{0.05\textwidth}{\centering \text{\;\;PAPR in}\\\text{\;\;\;Motion}}}          &   
    \includegraphics[width=\hsize,valign=m]{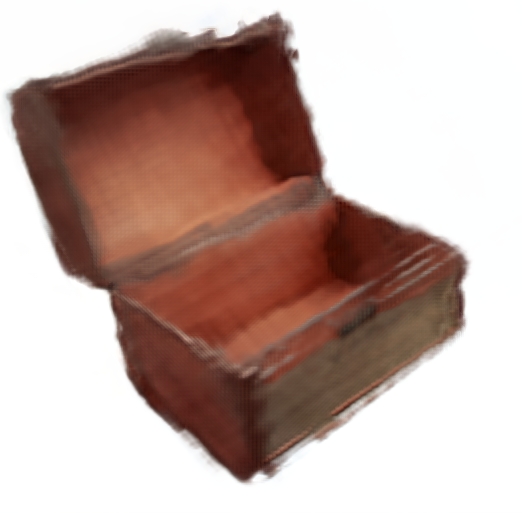}
    &   \includegraphics[width=\hsize,valign=m]{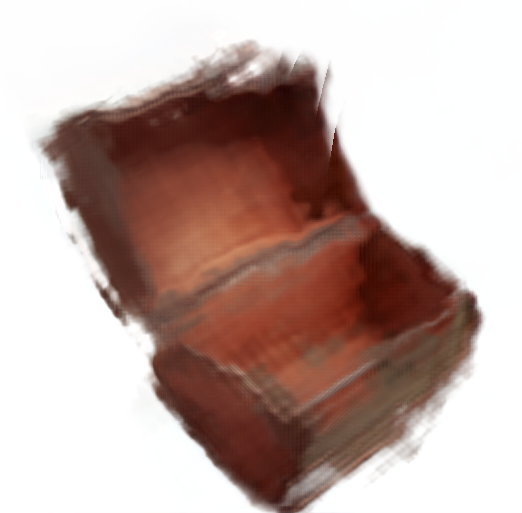}
    &   \includegraphics[width=\hsize,valign=m]{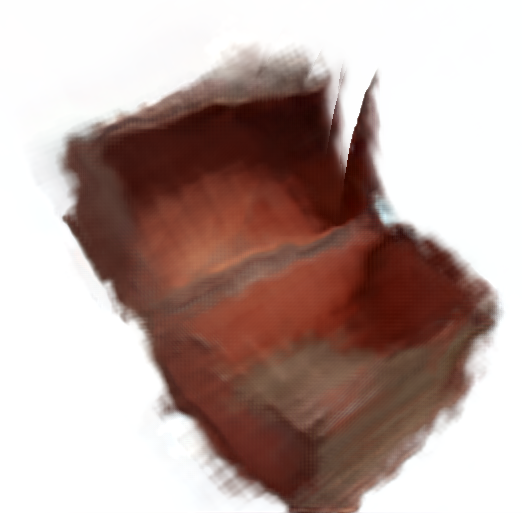}
    &   \includegraphics[width=\hsize,valign=m]{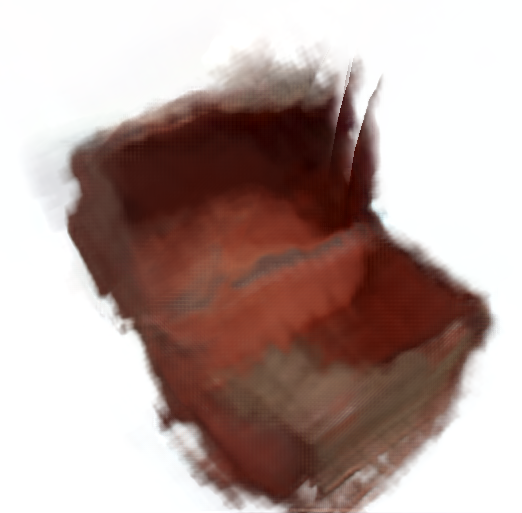}
    &   \includegraphics[width=\hsize,valign=m]{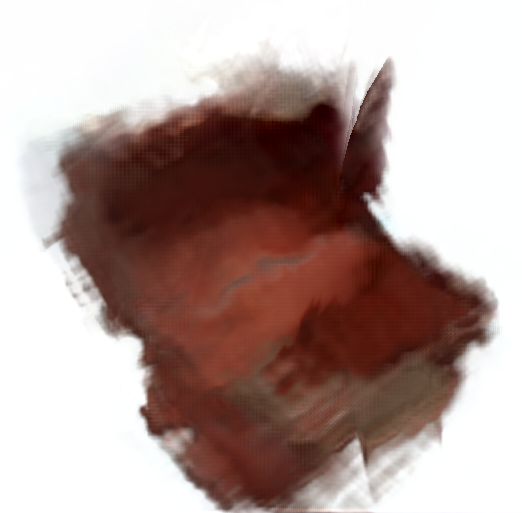}\\

    \\

\rotatebox[origin=c]{90}{\parbox[l]{0.05\textwidth}{\textbf{\;\;\;Our }\\\textbf{\;Method}}}          &   
    \includegraphics[width=\hsize,valign=m]{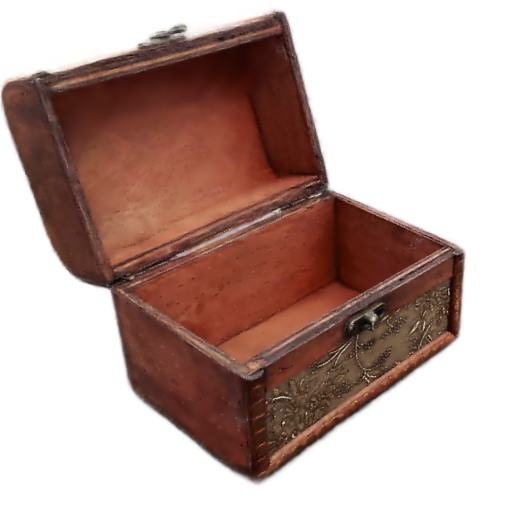}
    &   \includegraphics[width=\hsize,valign=m]{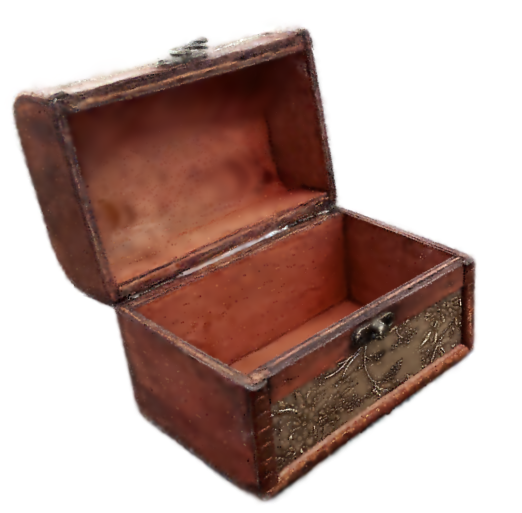}
    &   \includegraphics[width=\hsize,valign=m]{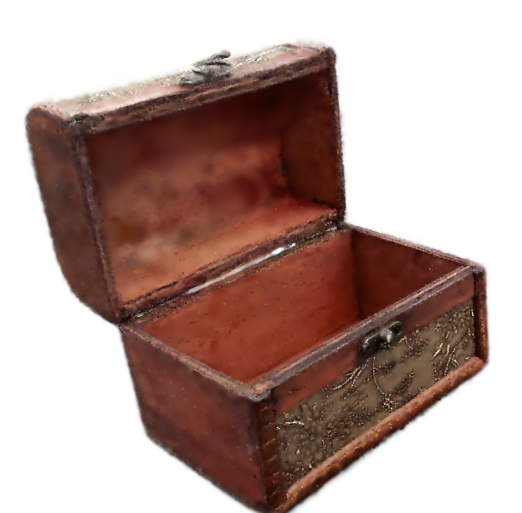}
    &   \includegraphics[width=\hsize,valign=m]{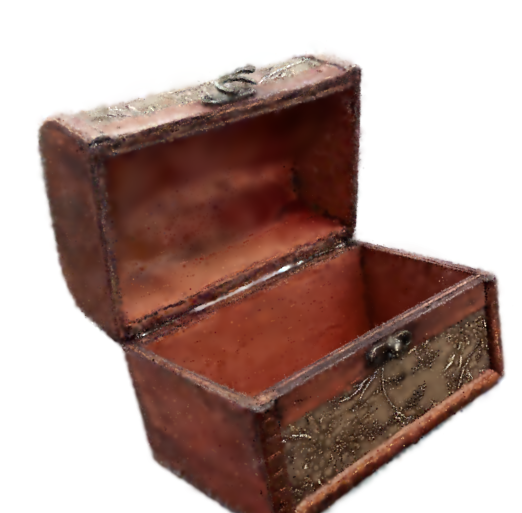}
    &   \includegraphics[width=\hsize,valign=m]{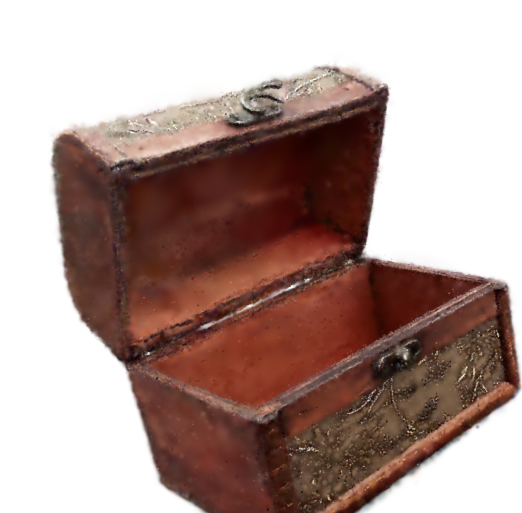}
    \\  
\end{tabularx}
\caption{\textbf{Real-World Scene Interpolation.} In the real-world \texttt{Box} scene, the box undergoes global motion while its lid exhibits local motion. 
Our method accurately captures both motion types and delivers realistic interpolation.} 
\label{fig:interpolation_real}
\end{figure}

%% file: tables/real-world_global-interp.tex
\begin{table}[t]
    \centering
    \begin{adjustbox}{max width=\linewidth}
    \begin{tabular}{lccccc}
    \toprule
    Metric & Method & \texttt{Box} & \texttt{Shoe} & \texttt{Tapeline} & Avg\\
    \midrule
    \multirow{5}{*}{SI-FID $\downarrow$} 
    & 4DGS~\cite{4DGS}   & 418.6   & 350.09   & -      & -       \\
    & Deformable 3DGS~\cite{Deformable-3DGS} & -       & -        & -      & -       \\
    & Dynamic Gaussian~\cite{luiten2023dynamic} & \underline{293.20} & \underline{302.67} & \underline{330.08} & \underline{308.65}  \\
    & PAPR in Motion~\cite{peng2024papr}    & 371.51  & 339.24   & 425.28 & 378.68  \\
    & \textbf{Ours}  & \textbf{182.13}  & \textbf{168.65}  & \textbf{148.2}  & \textbf{166.33}   \\
    
    \midrule
    \multirow{5}{*}{SI-EMD $\downarrow$} 
    & 4DGS~\cite{4DGS}   & \underline{67.73}   & \underline{57.68}   & -       & -       \\
    & Deformable 3DGS~\cite{Deformable-3DGS} & -       & -                & -       & -       \\
    & Dynamic Gaussian~\cite{luiten2023dynamic} & 368.93  & 524.19           & \underline{697.19}  & \underline{530.10}  \\
    & PAPR in Motion~\cite{peng2024papr}    & 724.65  & 531.33           & 1577.02 & 944.33 \\
    & \textbf{Ours}     & \underline{\textbf{81.44}}  & \textbf{46.43}    & \textbf{84.69}   & \textbf{70.85}   \\

    \midrule
    \multirow{5}{*}{SI-MPED $\downarrow$} 
    & 4DGS~\cite{4DGS}  & 748.67         & 345.71         & -         & -         \\
    & Deformable 3DGS~\cite{Deformable-3DGS}  & -              & -              & -         & -         \\
    & Dynamic Gaussian~\cite{luiten2023dynamic} & \underline{279.00} & 627.56         & \underline{290.62} & 399.06    \\
    & PAPR in Motion~\cite{peng2024papr}   & 566.24         & \underline{268.97} & 334.86    & \underline{390.02} \\
    & \textbf{Ours}  & \textbf{36.27}  & \textbf{20.38}  & \textbf{41.70}  & \textbf{32.78}  \\
    
    \bottomrule
  \end{tabular}
  \end{adjustbox}
  \vspace{2pt}
  \caption{\textbf{Real-World Global-Motion Evaluation.} Comparison of our method with the baselines on real-world scenes with global motion (``-'' indicates failure).
  Overall, our method outperforms the baselines in rendering quality and geometry fidelity.}
    \label{tab:global_motion_real}
\end{table}

%% file: tables/ablation.tex
\begin{table}[t]
    \centering
    \begin{adjustbox}{max width=\linewidth}
    \begin{tabular}{lccc}
    \toprule
    Model & SI-FID $\downarrow$ & SI-EMD $\downarrow$ & SI-MPED $\downarrow$ \\
    \midrule
    No DINO Input & 203.38 & 230.04 & 50.35 \\
    No Position Input & 198.72 & 233.68 & 54.29 \\
    No Local Isometry Loss & 252.09 & 219.81 & 69.38 \\
    No Joint Refinement & \underline{194.42} & \underline{217.70} & \underline{28.08} \\
    \textbf{Full} & \textbf{188.98} & \textbf{215.28} & \textbf{26.05}  \\

    \bottomrule
  \end{tabular}
  \end{adjustbox}
  \vspace{2pt}
  \caption{\textbf{Ablation Study.} Average metrics on synthetic global-motion scenes for various model variants are reported to reveal the impact of each method component on interpolation performance.}
    \label{tab:ablation}
\end{table}

%% file: sec/5_conclusion.tex
\section{Conclusion}
We present \Method (GMC), a novel method for 3D scene interpolation and extrapolation. We show that predicting large motion between timesteps is analogous to establishing smooth global correspondences between points across time. Next, we present our methods with the desired interpolation properties.
Experimental results demonstrate that GMC significantly outperforms prior work in interpolation tasks, especially when dealing with large changes between captures. Moreover, GMC enables extrapolation beyond the captured states, a capability lacking in prior work. 

%% file: sec/suppl/0_overview.tex
\clearpage
\setcounter{page}{1}
\setcounter{section}{0} 
\maketitlesupplementary

\renewcommand{\thesection}{\Alph{section}}

In this document, we include
\begin{itemize}
    \item more details about the implementation,
    \item more details about the SI-MPED metric,
    \item more results on motion interpolation and extrapolation,
    \item more results on the application of sparse view refinement,
    \item and more results on the ablation study.
\end{itemize}

%% file: sec/suppl/1_method.tex
\section{Implementation Details}
The input positions and PCA-DINO for the MLPs are normalized using scalars pre-calculated from the start-state 3DGS model. The position input is then scaled by a hyperparameter weight, selected from $\{0.1, 1.0\}$ based on the importance of the positional information. Correspondingly, dropout is applied to the position input to avoid trivial local minima, with a ratio of $0.1$ or $0.2$ depending on the previously chosen scale. 
To mitigate the issue of getting trapped in local minima, the Perturb-and-MAP strategy~\cite{perturb-and-map} is applied to the total energy, where the perturbations are sampled from a Gumbel distribution. The learning rate of training the MLPs is set to $0.0005$, using the Adam optimizer~\cite{Adam} with default parameters. For the RGB loss during the joint refinement, L1 and LPIPS~\cite{lpips} losses are combined with weights $1.0$ and $0.1$, respectively; gradients from the MLPs are not used to update the 3DGS models. Due to the large size of Gaussian sets, batches of Gaussians are sampled during each iteration when searching for the minimum energy between the two Gaussian sets. For most scenes, the batch size is set to $20, 000$, and it can be reduced accordingly if the total number of Gaussians is smaller. The FAISS library~\cite{faiss} is used to perform efficient nearest neighbor searches.

\section{SI-MPED Metric}
For each interpolation step, the Multiscale Potential Energy Discrepancy (MPED)~\cite{mped} is calculated between the interpolated point cloud and the ground-truth point cloud from the start and end states, respectively. The MPED is computed by aggregating distances from the neighborhoods comprising $0.1\%$, $0.5\%$, and $1\%$ of the total points, summing these values to obtain the overall MEPD. Following PAPR in Motion~\cite{peng2024papr}, the Scene Interpolation MPED (SI-MPED) is defined as a weighted sum of the MEPD at each interpolation step, where the weights are proportional to the average distance movement of points compared to the total movement from the start to the end states.

%% file: sec/suppl/2_results.tex
\section{Motion Interpolation and Extrapolation}
Qualitative results on motion interpolation and extrapolation for global-motion scenes are presented in Figure~\ref{fig:suppl_inter_extra_global}. Additionally, qualitative results on local-motion scenes~\cite{peng2024papr} are shown in Figure~\ref{fig:suppl_inter_extra_local}, and the quantitative interpolation evaluations for these scenes are provided in Table~\ref{tab:suppl_local_motion_interp}. Qualitative interpolation results on two real-world scenes from Dynamic Gaussian~\cite{luiten2023dynamic} are provided in Figure~\ref{fig:suppl_interpolation_dg_scenes}, in comparison with the Dynamic Gaussian~\cite{luiten2023dynamic} baseline.

\section{Ablation Study}
We find the following properties when removing one of the key components of our method:
\begin{enumerate}
    \item Removing DINO input can result in implausible interpolation (\texttt{Ball}), wrong global motion interpolation (\texttt{Boat}), or wrong local motion interpolation.
    \item Removing position input can result in wrong global matching (\texttt{Ball} and \texttt{Car}) or wrong local motion interpolation (\texttt{Butterfly}).
    \item Removing local isometry loss can result in noisy floaters (\texttt{Dolphin}) or blurry rendering (\texttt{Butterfly} and \texttt{Microwave}) during the interpolation.
    \item Removing local isometry loss can result in noisy rendering (\texttt{Ball} and \texttt{Microwave}) during the interpolation or suboptimal end status prediction (\texttt{Butterfly}).
\end{enumerate}

\input{tables/suppl_sparse_dense}

\section{Sparse View Refinement}
In addition to motion interpolation and extrapolation, GMC can also be used to improve reconstruction quality in sparse capture scenarios. Specifically, only five or ten views are available for sparse captures, and we consider two settings: (1) the start state has dense views and the end state has sparse views, and (2) both states have sparse views. When the input views are sparse, the reconstructed 3DGS will have bad geometry and thus will perform poorly in novel view synthesis. While a single state might not have enough views for good 3D reconstruction, we can borrow the information from the other state so that it can refine the self geometry and thus improve the novel view synthesis. Specifically, through the rendering loss $\mathcal{L}_{\mathrm{RGB}}(\boldsymbol{I}_{f}, \hat{\boldsymbol{I}}_{f})$ in Eq.~\ref{eq:joint_training_render_loss}, the sparse-view 3DGS can use the training views from the other state, and thus improve itself. 

\input{figures/sparse_view}

\paragraph{Results.}
For the sparse-view setting, we set $\beta=5$, because in this setting, the ground-truth training views are more reliable than the "borrowed" information based on Gaussian matching. Otherwise, each scene has the same setting as the dense-dense setting studied for interpolation and extrapolation tasks. We showcase three examples of the application in sparse-view refinement in Figure~\ref{fig:sparse_view}, and we report quantitative results, average PSNR, SSIM~\cite{ssim}, and LPIPS~\cite{lpips} of novel-view synthesis of 200 views. Quantitative results on sparse-view refinement with the sparse-sparse view setting are presented in Table~\ref{tab:suppl_sparse_sparse}, and results with the sparse-dense view setting are shown in Table~\ref{tab:suppl_sparse_dense}.

Both qualitative and quantitative results show that our method significantly improves upon the vanilla 3DGS trained on sparse views. When training views are few and sparse, two significant issues arise: (1) the presence of floaters, and (2) a lack of details in under-observed regions. The qualitative results show that our method is able to reduce (1) and handle (2). Our method reduces floaters because they have a poor match in the other state, and thus, when transformed and rendered, they can be removed by the rendering loss. Our method improves details in under-observed regions because the other state may have more information on appearance details, which can be borrowed to enhance the current state.

\input{tables/suppl_sparse_sparse}

\input{figures/suppl_interpolation_dg_scenes}
\input{tables/suppl_local_motion_interp_full}

\input{figures/suppl_inter_extra}

\input{figures/suppl_inter_extra_papr}

%% file: tables/suppl_sparse_dense.tex
\begin{table*}[t]
    \centering
    \begin{adjustbox}{max width=\linewidth}
    \begin{tabular}{lc|cccccccc|ccc|c}
    \toprule
    \multicolumn{1}{c}{} & \multicolumn{1}{c}{} & \multicolumn{12}{c}{Sparse + Dense} \\   
    \midrule
    \multicolumn{1}{c}{} & \multicolumn{1}{c}{} & \multicolumn{8}{c}{Synthetic Scenes} & \multicolumn{3}{c}{Real-world Scenes} \\
     \midrule
    \multirow{2}{*}{Metric} & \multirow{2}{*}{Method} & \multirow{2}{*}{\texttt{Ball}} & \multirow{2}{*}{\texttt{Boat}} & \multirow{2}{*}{\texttt{Butterfly}} & \multirow{2}{*}{\texttt{Car}} & \multirow{2}{*}{\texttt{Dolphin}} & \multirow{2}{*}{\texttt{Knight}} & \multirow{2}{*}{\texttt{Microwave}} & \multirow{2}{*}{\texttt{Seagull}} & \multirow{2}{*}{\texttt{Box}} & \multirow{2}{*}{\texttt{Shoe}} & \multirow{2}{*}{\texttt{Tapeline}} & \multirow{2}{*}{Avg}\\
    & &  &  &  & &  &  &  &  &  &  & & \\
    
    \midrule
    
    \multirow{2}{*}{PSNR $\uparrow$} 
    & 3DGS~\cite{kerbl20233dgaussiansplattingrealtime} & 30.39 & 31.64 & 28.94 & 24.42 & 34.50 & 26.82 & 31.98 & 31.12 & 23.50 & 26.10 & 26.39 &  28.71 \\
    & \textbf{Ours} & \textbf{38.18} & \textbf{36.25} & \textbf{31.18} & \textbf{33.63} & \textbf{37.59} & \textbf{33.49} & \textbf{37.63} & \textbf{35.76} & \textbf{26.31} & \textbf{26.94} & \textbf{26.81} & \textbf{33.07} \\
    
    \midrule
    \multirow{2}{*}{SSIM $\uparrow$} 
    & 3DGS~\cite{kerbl20233dgaussiansplattingrealtime} & 0.978 & 0.970 & 0.973 & 0.946 & 0.992 & 0.965 & 0.980 & 0.964 & 0.890 & 0.930 & 0.959 &  0.959  \\
    & \textbf{Ours} & \textbf{0.992} & \textbf{0.984} & \textbf{0.983} & \textbf{0.981} & \textbf{0.995} & \textbf{0.985} & \textbf{0.989} & \textbf{0.981} & \textbf{0.912} & \textbf{0.940} & \textbf{0.964} &  \textbf{0.973} \\

    \midrule
    \multirow{2}{*}{LPIPS $\downarrow$} 
    & 3DGS~\cite{kerbl20233dgaussiansplattingrealtime} & 0.045 & 0.047 & 0.059 & 0.086 & 0.018 & 0.063 & 0.042 & 0.051 & 0.107 & 0.087 & 0.046 &  0.059 \\
    & \textbf{Ours} & \textbf{0.006} & \textbf{0.011} & \textbf{0.013} & \textbf{0.016} & \textbf{0.005} & \textbf{0.008} & \textbf{0.012} & \textbf{0.013} & \textbf{0.064} & \textbf{0.066} & \textbf{0.033} &  \textbf{0.022} \\
    
    \bottomrule
  \end{tabular}
  \end{adjustbox}
  \vspace{2pt}
  \caption{\textbf{Novel View Synthesis for Sparse-Dense View Setting.} For the synthetic scenes, the start state has 100 dense training views, while the end state has 10 sparse training views. For real-world scenes (\texttt{Shoe}, \texttt{tapeline}, and \texttt{Box}), the end state has 5 sparse training views. The results are reported as the mean value of test views for each scene.}
    \label{tab:suppl_sparse_dense}
\end{table*}

%% file: figures/sparse_view.tex
\begin{figure}[t]
\setlength\tabcolsep{1pt}
\footnotesize
\begin{tabularx}{\linewidth}{l@{\hskip 2em}YYYY}

\rotatebox[origin=c]{90}{\parbox[l]{0.05\textwidth}{\centering 3DGS~\cite{kerbl20233dgaussiansplattingrealtime}}}          &   
    \includegraphics[width=\hsize,valign=m]{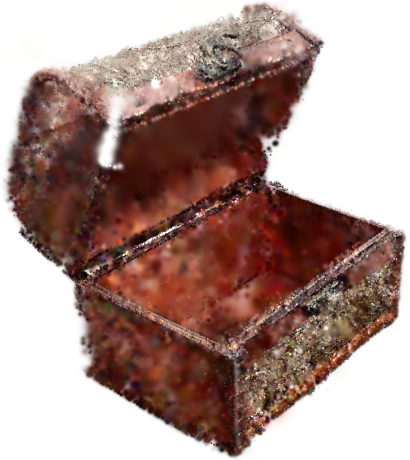}
    &   \includegraphics[width=\hsize,valign=m]{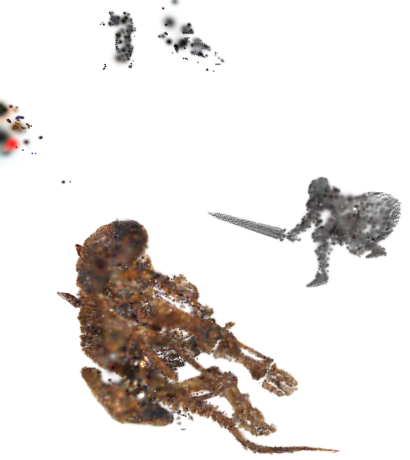}
    &   \includegraphics[width=\hsize,valign=m]{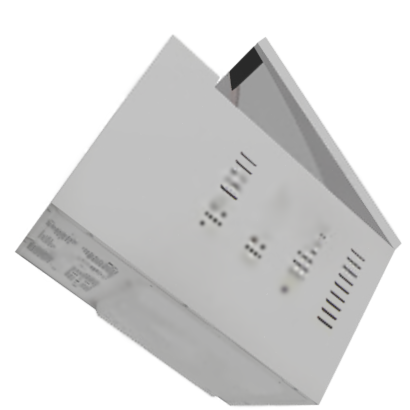}
    &   \includegraphics[width=\hsize,valign=m]{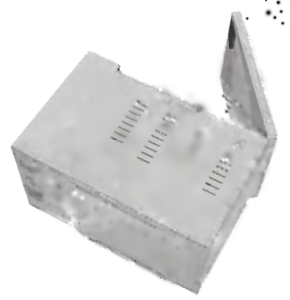}\\
        
\rotatebox[origin=c]{90}{\parbox[l]{0.05\textwidth}{\textbf{Ours}}}          &  
    \includegraphics[width=\hsize,valign=m]{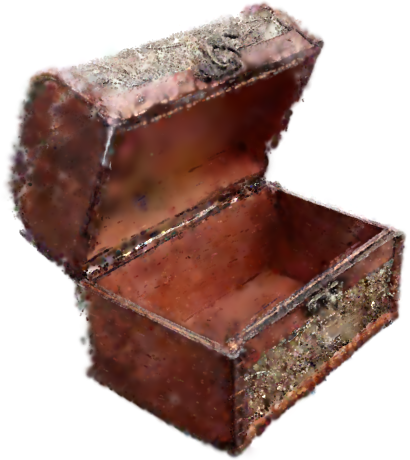}
    &   \includegraphics[width=\hsize,valign=m]{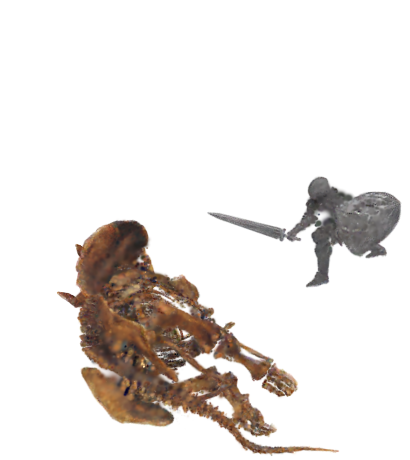}
    &   \includegraphics[width=\hsize,valign=m]{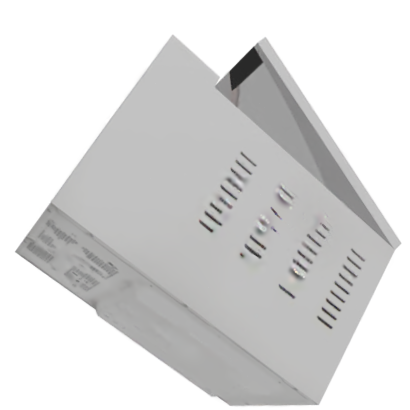}
    &   \includegraphics[width=\hsize,valign=m]{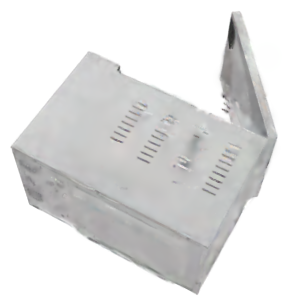}\\

\rotatebox[origin=c]{90}{\parbox[l]{0.05\textwidth}{\text{GT}}}          &   
    \includegraphics[width=\hsize,valign=m]{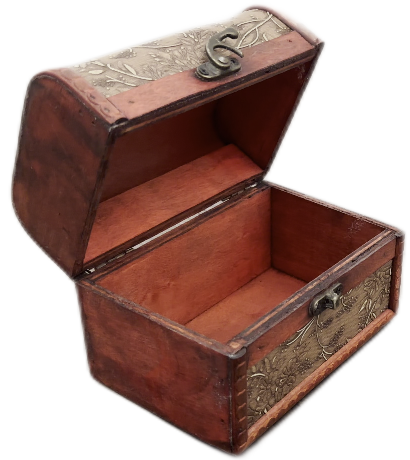}
    &   \includegraphics[width=\hsize,valign=m]{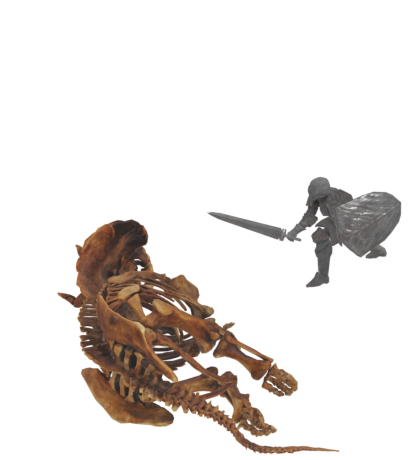}
    &   \includegraphics[width=\hsize,valign=m]{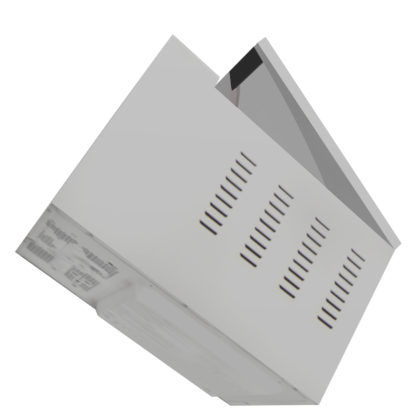}
    &   \includegraphics[width=\hsize,valign=m]{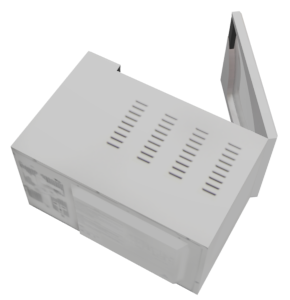}
    \\  
\end{tabularx}
\caption{\textbf{Novel-View Synthesis in Sparse-View Setting.} This figure demonstrate novel-view syhnthesis results in sparse-view setting. From top to bottom, the rows show results from vanilla 3DGS~\cite{kerbl20233dgaussiansplattingrealtime}, 3DGS refined by our method, and the ground truth. The first column displays the real-world \texttt{Box} scene, where the start state has $100$ dense training views and the end state has only $5$ sparse views. The second column shows the synthetic \texttt{Knight} scene, with the start state having $100$ dense views and the end state having $10$ sparse views. The last two columns present the \texttt{Microwave} scene, where both start and end states have $10$ sparse views, showing novel view synthesis for the start state (left) and end state (right). The proposed method effectively transfers information between states, improving texture quality in the \texttt{Box} and \texttt{Microwave} scenes, and enhancing geometry by remvinng floaters in the \texttt{Knight} scene.}
\label{fig:sparse_view}
\end{figure}

%% file: tables/suppl_sparse_sparse.tex
\begin{table}[t]
    \centering
    \begin{adjustbox}{max width=\linewidth}
    \begin{tabular}{lc|cccc|cc|c}
    \toprule
    \multicolumn{1}{c}{} & \multicolumn{1}{c}{} & \multicolumn{7}{c}{Sparse + Sparse} \\   
    \midrule
    \multicolumn{1}{c}{} & \multicolumn{1}{c}{} & \multicolumn{4}{c}{Synthetic Scenes} & \multicolumn{2}{c}{Real-world Scene} \\
     \midrule
    \multirow{2}{*}{Metric} & \multirow{2}{*}{Method} & \multicolumn{2}{c}{\texttt{Car}} & \multicolumn{2}{c}{\texttt{Microwave}} & \multicolumn{2}{c}{\texttt{Box}} &  \multirow{2}{*}{Avg}\\
    & &  start & end & start & end & start & end &  \\
    
    \midrule
    
    \multirow{2}{*}{PSNR $\uparrow$} 
    & 3DGS~\cite{kerbl20233dgaussiansplattingrealtime} & 23.54 & 24.37 & 26.25 & 31.94 & 23.80 & 23.48 & 25.56 \\
    & \textbf{Ours} & \textbf{29.07} & \textbf{29.96} & \textbf{33.60} & \textbf{34.83} & \textbf{25.19} & \textbf{25.36} & \textbf{29.67}\\
    
    \midrule
    \multirow{2}{*}{SSIM $\uparrow$} 
    & 3DGS~\cite{kerbl20233dgaussiansplattingrealtime} & 0.943 & 0.946 & 0.962 & 0.980 & 0.900 & 0.890 & 0.937 \\
    & \textbf{Ours}  & \textbf{0.965} & \textbf{0.967} & \textbf{0.983} & \textbf{0.985} & \textbf{0.915} & \textbf{0.907} & \textbf{0.953}\\

    \midrule
    \multirow{2}{*}{LPIPS $\downarrow$} 
    & 3DGS~\cite{kerbl20233dgaussiansplattingrealtime} & 0.097 & 0.086 & 0.079 & 0.042 & 0.105 & 0.107 & 0.086\\
    & \textbf{Ours} & \textbf{0.041} & \textbf{0.040} & \textbf{0.028} & \textbf{0.021} & \textbf{0.069} & \textbf{0.075} & \textbf{0.046} \\
    
    \bottomrule
  \end{tabular}
  \end{adjustbox}
  \vspace{2pt}
  \caption{\textbf{Novel View Synthesis in Sparse + Sparse View Setting.} For the scenes \texttt{Car} and \texttt{Microwave}, both states have 10 training views; for \texttt{Box}, both states have 5 training views. The results are reported as the mean value of test views for each scene.}
    \label{tab:suppl_sparse_sparse}
\end{table}

%% file: figures/suppl_interpolation_dg_scenes.tex
\begin{figure*}[h]
\setlength\tabcolsep{1pt}
\footnotesize
\begin{tabularx}{\linewidth}{l@{\hskip 2em}YYYYY}
& \cellcolor[HTML]{FFCCC9}{Start} & \multicolumn{3}{c}{\cellcolor[HTML]{ebdbe5}{Intermediate}} & \cellcolor[HTML]{DAE8FC}{End} \\
\\

\rotatebox[origin=c]{90}{\parbox[l]{0.05\textwidth}{\text{\;{Dynamic} }\\\text{{Gaussians}}}}          &   
    \includegraphics[width=\hsize,valign=m]{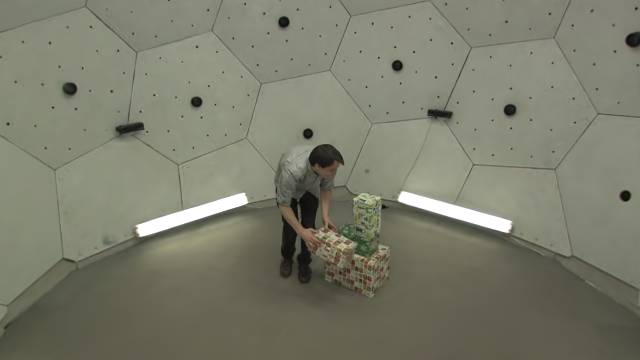}
    &   \includegraphics[width=\hsize,valign=m]{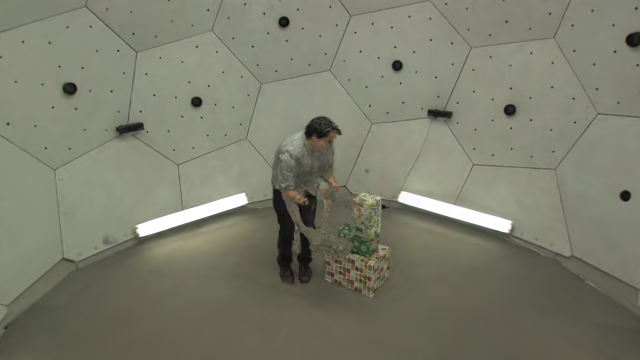}
    &   \includegraphics[width=\hsize,valign=m]{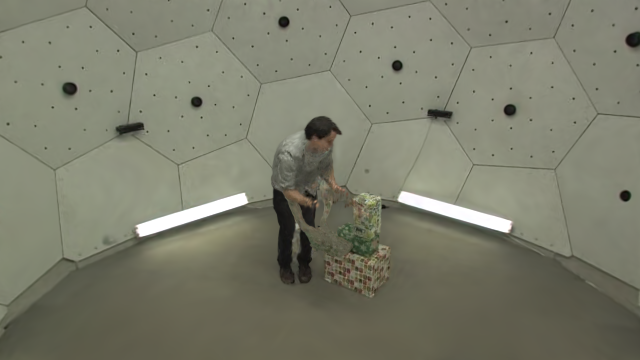}
    &   \includegraphics[width=\hsize,valign=m]{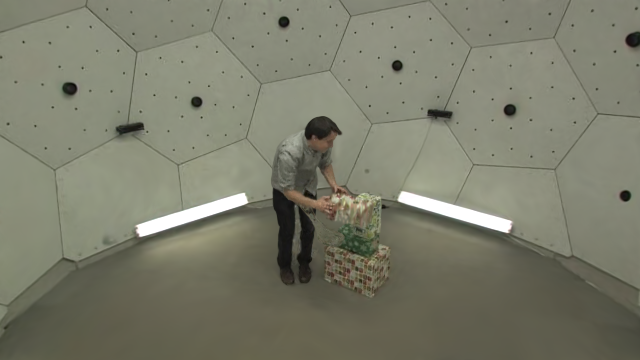}
    &   \includegraphics[width=\hsize,valign=m]{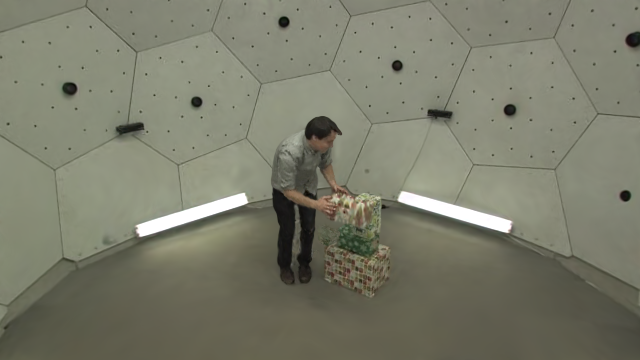}
    \\

\rotatebox[origin=c]{90}{\parbox[l]{0.05\textwidth}{\text{\;\;\;\;\textbf{Our} }\\\text{\;\textbf{Method}}}}          &   
    \includegraphics[width=\hsize,valign=m]{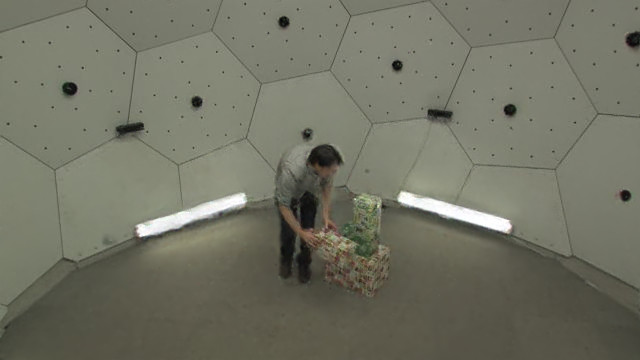}
    &   \includegraphics[width=\hsize,valign=m]{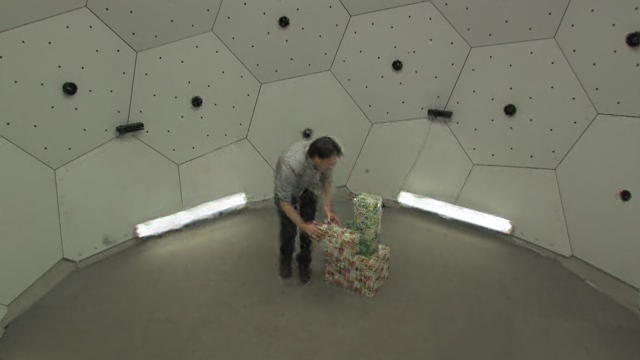}
    &   \includegraphics[width=\hsize,valign=m]{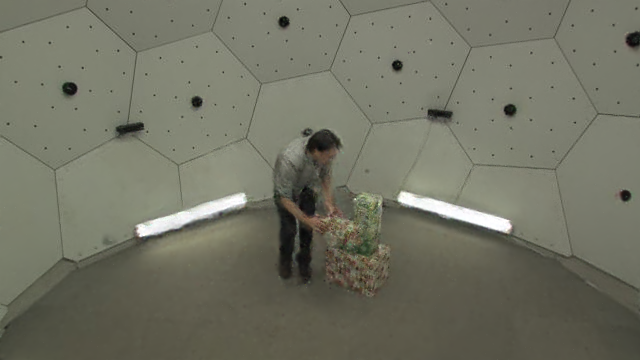}
    &   \includegraphics[width=\hsize,valign=m]{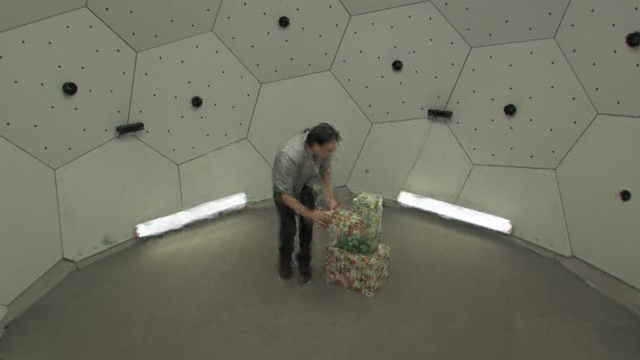}
    &   \includegraphics[width=\hsize,valign=m]{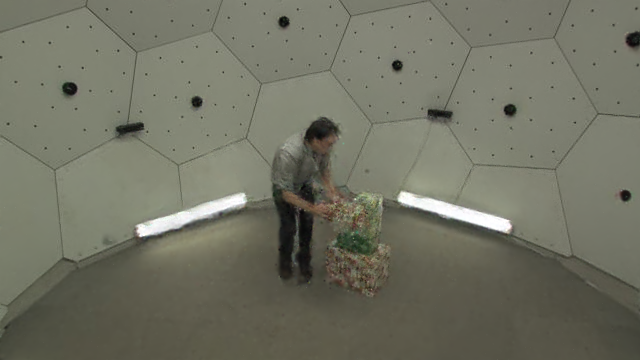}
    \\  

\midrule

\rotatebox[origin=c]{90}{\parbox[l]{0.05\textwidth}{\text{\;{Dynamic} }\\\text{{Gaussians}}}}          &   
    \includegraphics[width=\hsize,valign=m]{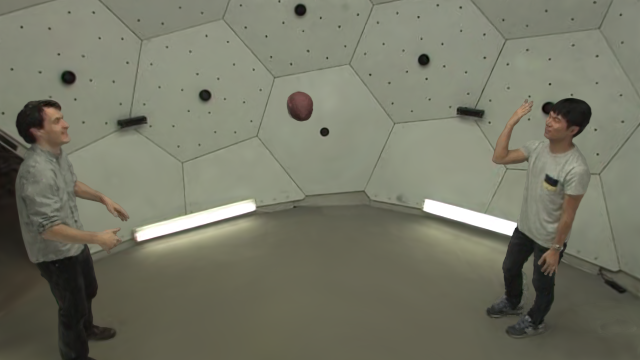}
    &   \includegraphics[width=\hsize,valign=m]{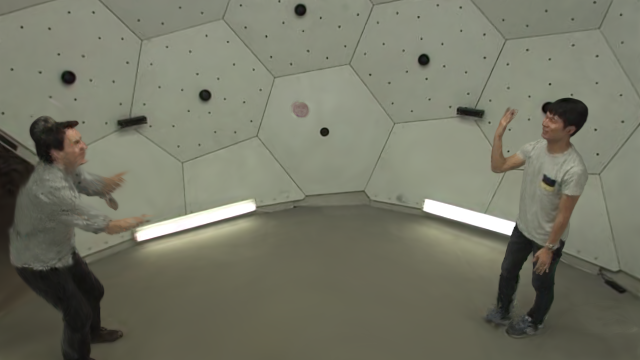}
    &   \includegraphics[width=\hsize,valign=m]{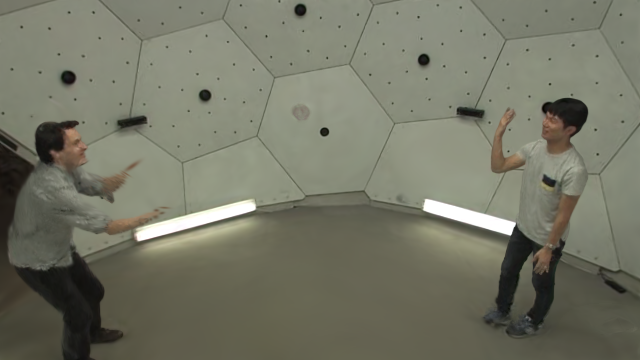}
    &   \includegraphics[width=\hsize,valign=m]{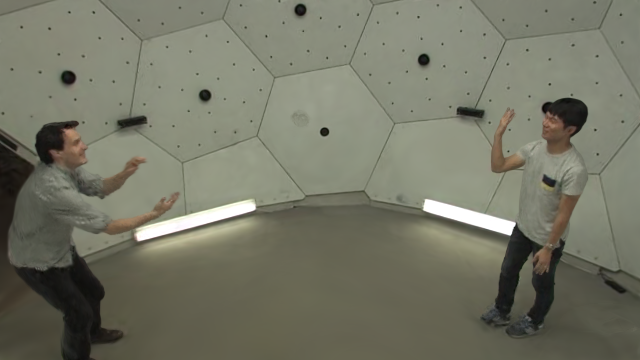}
    &   \includegraphics[width=\hsize,valign=m]{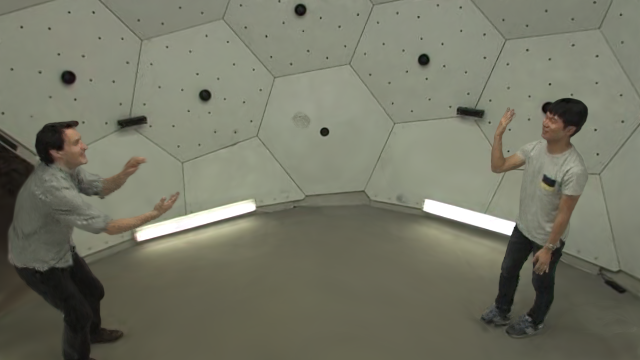}
    \\

\rotatebox[origin=c]{90}{\parbox[l]{0.05\textwidth}{\text{\;\;\;\;\textbf{Our} }\\\text{\;\textbf{Method}}}}          &   
    \includegraphics[width=\hsize,valign=m]{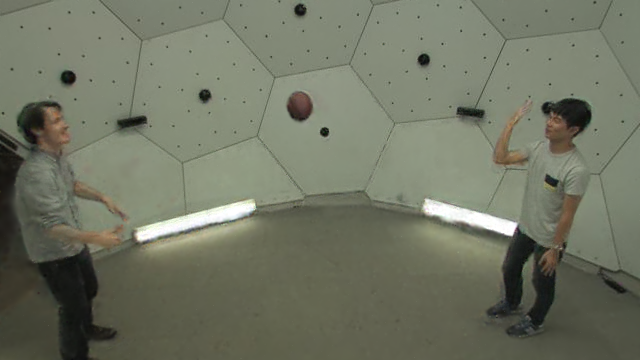}
    &   \includegraphics[width=\hsize,valign=m]{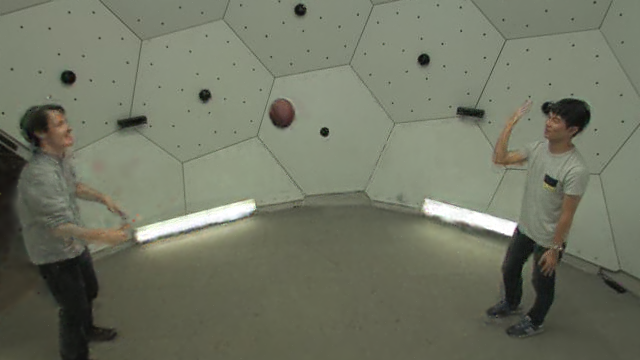}
    &   \includegraphics[width=\hsize,valign=m]{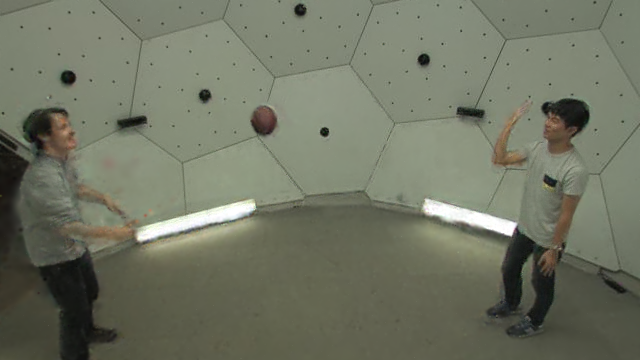}
    &   \includegraphics[width=\hsize,valign=m]{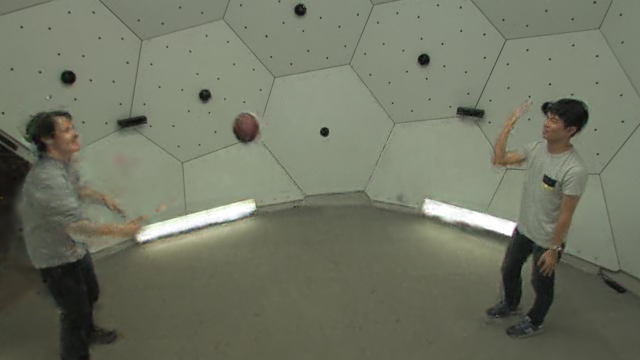}
    &   \includegraphics[width=\hsize,valign=m]{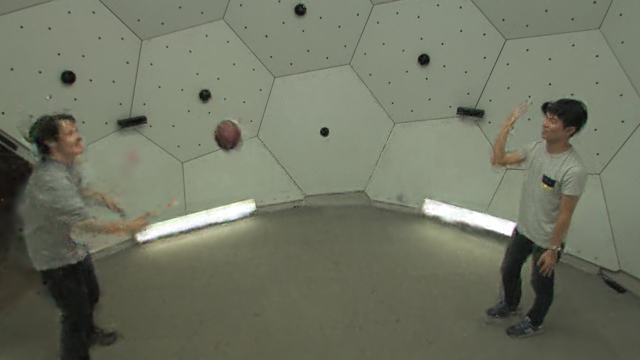}
    \\  
    
\end{tabularx}
\caption{\textbf{Additional Interpolation Results.}  The figure presents interpolation results using our method on the \texttt{Bxoes} and \texttt{Football} scene from Dynamic Gaussian~\cite{luiten2023dynamic}. The five columns correspond to five timesteps: {0.00, 0.25, 0.50, 0.75, 1.00}.}
\label{fig:suppl_interpolation_dg_scenes}
\end{figure*}

%% file: tables/suppl_local_motion_interp_full.tex
\begin{table*}[ht]
    \centering
    \begin{adjustbox}{max width=\textwidth}
    \begin{tabular}{lc|ccccccc|ccc}
    \toprule
     \multicolumn{1}{c}{} & \multicolumn{1}{c}{} & \multicolumn{7}{c}{Synthetic Scenes} & \multicolumn{3}{c}{Real-world Scenes} \\
     \midrule
    Metric & Method & \texttt{Butterfly} & \texttt{Crab} & \texttt{Dolphin} & \texttt{Giraffe} & \texttt{Lego Bulldozer} & \texttt{Lego Man} & Avg & \texttt{Stand} & \texttt{Lamp} & Avg \\
    
    \midrule
    \multirow{5}{*}{SI-FID $\downarrow$} 
    & 4DGS~\cite{4DGS}  & 130.98  & 131.34  & \textbf{107.55}  & \textbf{166.66}  & 229.14  & \underline{136.82}  & 150.42  & 431.64  & 380.52  & 406.08 \\ 
    & Deformable 3DGS~\cite{Deformable-3DGS} & 569.21  & 179.67  & 365.49         & 412.67         & 254.60  & 645.14         & 404.46  & -       & -       & - \\  
    & Dynamic Gaussian~\cite{luiten2023dynamic} & 328.69  & 129.52  & 165.78         & 215.83         & 197.68  & 330.94         & 228.07  & 302.40  & \underline{248.49} & 275.45 \\
    & PAPR in Motion~\cite{peng2024papr}   & \underline{90.89}  & \underline{73.86}  & 112.92         & 174.73         & \textbf{103.34}  & 151.89         & \underline{117.94}  & \underline{203.56}  & 265.08  & \underline{234.32} \\
    & \textbf{Ours}                     & \textbf{87.13}   & \textbf{65.24}   & \underline{112.34} & \underline{169.65} & \underline{110.37} & \textbf{131.01} & \textbf{112.62}  & \textbf{165.36}  & \textbf{181.22}  & \textbf{173.29} \\
    
    \midrule
    \multirow{5}{*}{SI-EMD $\downarrow$} 
    & 4DGS~\cite{4DGS} & \underline{22.94}  & 12.99  & \textbf{1.81}  & 5.72   & 37.60  & \textbf{11.06} & 15.35  & 97.71  & 88.11  & 92.91 \\  
    & Deformable 3DGS~\cite{Deformable-3DGS} & 45.84  & 29.34  & 5.96   & 22.22  & 24.50  & 36.24  & 27.35  & -      & -      & - \\  
    & Dynamic Gaussian~\cite{luiten2023dynamic} & 104.57 & \underline{10.19} & 50.47  & 11.13  & 62.60  & 146.65 & 64.27  & 84.69  & 103.58 & 94.14 \\  
    & PAPR in Motion~\cite{peng2024papr} & 34.93  & \textbf{9.87}  & \underline{2.17}  & \textbf{5.03}  & \underline{13.34}  & \underline{12.61} & \underline{12.99} & \underline{29.77} & \underline{63.12} & \underline{46.45} \\  
    & \textbf{Ours} & \underline{\textbf{32.59}} & 13.26  & 2.78   & \underline{5.40}  & \textbf{9.52}  & 14.23  & \textbf{12.96} & \textbf{17.21} & \textbf{56.92} & \textbf{37.07} \\

    \midrule
    \multirow{5}{*}{SI-MPED $\downarrow$} 
    & 4DGS~\cite{4DGS} & 140.40  & 127.8   & 30.71   & 50.61   & 500.84  & 89.47   & 156.64  & 620.66  & 769.12  & 694.89 \\ 
    & Deformable 3DGS~\cite{Deformable-3DGS} & 81.87   & 32.26   & 12.53   & 14.78   & 47.27   & 53.59   & 40.38   & -       & -       & - \\  
    & Dynamic Gaussian~\cite{luiten2023dynamic} & 143.99  & 44.60   & 79.09   & 24.26   & 260.85  & 136.67  & 114.91  & 260.68  & 166.58  & 213.63 \\
    & PAPR in Motion~\cite{peng2024papr}   & \underline{11.57}  & \textbf{7.16}   & \textbf{4.05}   & \underline{5.89}  & \underline{19.13}  & \underline{8.50}   & \underline{9.38}   & \textbf{26.72}  & \underline{22.00}  & \textbf{24.36} \\
    & \textbf{Ours}                       & \textbf{11.02}   & \underline{7.85}   & \underline{5.07}   & \textbf{4.70}   & \textbf{18.55}   & \textbf{8.40}   & \textbf{9.27}   & \underline{30.27}  & \textbf{18.95}   & \underline{24.61} \\
    
    \bottomrule
  \end{tabular}
  \end{adjustbox}
  \vspace{2pt}
  \caption{\textbf{Scene Interpolation Evaluation on Local-Motion Scenes~\cite{peng2024papr}.} The table compares our method with the baseline methods on scenes with local motion~\cite{peng2024papr}, where ``-'' represents failure of a method. Rendering quality is evaluated using Scene Interpolation FID (SI-FID), while geometry quality is assessed using Scene Interpolation Earth Mover’s Distance (SI-EMD) and Scene Interpolation Multiscale Potential Energy Discrepancy\cite{mped} (SI-MPED).}

    \label{tab:suppl_local_motion_interp}
\end{table*}

%% file: figures/suppl_inter_extra.tex
\begin{figure*}[t]
\centering
\footnotesize
\begin{tabularx}{\linewidth}{Y|Y|Y|Y|Y|Y|Y|Y|Y}
    \multicolumn{2}{c}{\cellcolor[HTML]{e6b8b5}{Extrapolation}} & \multicolumn{1}{c}{\cellcolor[HTML]{FFCCC9}{Start}} & \multicolumn{3}{c}{\cellcolor[HTML]{ebdbe5}{Interpolation}} & \multicolumn{1}{c}{\cellcolor[HTML]{DAE8FC}{End}} & \multicolumn{2}{c}{\cellcolor[HTML]{cfdcef}{Extrapolation}} 

    \vspace{1em}
    \\

    \includegraphics[width=\hsize,valign=m]{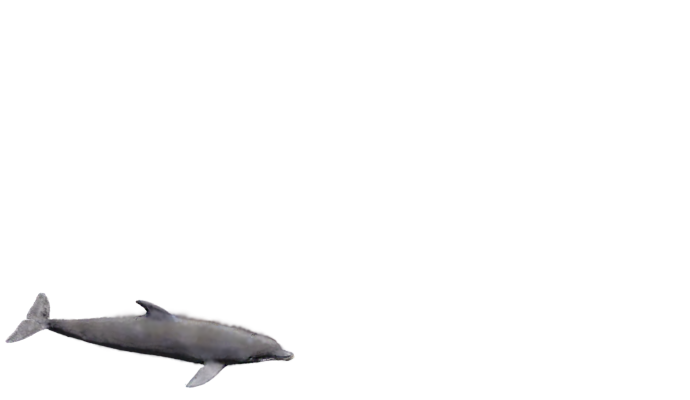}
    &   \includegraphics[width=\hsize,valign=m]{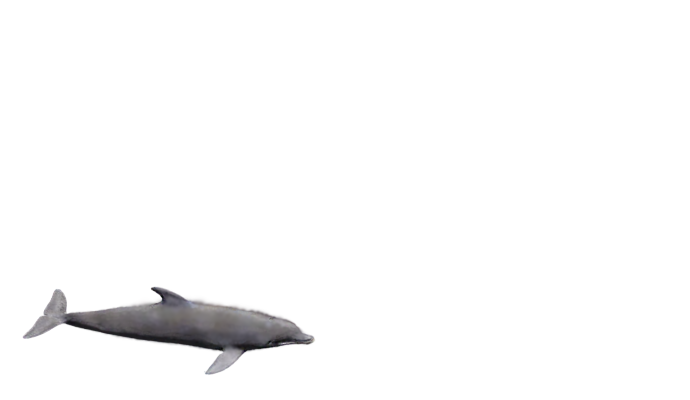}
    &   \includegraphics[width=\hsize,valign=m]{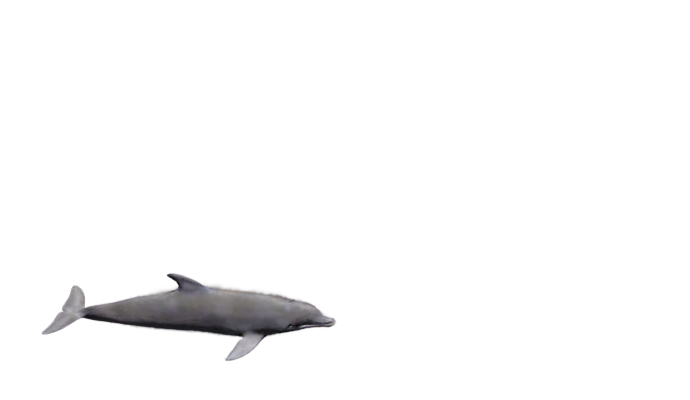}
    &   \includegraphics[width=\hsize,valign=m]{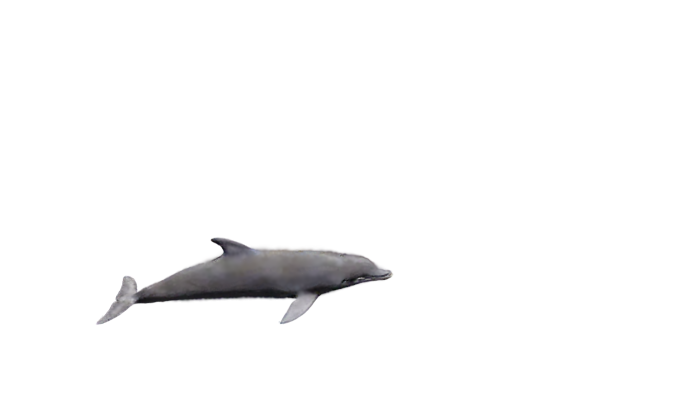}
     &   \includegraphics[width=\hsize,valign=m]{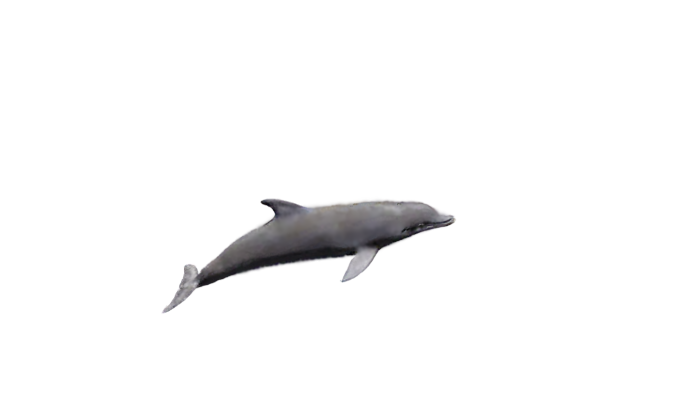}
    &   \includegraphics[width=\hsize,valign=m]{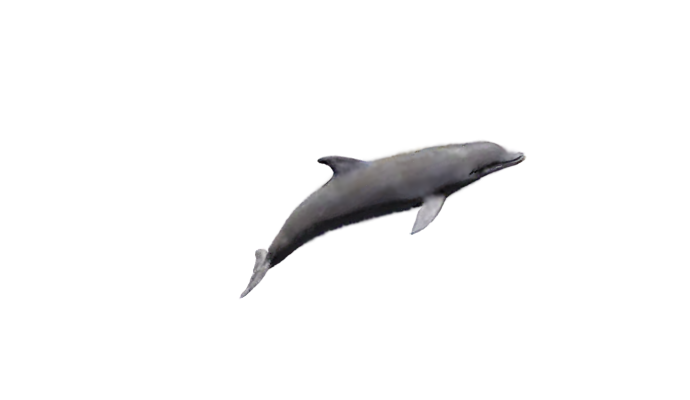}
    &   \includegraphics[width=\hsize,valign=m]{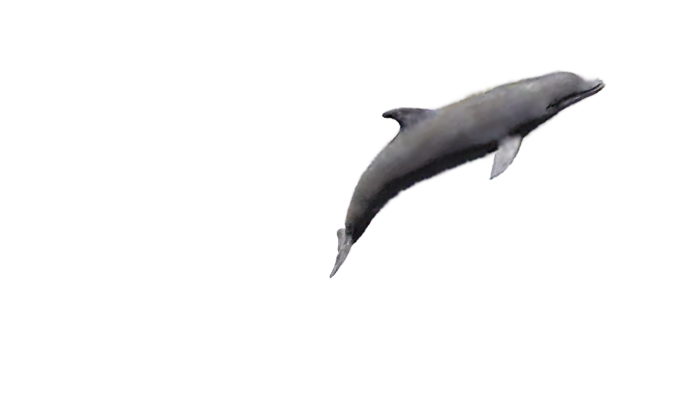}
    &   \includegraphics[width=\hsize,valign=m]{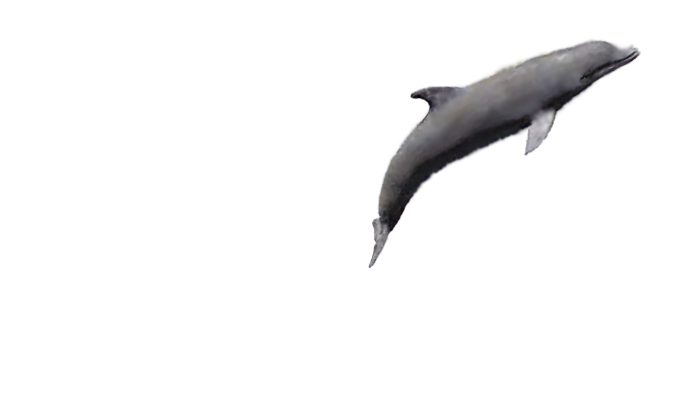}
    &   \includegraphics[width=\hsize,valign=m]{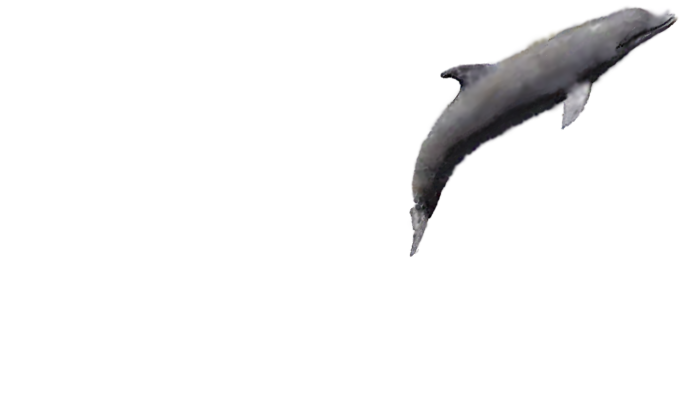}
    \\

    \midrule

    \includegraphics[width=\hsize,valign=m]{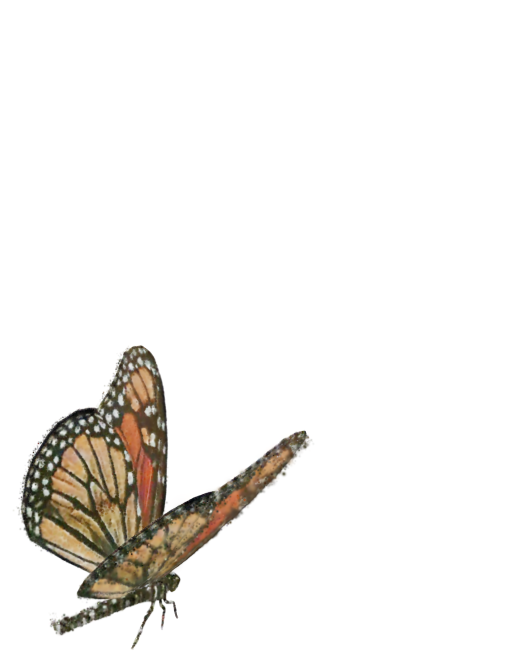}
    &   \includegraphics[width=\hsize,valign=m]{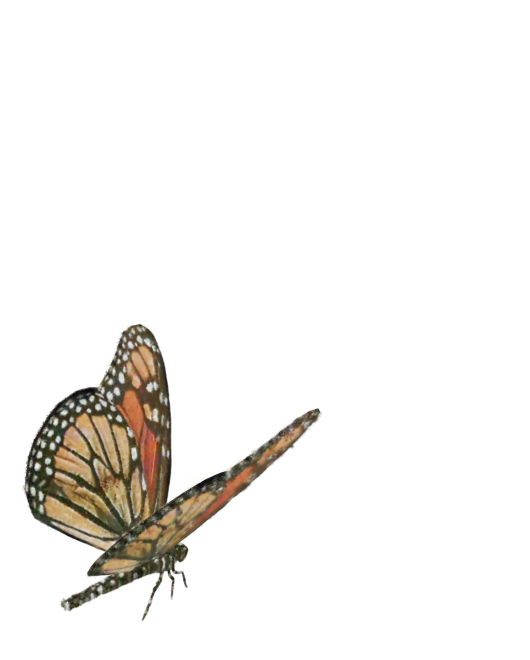}
    &   \includegraphics[width=\hsize,valign=m]{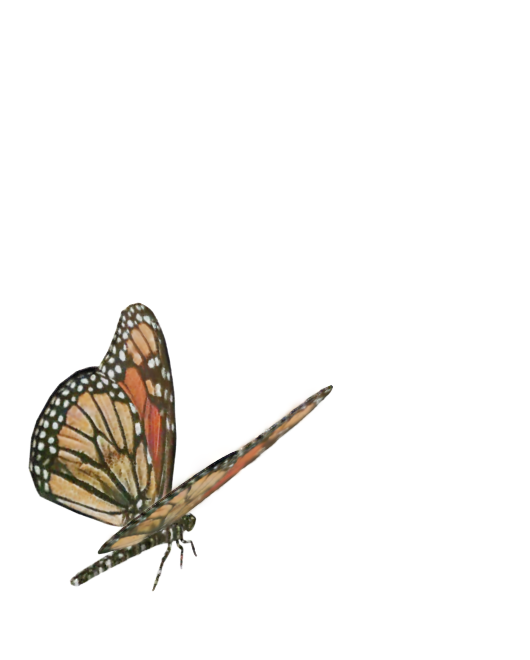}
    &   \includegraphics[width=\hsize,valign=m]{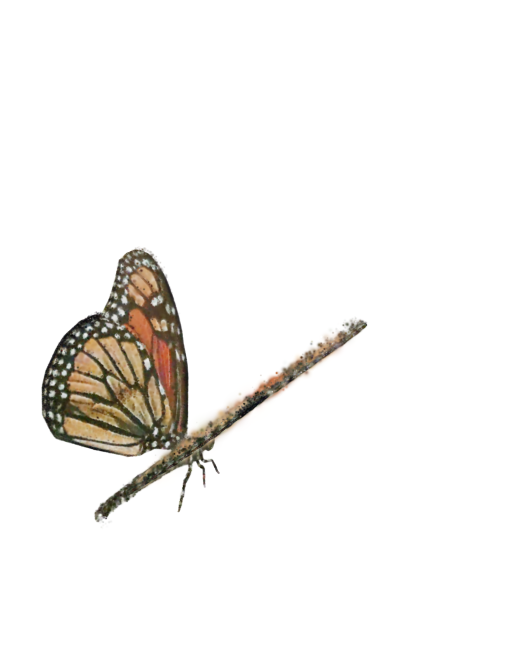}
     &   \includegraphics[width=\hsize,valign=m]{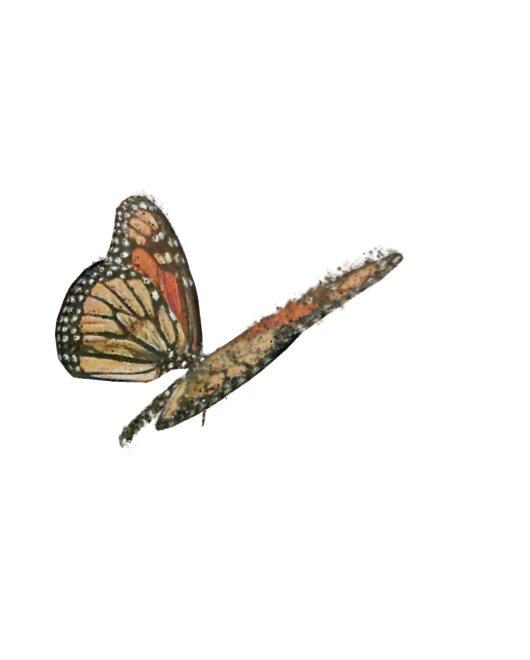}
    &   \includegraphics[width=\hsize,valign=m]{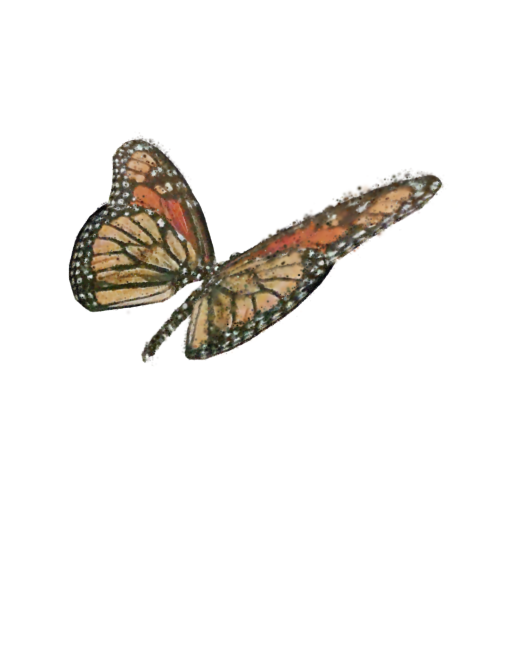}
    &   \includegraphics[width=\hsize,valign=m]{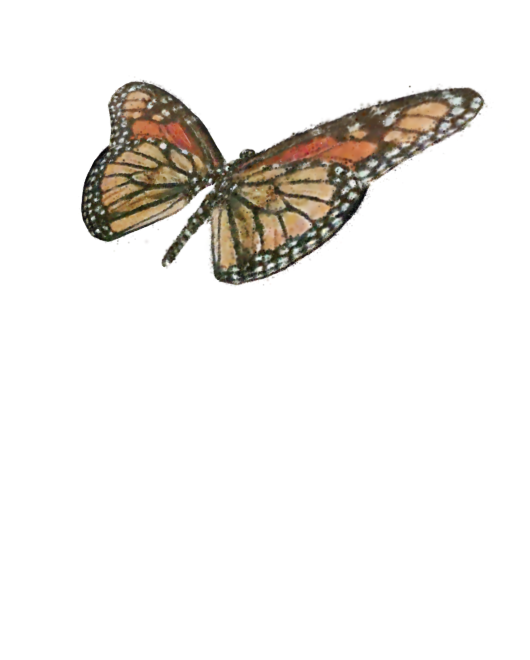}
    &   \includegraphics[width=\hsize,valign=m]{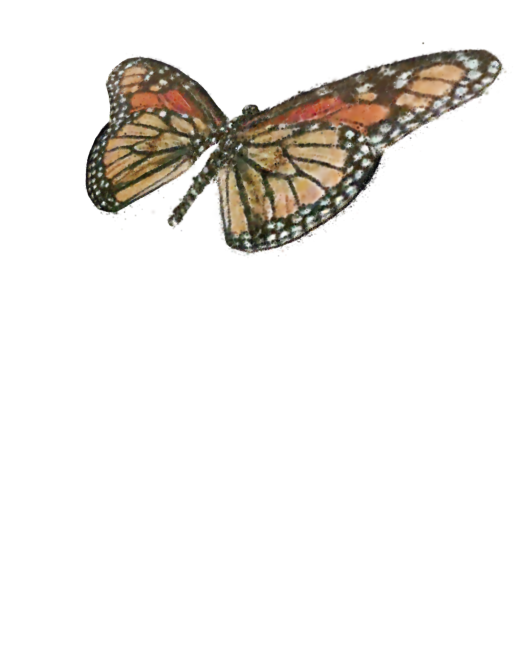}
    &   \includegraphics[width=\hsize,valign=m]{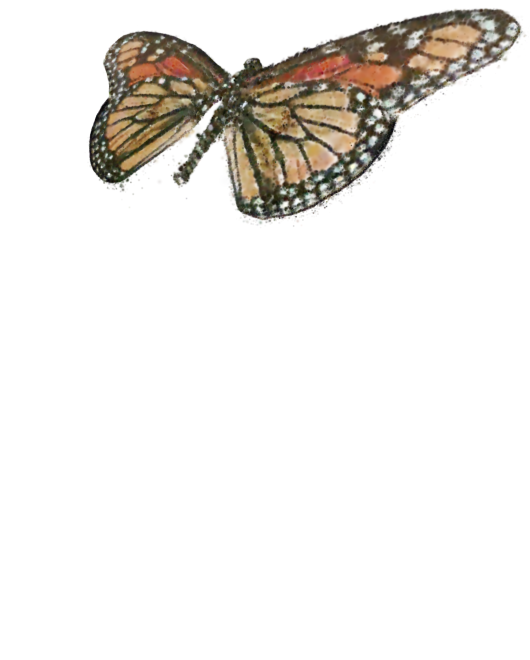}
    \\

    \midrule

    \includegraphics[width=\hsize,valign=m]{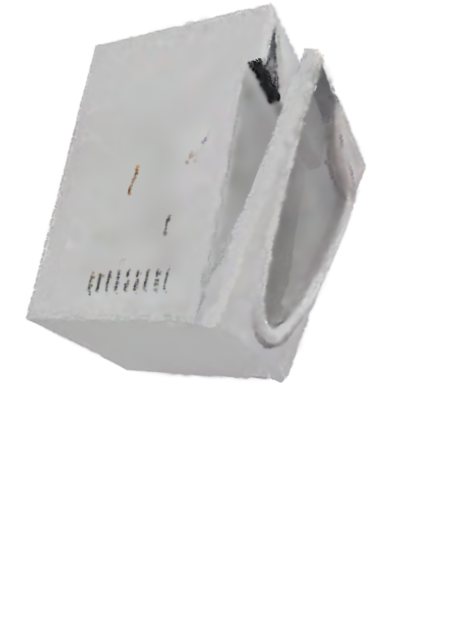}
    &   \includegraphics[width=\hsize,valign=m]{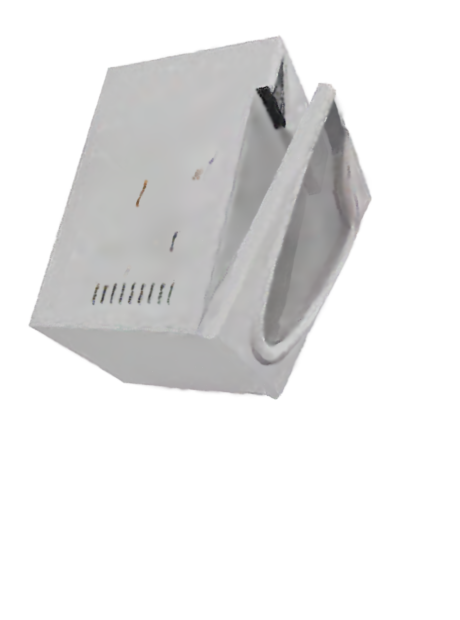}
    &   \includegraphics[width=\hsize,valign=m]{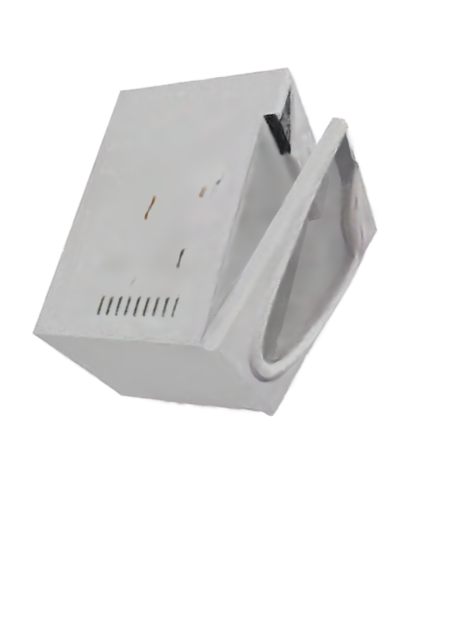}
    &   \includegraphics[width=\hsize,valign=m]{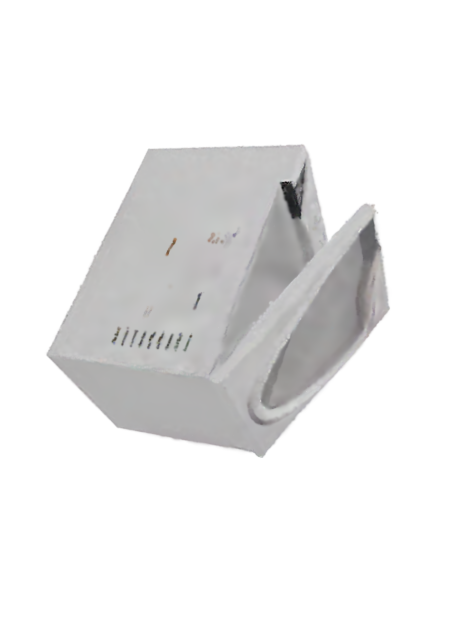}
     &   \includegraphics[width=\hsize,valign=m]{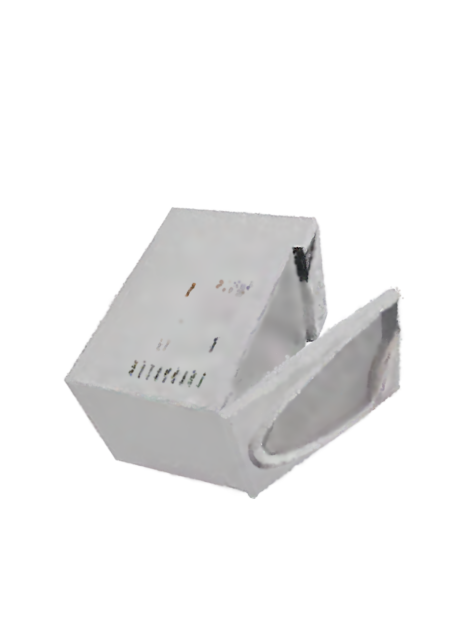}
    &   \includegraphics[width=\hsize,valign=m]{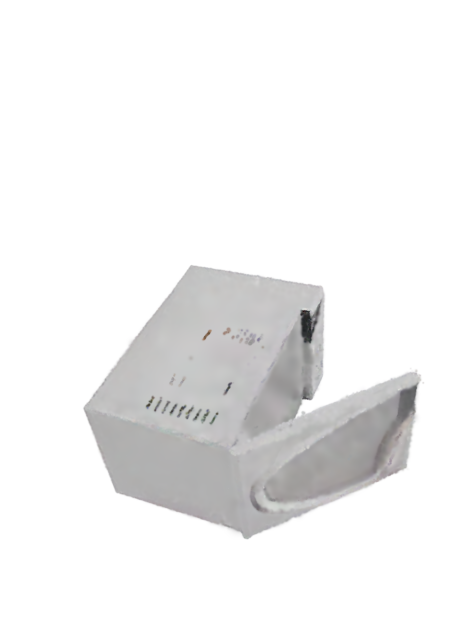}
    &   \includegraphics[width=\hsize,valign=m]{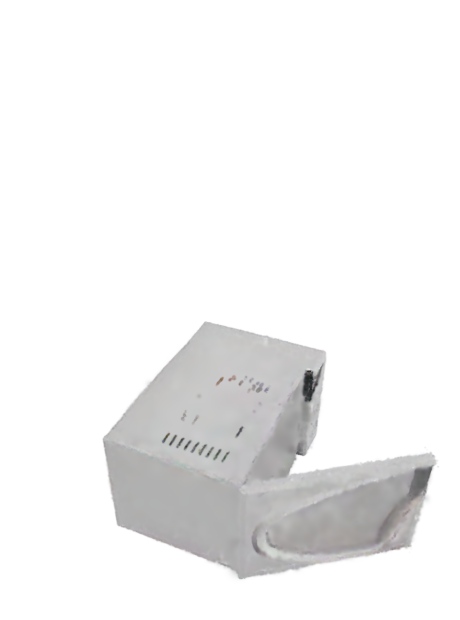}
    &   \includegraphics[width=\hsize,valign=m]{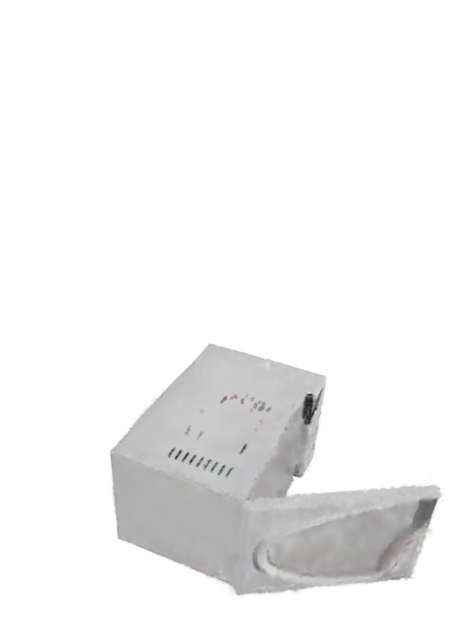}
    &   \includegraphics[width=\hsize,valign=m]{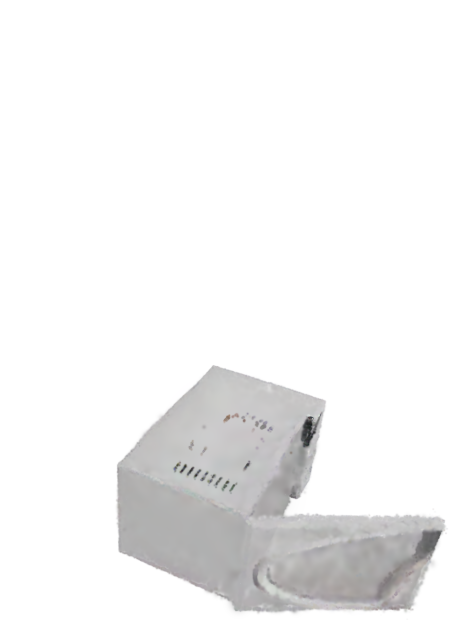}
    \\

    \midrule

    \includegraphics[width=\hsize,valign=m]{images/inter_extra/car/00009_-0.20.png}
    &   \includegraphics[width=\hsize,valign=m]{images/inter_extra/car/00009_-0.10.png}
    &   \includegraphics[width=\hsize,valign=m]{images/inter_extra/car/00009_0.00.png}
    &   \includegraphics[width=\hsize,valign=m]{images/inter_extra/car/00009_0.25.png}
     &   \includegraphics[width=\hsize,valign=m]{images/inter_extra/car/00009_0.50.png}
    &   \includegraphics[width=\hsize,valign=m]{images/inter_extra/car/00009_0.75.png}
    &   \includegraphics[width=\hsize,valign=m]{images/inter_extra/car/00009_1.00.png}
    &   \includegraphics[width=\hsize,valign=m]{images/inter_extra/car/00009_1.10.png}
    &   \includegraphics[width=\hsize,valign=m]{images/inter_extra/car/00009_1.20.png}
    \\ 

    \midrule

    \includegraphics[width=\hsize,valign=m]{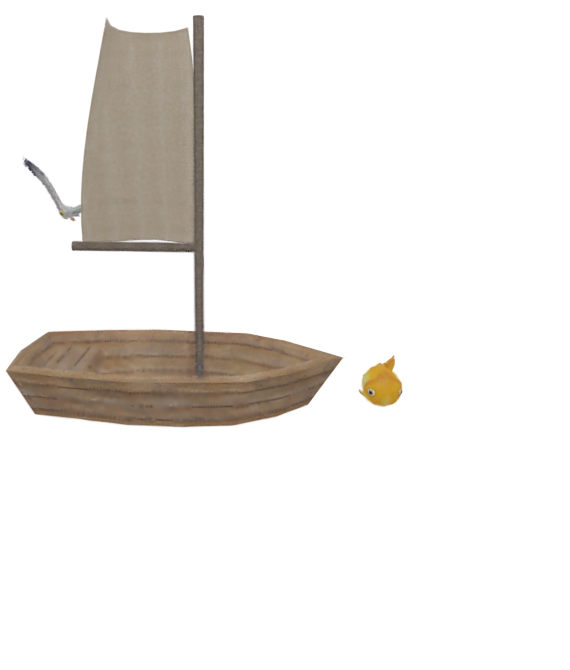} &
        \includegraphics[width=\hsize,valign=m]{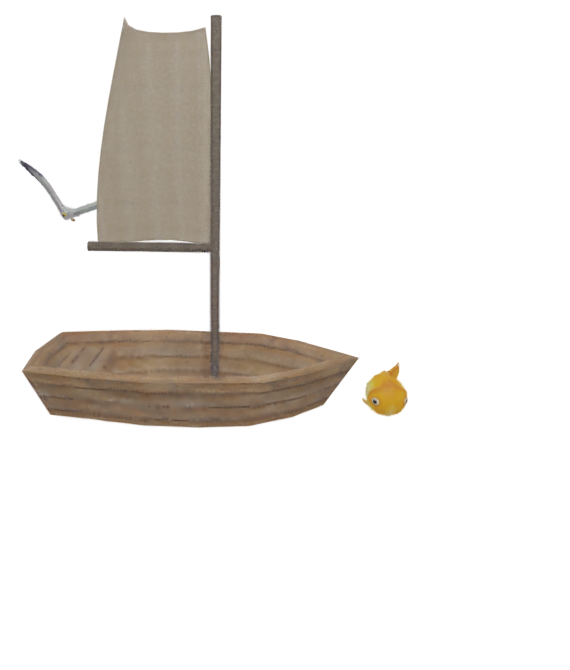} &
        \includegraphics[width=\hsize,valign=m]{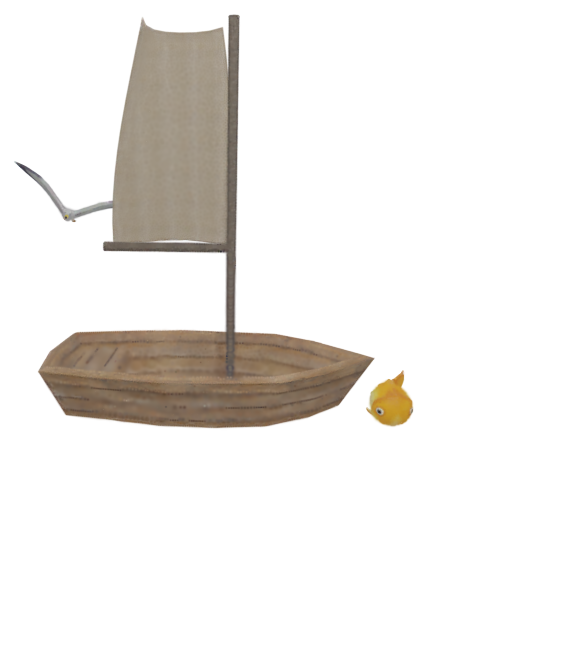} &
        \includegraphics[width=\hsize,valign=m]{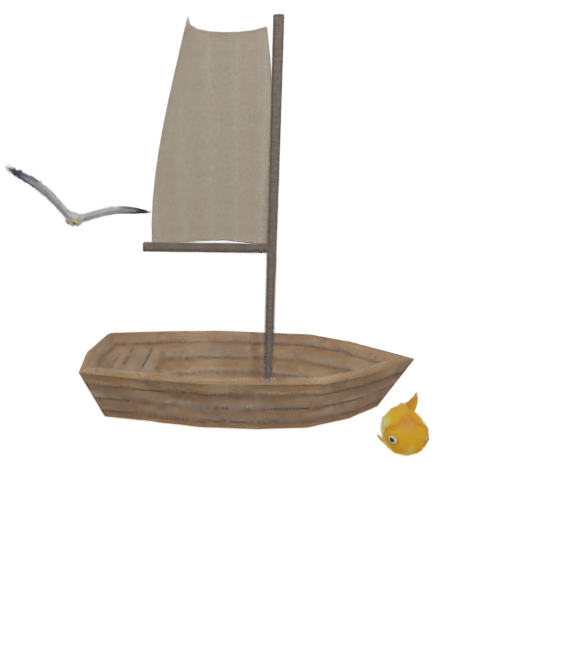} &
        \includegraphics[width=\hsize,valign=m]{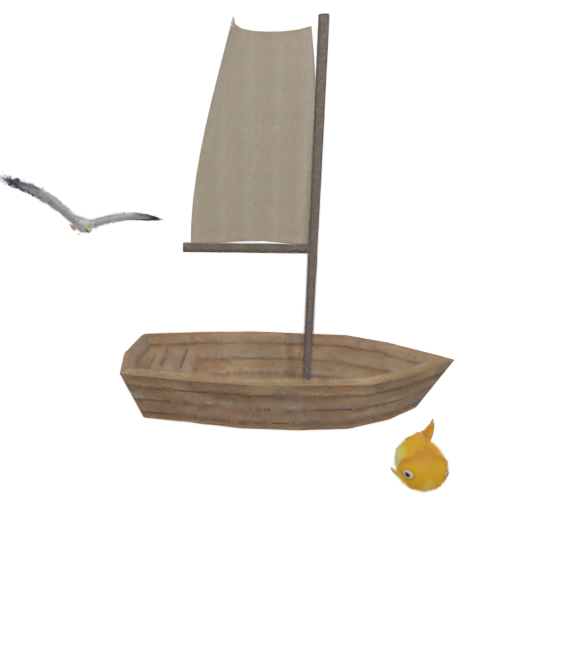} &
        \includegraphics[width=\hsize,valign=m]{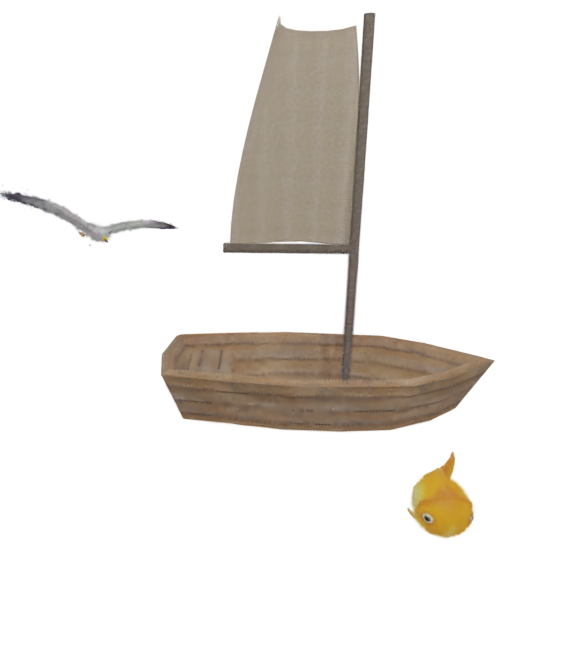} &
        \includegraphics[width=\hsize,valign=m]{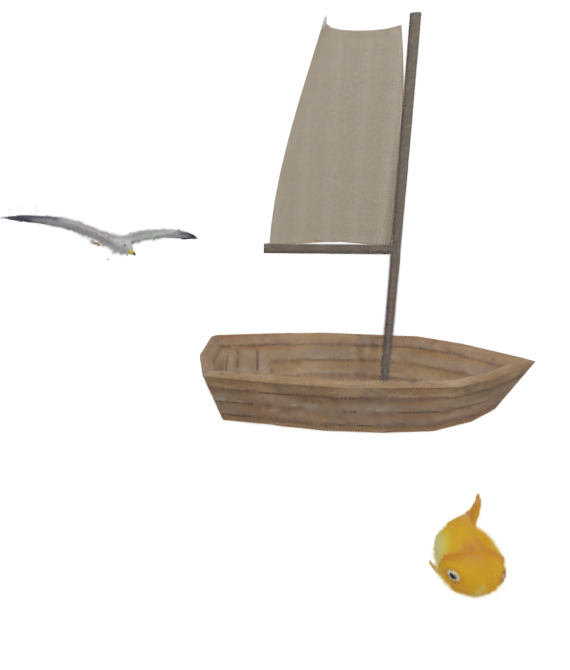} &
        \includegraphics[width=\hsize,valign=m]{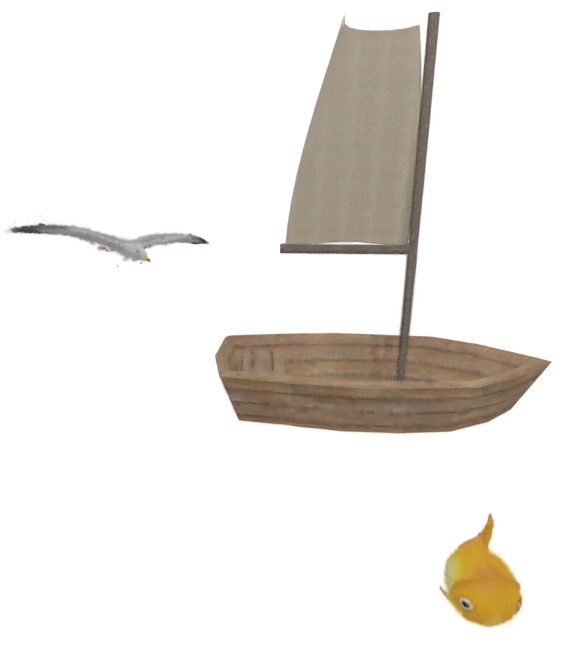} &
        \includegraphics[width=\hsize,valign=m]{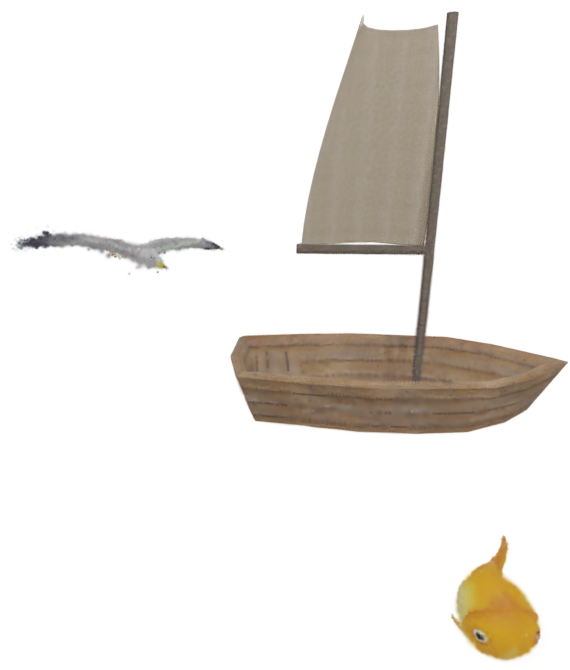} \\

    \midrule

    \includegraphics[width=\hsize,valign=m]{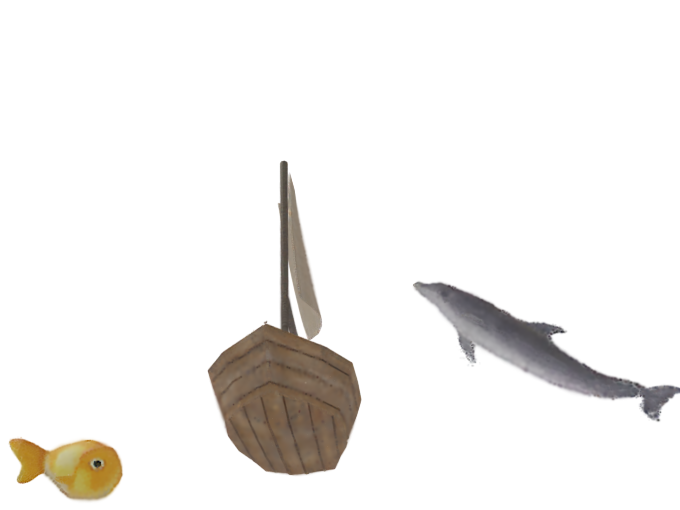}
    &   \includegraphics[width=\hsize,valign=m]{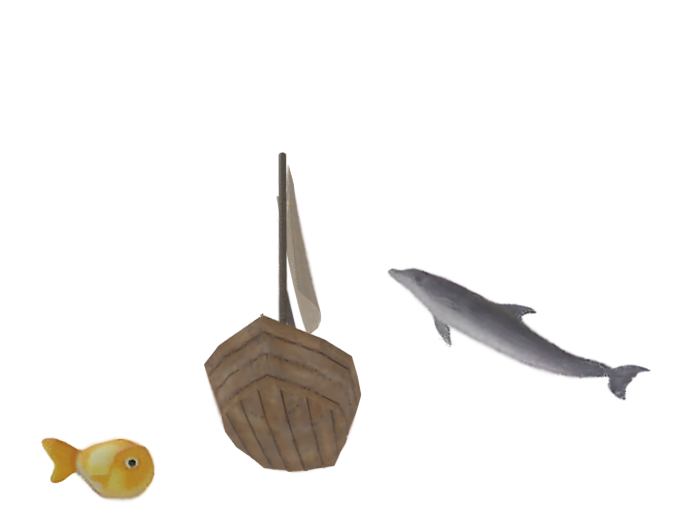}
    &   \includegraphics[width=\hsize,valign=m]{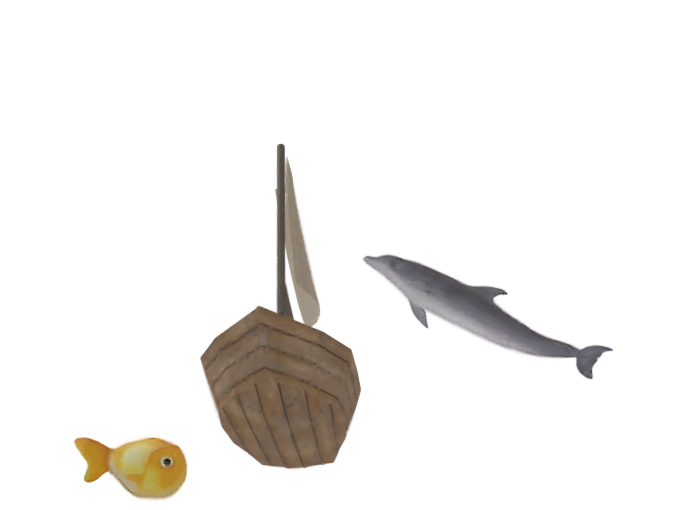}
    &   \includegraphics[width=\hsize,valign=m]{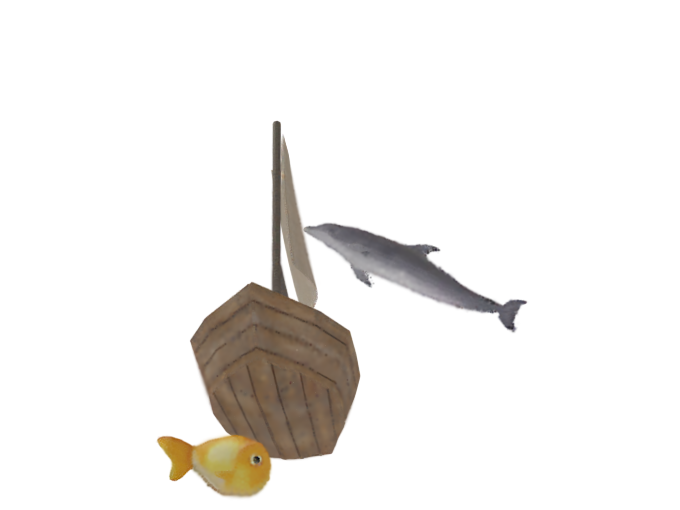}
     &   \includegraphics[width=\hsize,valign=m]{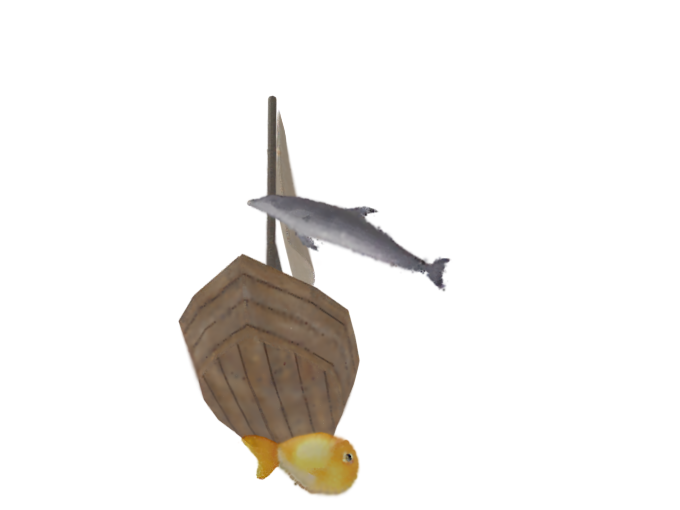}
    &   \includegraphics[width=\hsize,valign=m]{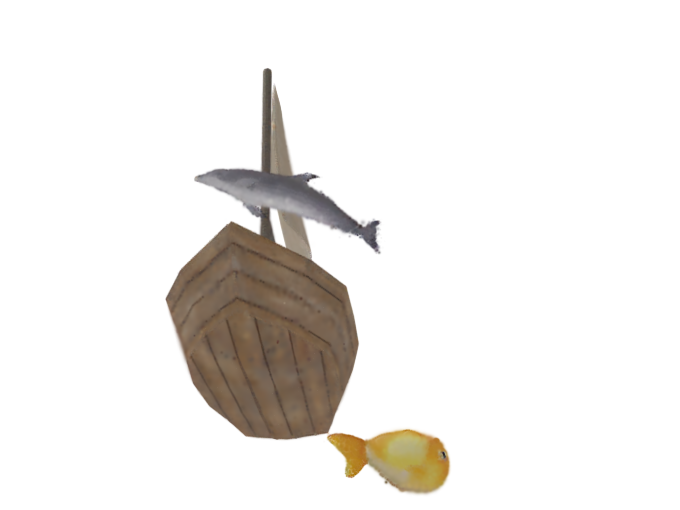}
    &   \includegraphics[width=\hsize,valign=m]{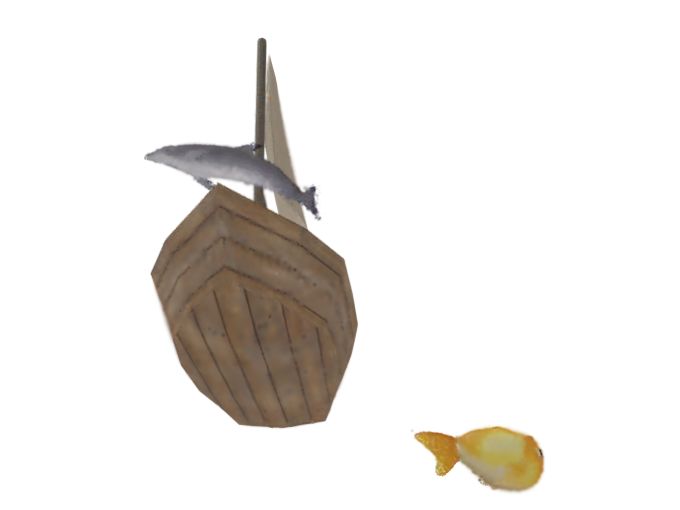}
    &   \includegraphics[width=\hsize,valign=m]{images/suppl_inter_extra/boat4/00049_1.00.png}
    &   \includegraphics[width=\hsize,valign=m]{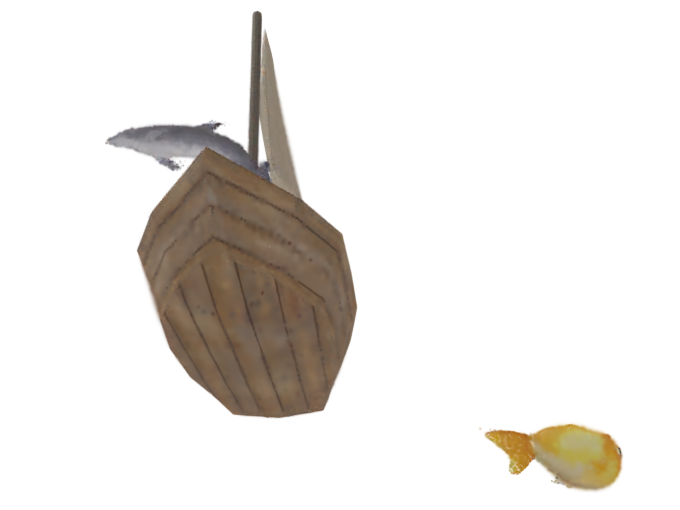}
    \\

    \midrule

    \includegraphics[width=\hsize,valign=m]{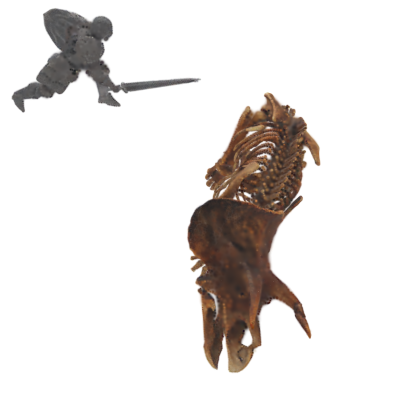}
    &   \includegraphics[width=\hsize,valign=m]{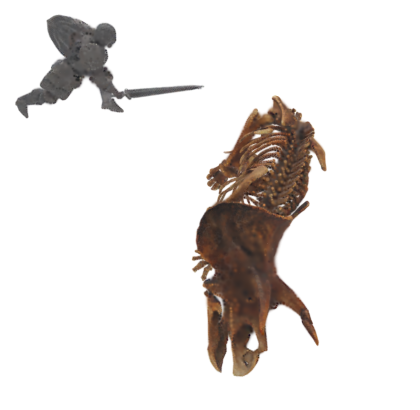}
    &   \includegraphics[width=\hsize,valign=m]{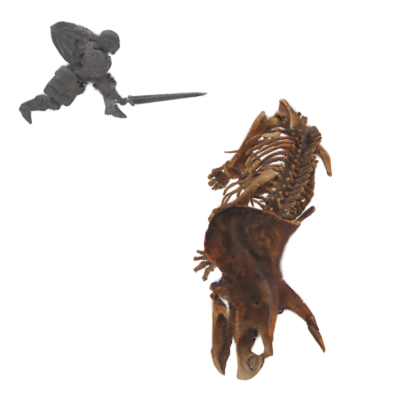}
    &   \includegraphics[width=\hsize,valign=m]{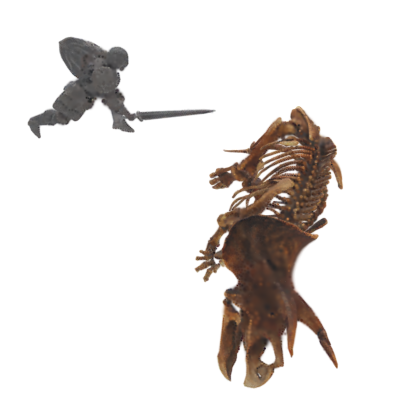}
     &   \includegraphics[width=\hsize,valign=m]{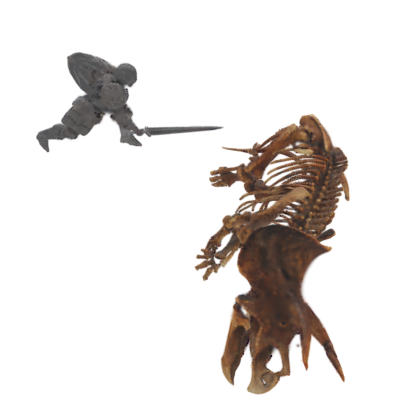}
    &   \includegraphics[width=\hsize,valign=m]{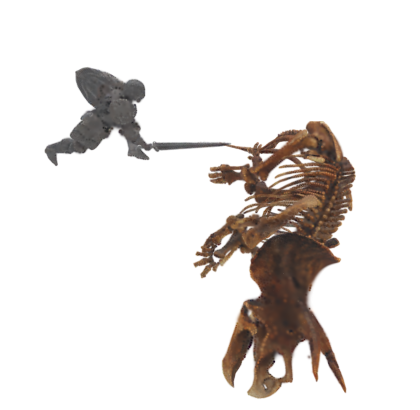}
    &   \includegraphics[width=\hsize,valign=m]{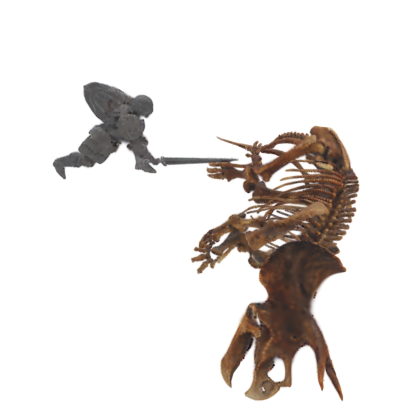}
    &   \includegraphics[width=\hsize,valign=m]{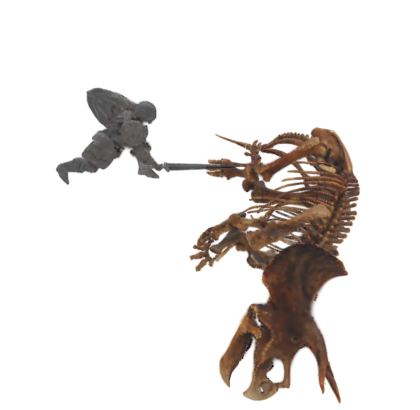}
    &   \includegraphics[width=\hsize,valign=m]{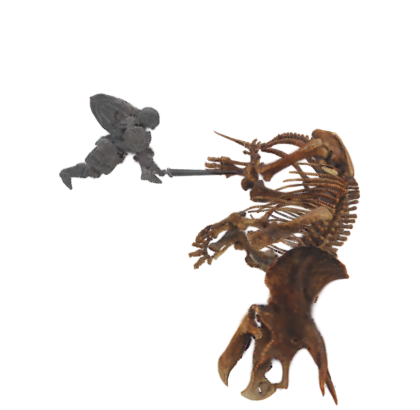}
    \\

    \midrule

    \includegraphics[width=\hsize,valign=m]{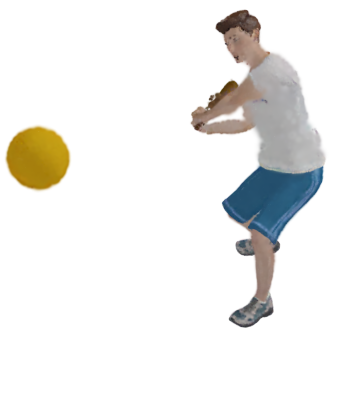}
    &   \includegraphics[width=\hsize,valign=m]{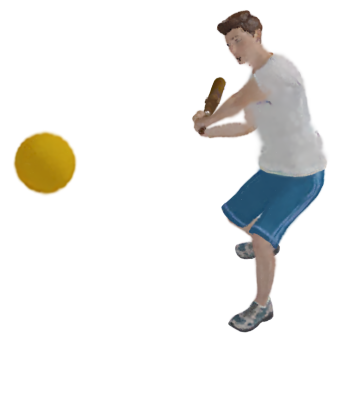}
    &   \includegraphics[width=\hsize,valign=m]{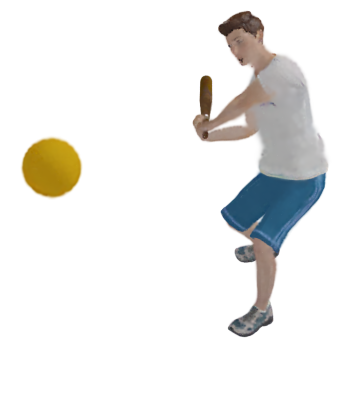}
    &   \includegraphics[width=\hsize,valign=m]{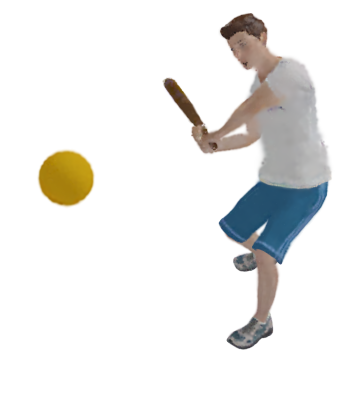}
     &   \includegraphics[width=\hsize,valign=m]{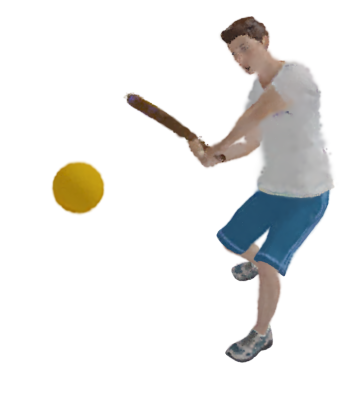}
    &   \includegraphics[width=\hsize,valign=m]{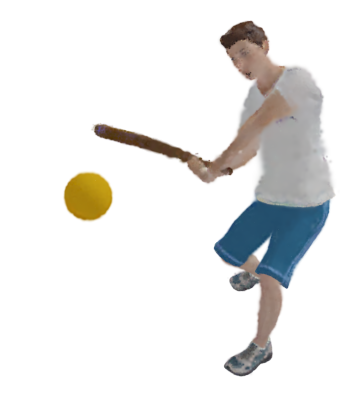}
    &   \includegraphics[width=\hsize,valign=m]{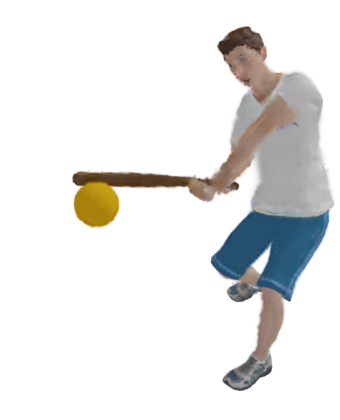}
    &   \includegraphics[width=\hsize,valign=m]{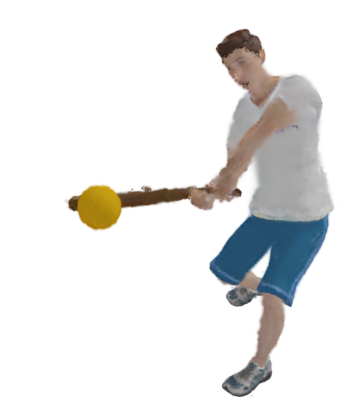}
    &   \includegraphics[width=\hsize,valign=m]{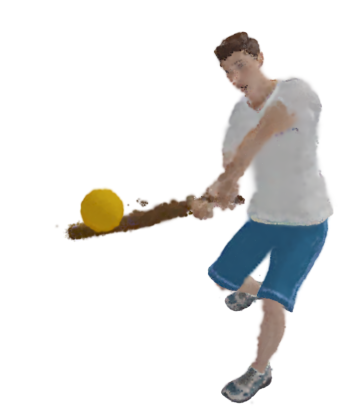}
    \\  

    \midrule

    \includegraphics[width=\hsize,valign=m]{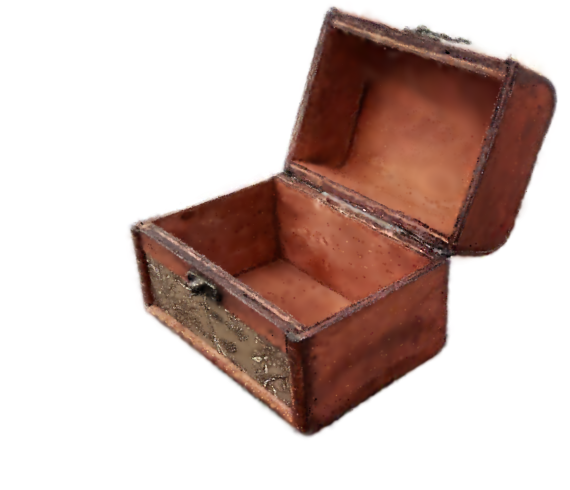}
    &   \includegraphics[width=\hsize,valign=m]{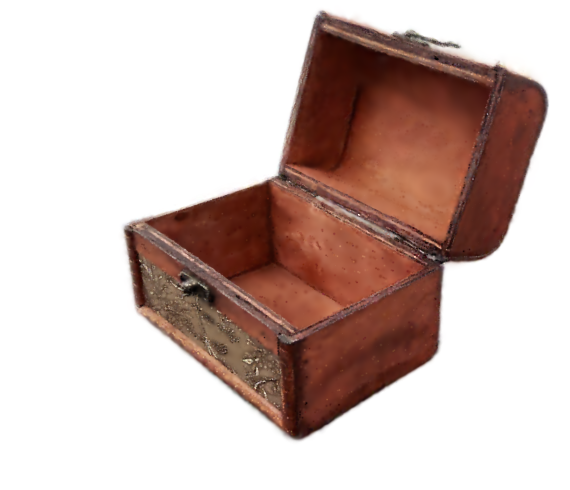}
    &   \includegraphics[width=\hsize,valign=m]{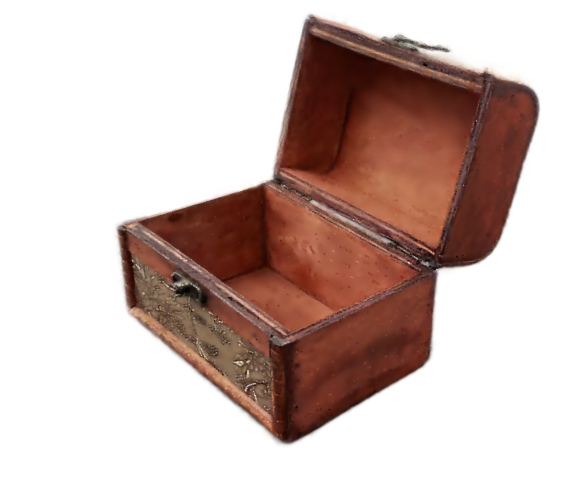}
    &   \includegraphics[width=\hsize,valign=m]{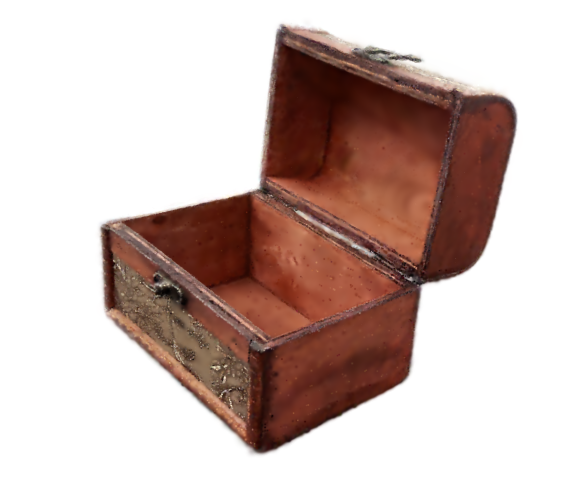}
     &   \includegraphics[width=\hsize,valign=m]{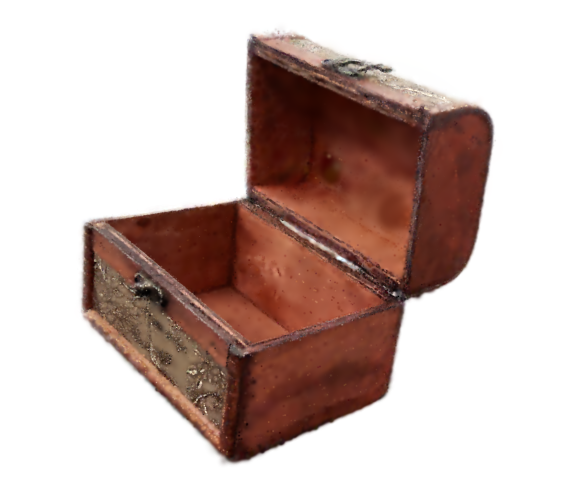}
    &   \includegraphics[width=\hsize,valign=m]{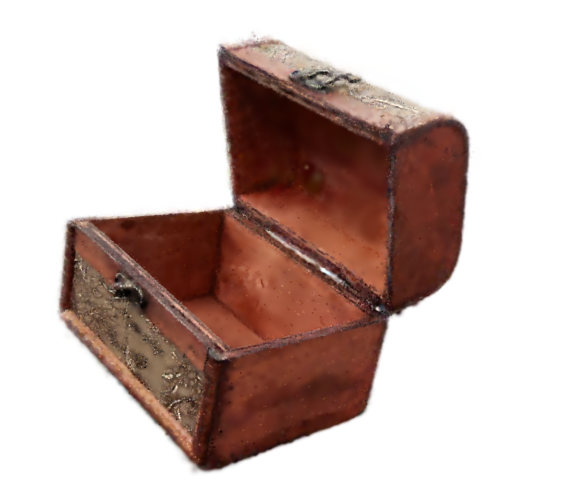}
    &   \includegraphics[width=\hsize,valign=m]{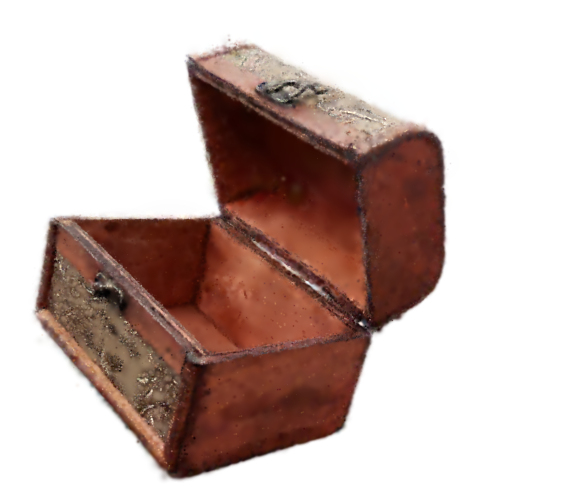}
    &   \includegraphics[width=\hsize,valign=m]{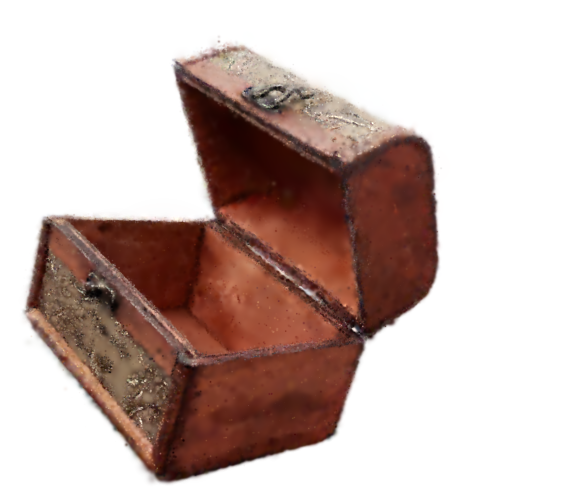}
    &   \includegraphics[width=\hsize,valign=m]{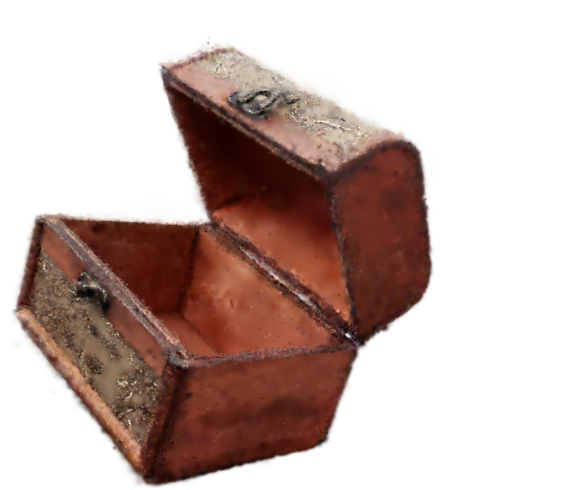}
    \\
    
    \midrule

    \includegraphics[width=\hsize,valign=m]{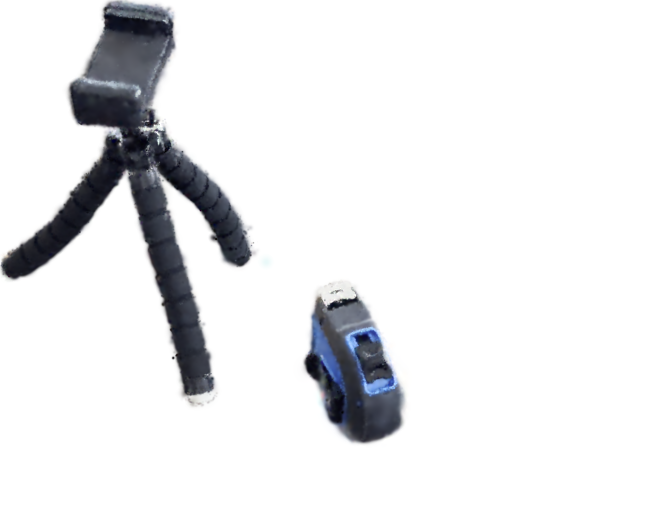}
    &   \includegraphics[width=\hsize,valign=m]{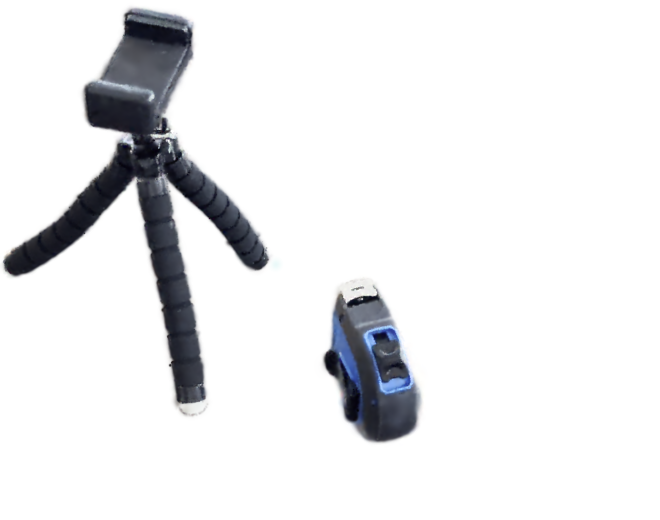}
    &   \includegraphics[width=\hsize,valign=m]{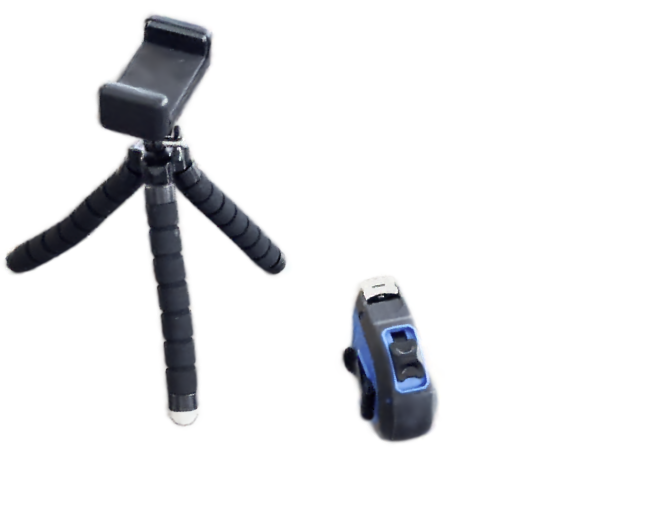}
    &   \includegraphics[width=\hsize,valign=m]{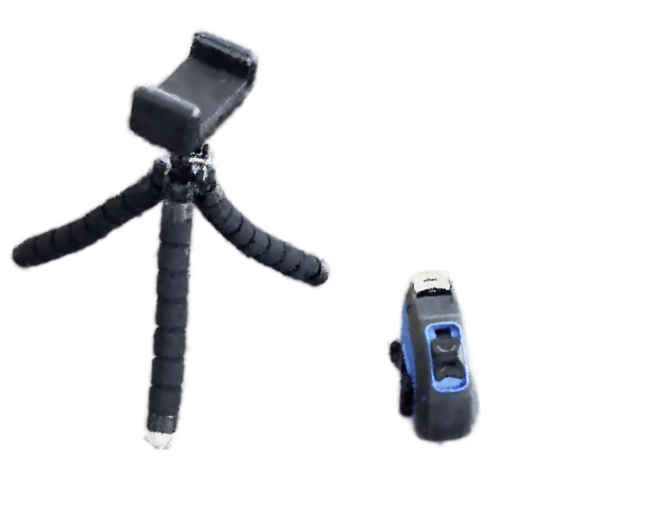}
     &   \includegraphics[width=\hsize,valign=m]{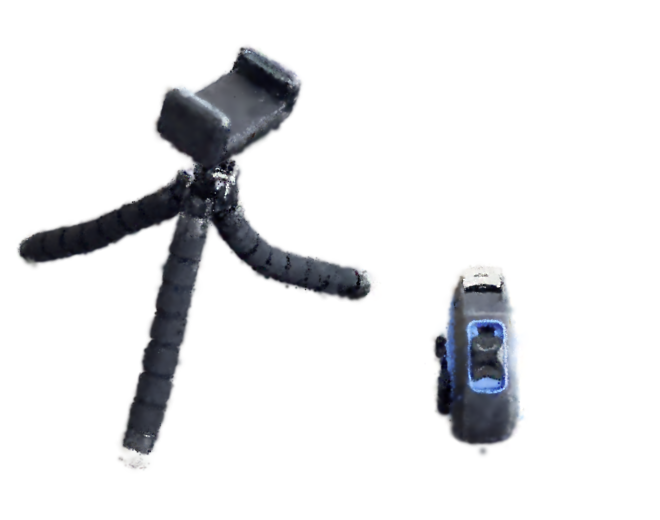}
    &   \includegraphics[width=\hsize,valign=m]{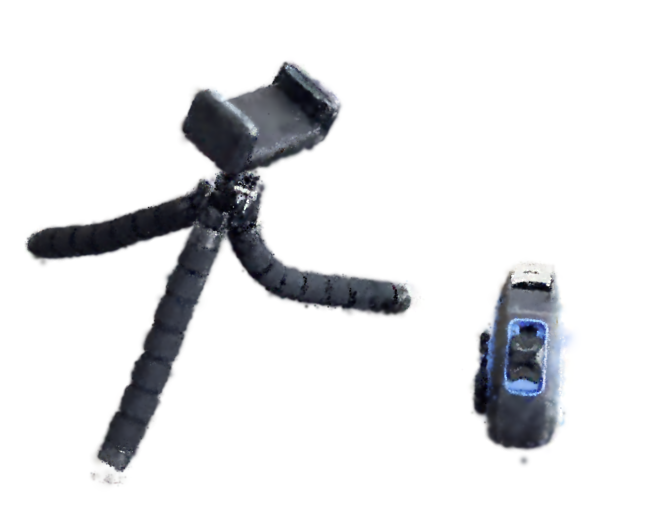}
    &   \includegraphics[width=\hsize,valign=m]{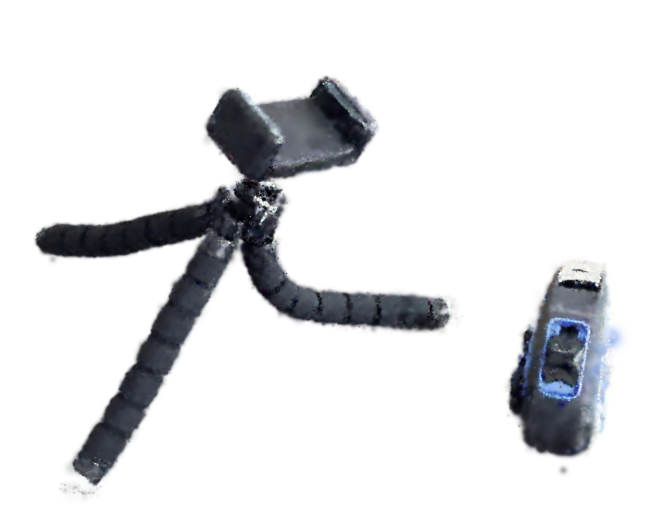}
    &   \includegraphics[width=\hsize,valign=m]{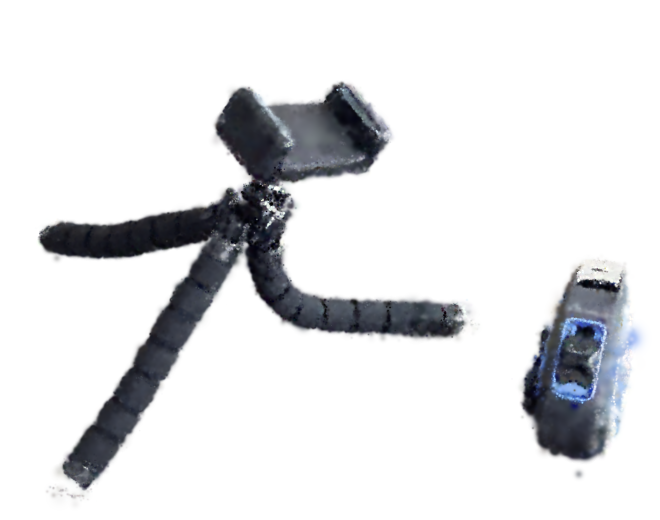}
    &   \includegraphics[width=\hsize,valign=m]{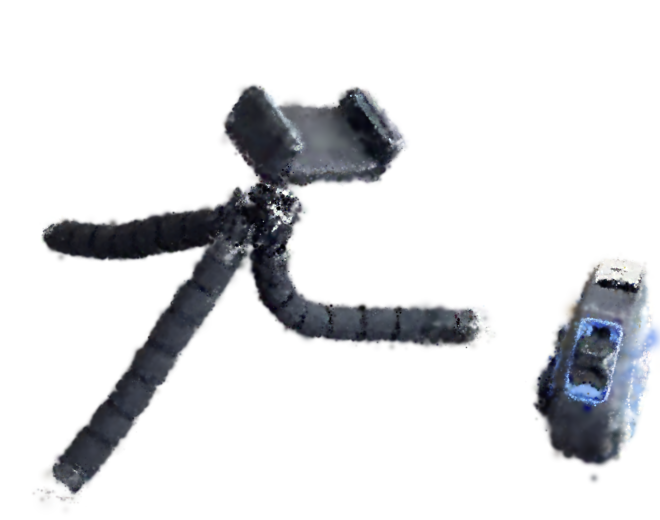}
    \\
    
    \midrule

    \includegraphics[width=\hsize,valign=m]{images/inter_extra/shoe/00029_-0.20.png}
    &   \includegraphics[width=\hsize,valign=m]{images/inter_extra/shoe/00029_-0.10.png}
    &   \includegraphics[width=\hsize,valign=m]{images/inter_extra/shoe/00029_0.00.png}
    &   \includegraphics[width=\hsize,valign=m]{images/inter_extra/shoe/00029_0.25.png}
     &   \includegraphics[width=\hsize,valign=m]{images/inter_extra/shoe/00029_0.50.png}
    &   \includegraphics[width=\hsize,valign=m]{images/inter_extra/shoe/00029_0.75.png}
    &   \includegraphics[width=\hsize,valign=m]{images/inter_extra/shoe/00029_1.00.png}
    &   \includegraphics[width=\hsize,valign=m]{images/inter_extra/shoe/00029_1.10.png}
    &   \includegraphics[width=\hsize,valign=m]{images/inter_extra/shoe/00029_1.20.png}
    \\

\end{tabularx}
\caption{\textbf{Additional Interpolation and Extrapolation Results.} The figure presents interpolation and extrapolation novel-view synthesis results using our method on the global-motion dataset. From top to bottom, the scenes displayed are \texttt{Dolphin}, \texttt{Butterfly}, \texttt{Microwave}, \texttt{Car}, \texttt{Seagull}, \texttt{Boat}, \texttt{Knight}, \texttt{Ball}, \texttt{Box}, \texttt{tapeline}, and \texttt{Shoe}. The top nine scenes are synthetic, and the bottom three are real-world. The nine columns correspond to nine timesteps: $\{-0.20, -0.10, 0.00, 0.25, 0.50, 0.75, 1.00, 1.10. 1.20\}$.}
\label{fig:suppl_inter_extra_global}
\end{figure*}

%% file: figures/suppl_inter_extra_papr.tex
\begin{figure*}[t]
\centering
\footnotesize
\begin{tabularx}{\linewidth}{Y|Y|Y|Y|Y|Y|Y|Y|Y}
    \multicolumn{2}{c}{\cellcolor[HTML]{e6b8b5}{Extrapolation}} & \multicolumn{1}{c}{\cellcolor[HTML]{FFCCC9}{Start}} & \multicolumn{3}{c}{\cellcolor[HTML]{ebdbe5}{Interpolation}} & \multicolumn{1}{c}{\cellcolor[HTML]{DAE8FC}{End}} & \multicolumn{2}{c}{\cellcolor[HTML]{cfdcef}{Extrapolation}} 

    \vspace{1em}
    \\

    \includegraphics[width=\hsize,valign=m]{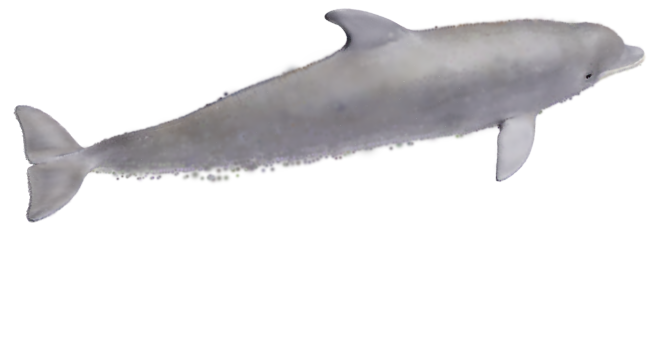}
    &   \includegraphics[width=\hsize,valign=m]{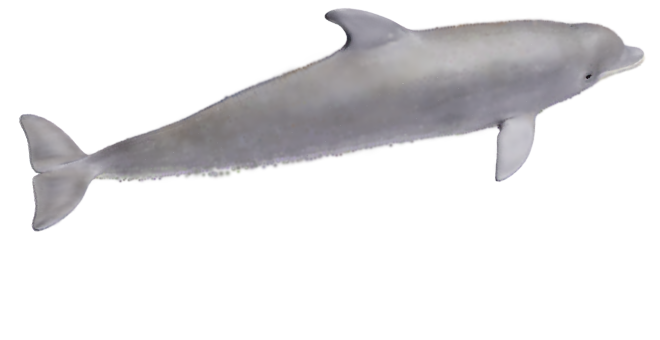}
    &   \includegraphics[width=\hsize,valign=m]{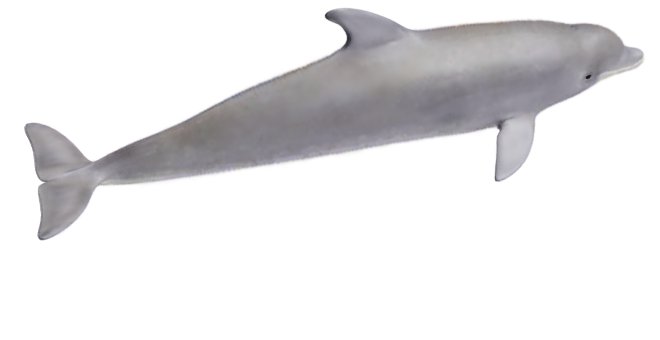}
    &   \includegraphics[width=\hsize,valign=m]{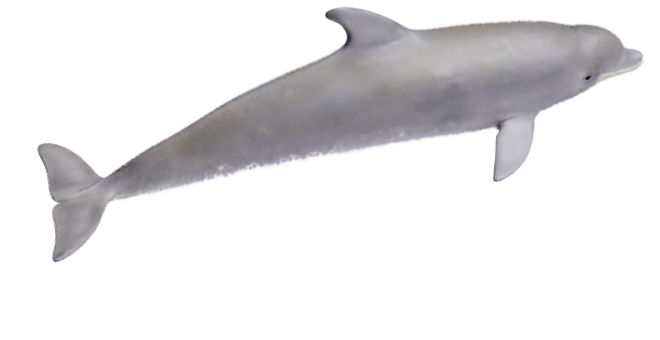}
     &   \includegraphics[width=\hsize,valign=m]{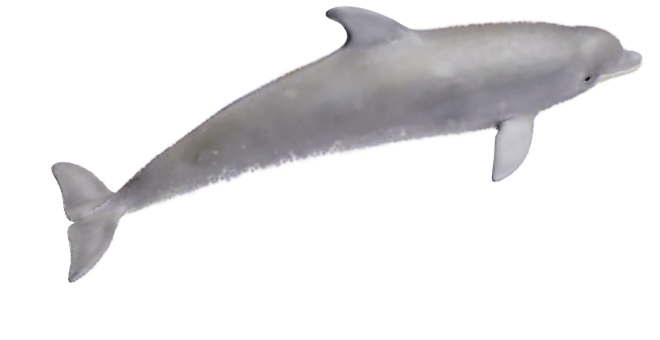}
    &   \includegraphics[width=\hsize,valign=m]{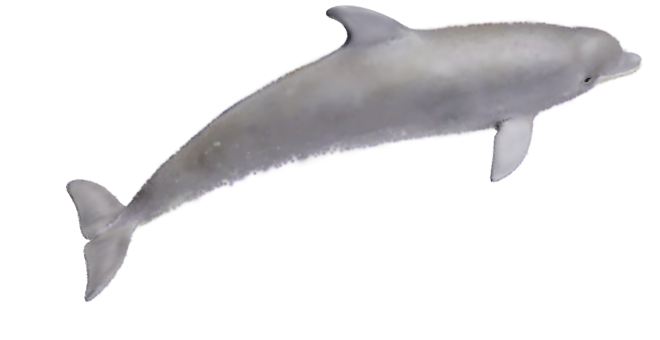}
    &   \includegraphics[width=\hsize,valign=m]{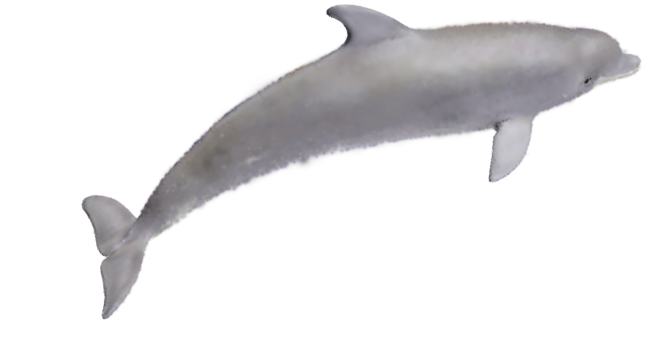}
    &   \includegraphics[width=\hsize,valign=m]{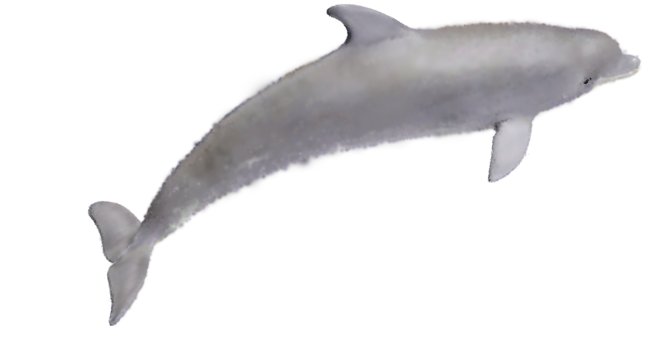}
    &   \includegraphics[width=\hsize,valign=m]{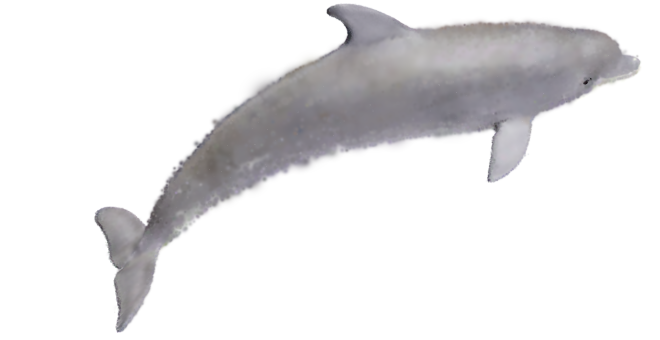}
    \\

    \midrule

    \includegraphics[width=\hsize,valign=m]{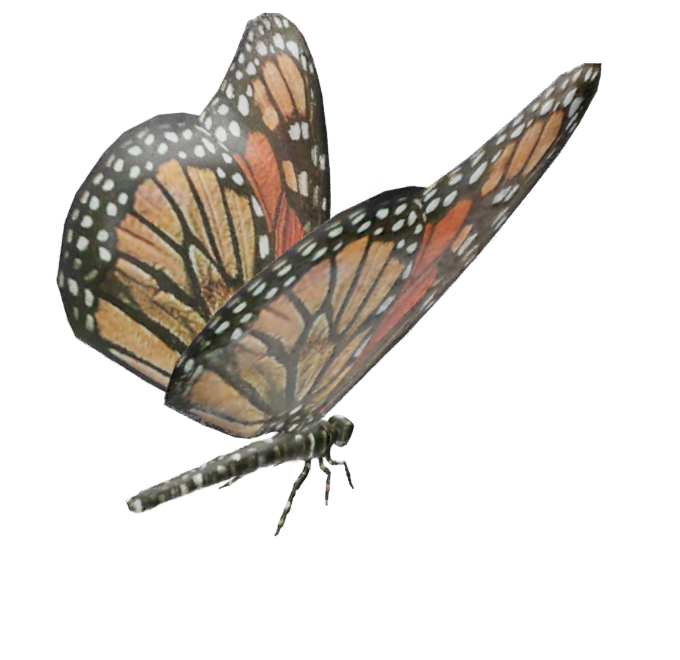}
    &   \includegraphics[width=\hsize,valign=m]{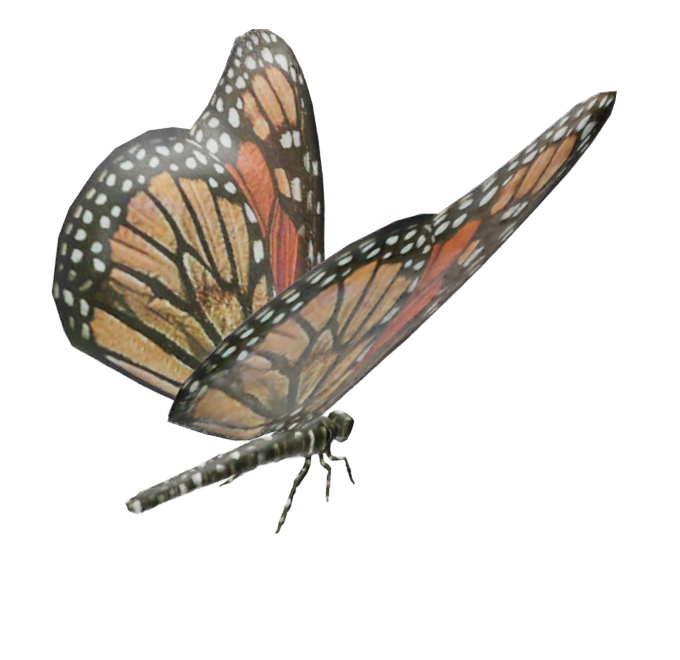}
    &   \includegraphics[width=\hsize,valign=m]{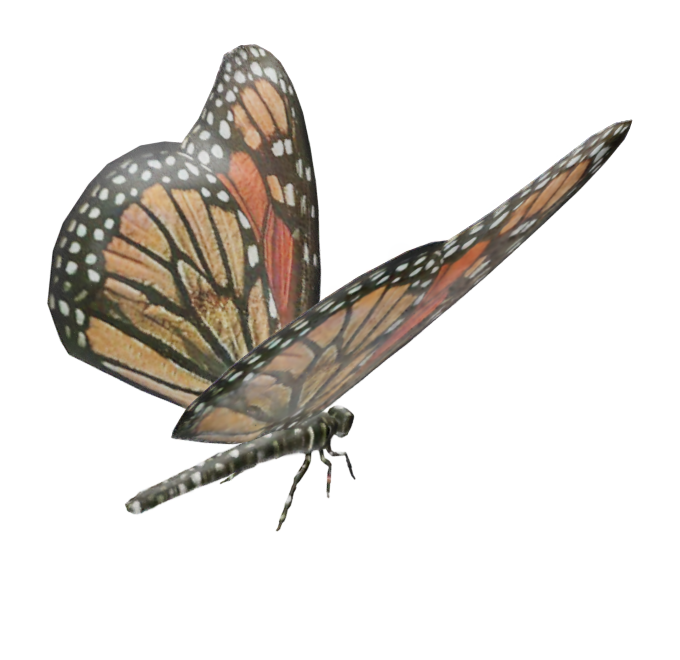}
    &   \includegraphics[width=\hsize,valign=m]{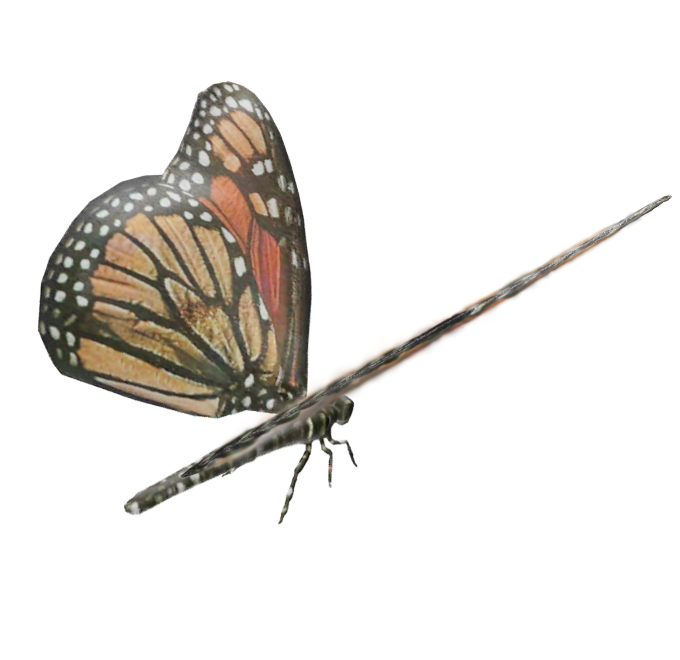}
     &   \includegraphics[width=\hsize,valign=m]{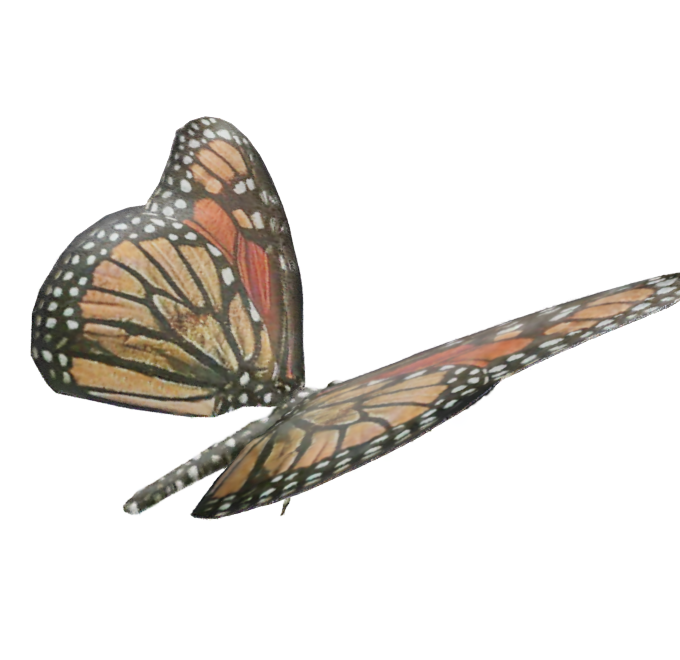}
    &   \includegraphics[width=\hsize,valign=m]{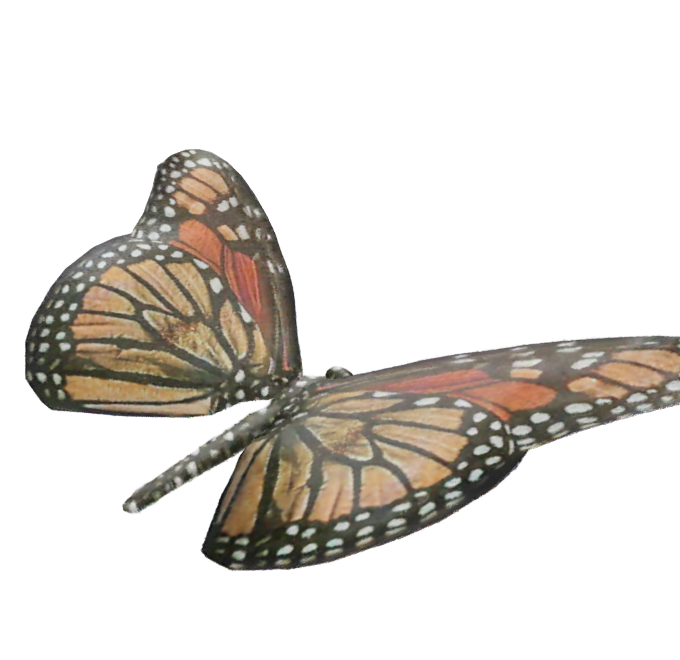}
    &   \includegraphics[width=\hsize,valign=m]{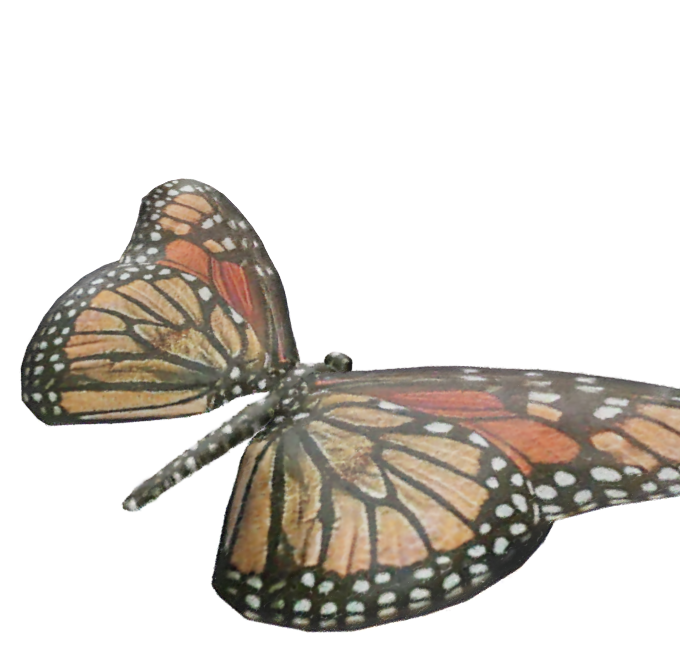}
    &   \includegraphics[width=\hsize,valign=m]{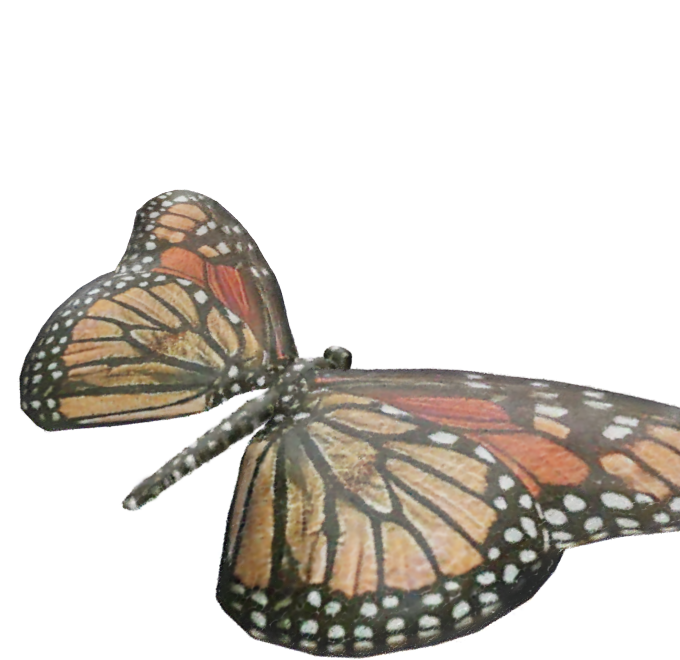}
    &   \includegraphics[width=\hsize,valign=m]{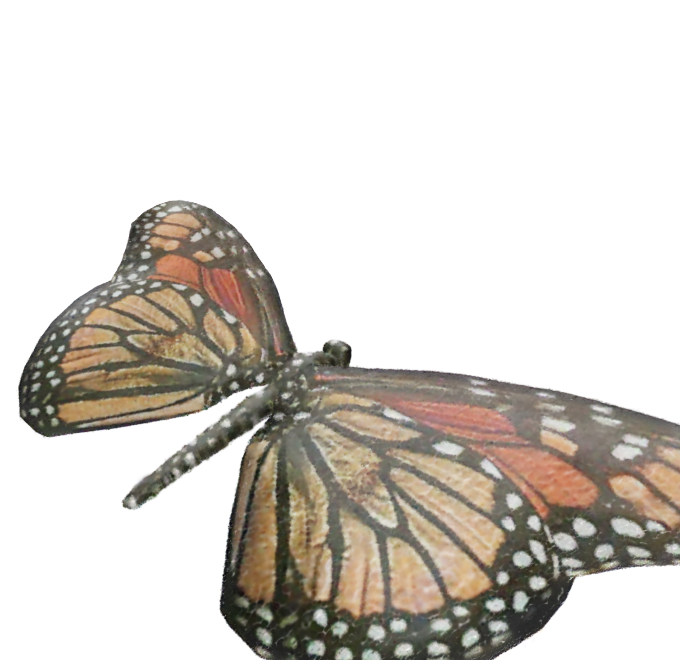}
    \\

    \midrule

    \includegraphics[width=\hsize,valign=m]{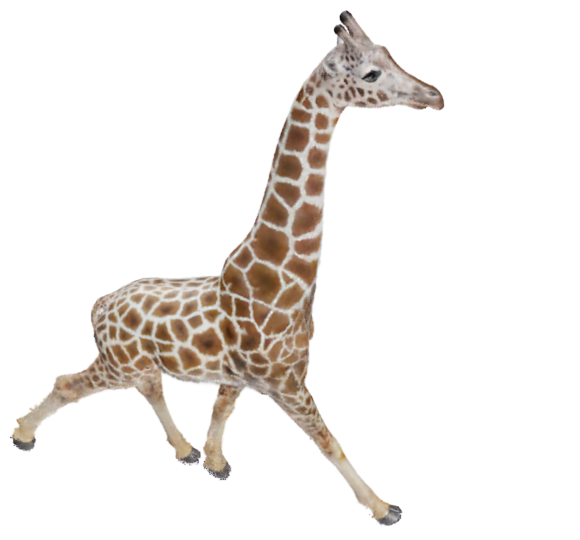}
    &   \includegraphics[width=\hsize,valign=m]{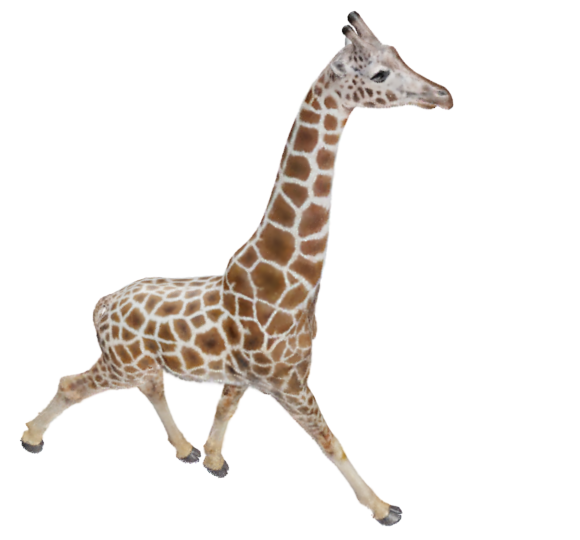}
    &   \includegraphics[width=\hsize,valign=m]{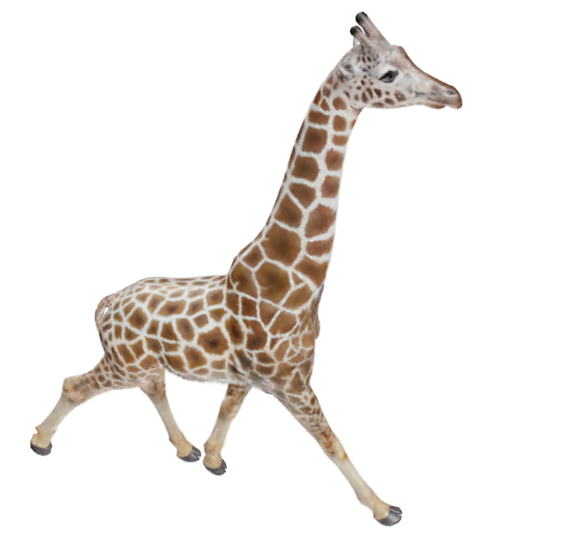}
    &   \includegraphics[width=\hsize,valign=m]{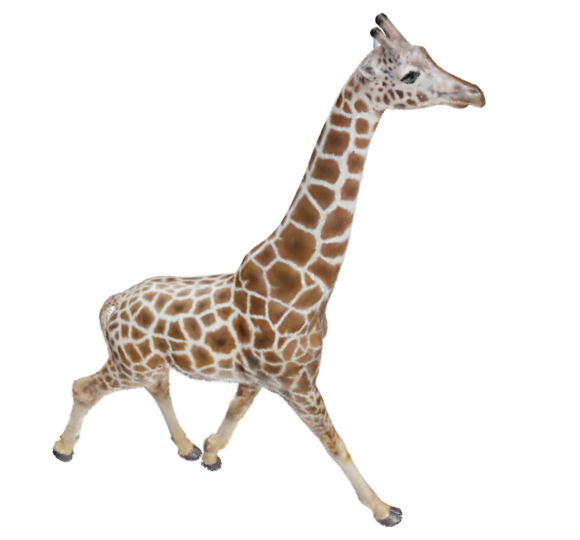}
     &   \includegraphics[width=\hsize,valign=m]{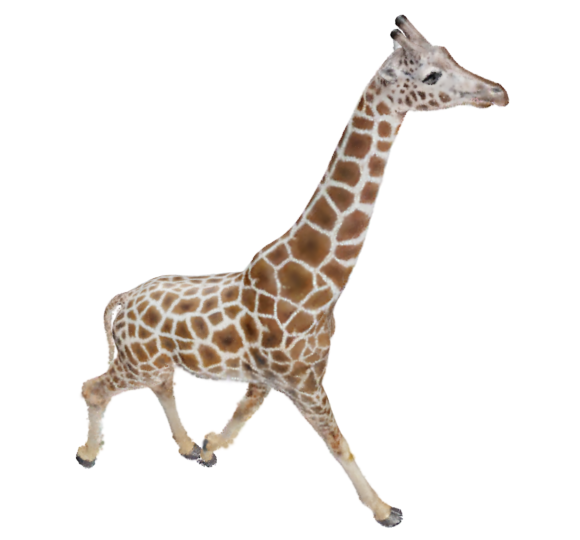}
    &   \includegraphics[width=\hsize,valign=m]{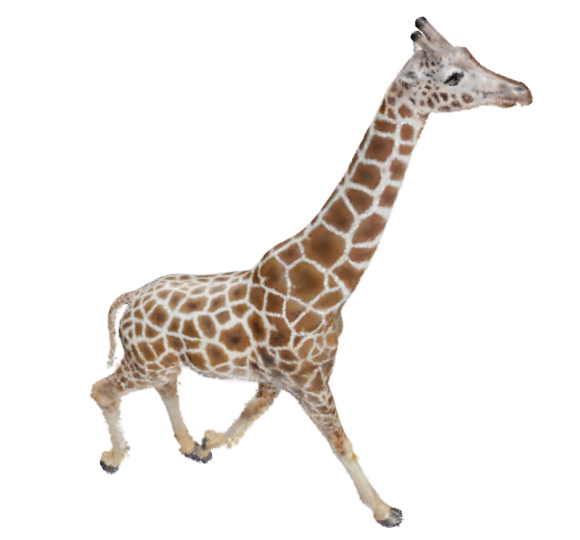}
    &   \includegraphics[width=\hsize,valign=m]{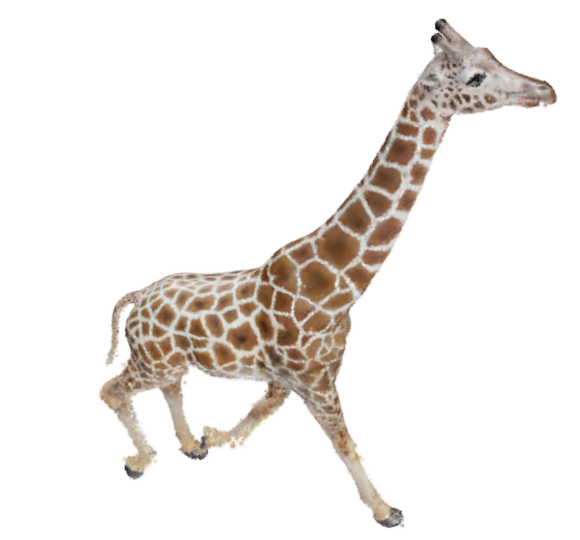}
    &   \includegraphics[width=\hsize,valign=m]{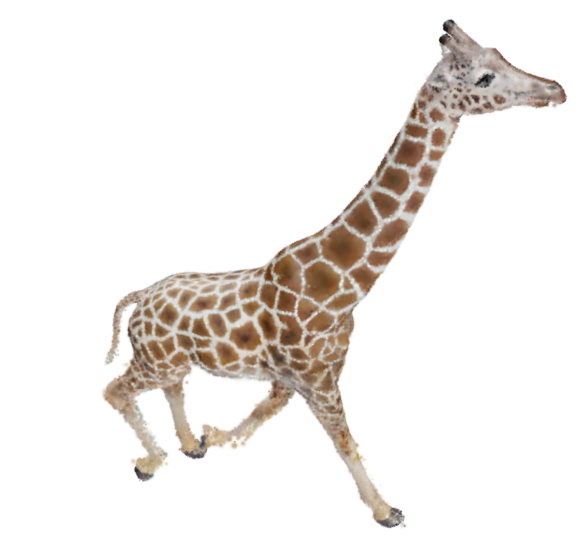}
    &   \includegraphics[width=\hsize,valign=m]{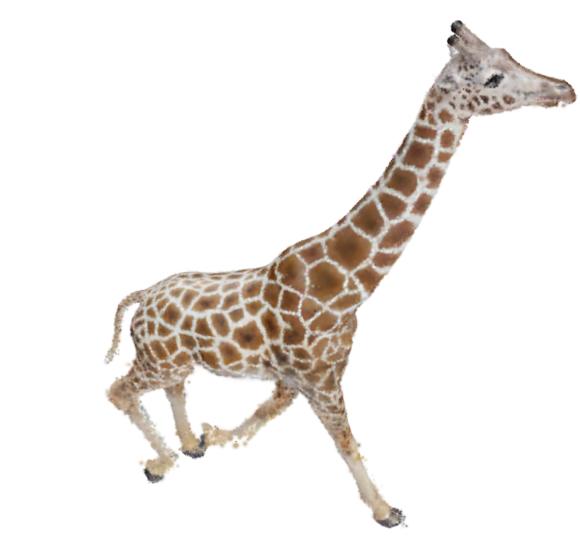}
    \\

    \midrule

    \includegraphics[width=\hsize,valign=m]{images/inter_extra/crab/00009_-0.40.png}
    &   \includegraphics[width=\hsize,valign=m]{images/inter_extra/crab/00009_-0.20.png}
    &   \includegraphics[width=\hsize,valign=m]{images/inter_extra/crab/00009_0.00.png}
    &   \includegraphics[width=\hsize,valign=m]{images/inter_extra/crab/00009_0.25.png}
     &   \includegraphics[width=\hsize,valign=m]{images/inter_extra/crab/00009_0.50.png}
    &   \includegraphics[width=\hsize,valign=m]{images/inter_extra/crab/00009_0.75.png}
    &   \includegraphics[width=\hsize,valign=m]{images/inter_extra/crab/00009_1.00.png}
    &   \includegraphics[width=\hsize,valign=m]{images/inter_extra/crab/00009_1.20.png}
    &   \includegraphics[width=\hsize,valign=m]{images/inter_extra/crab/00009_1.40.png}
    \\  

    \midrule

    \includegraphics[width=\hsize,valign=m]{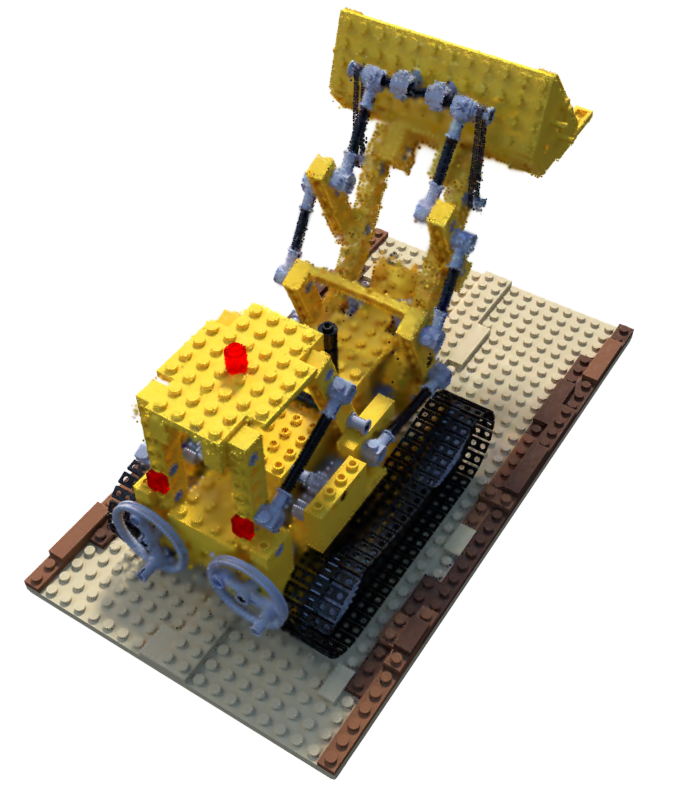}
    &   \includegraphics[width=\hsize,valign=m]{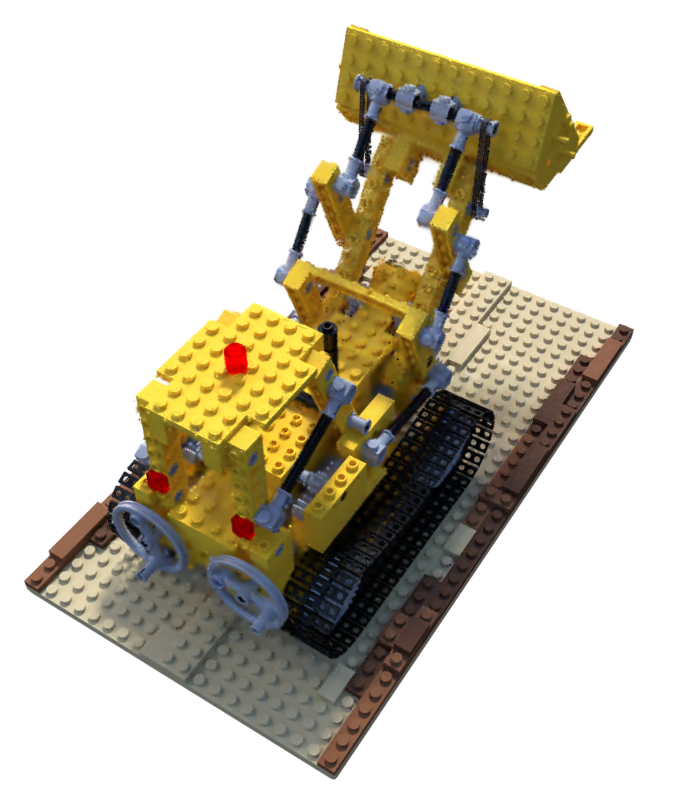}
    &   \includegraphics[width=\hsize,valign=m]{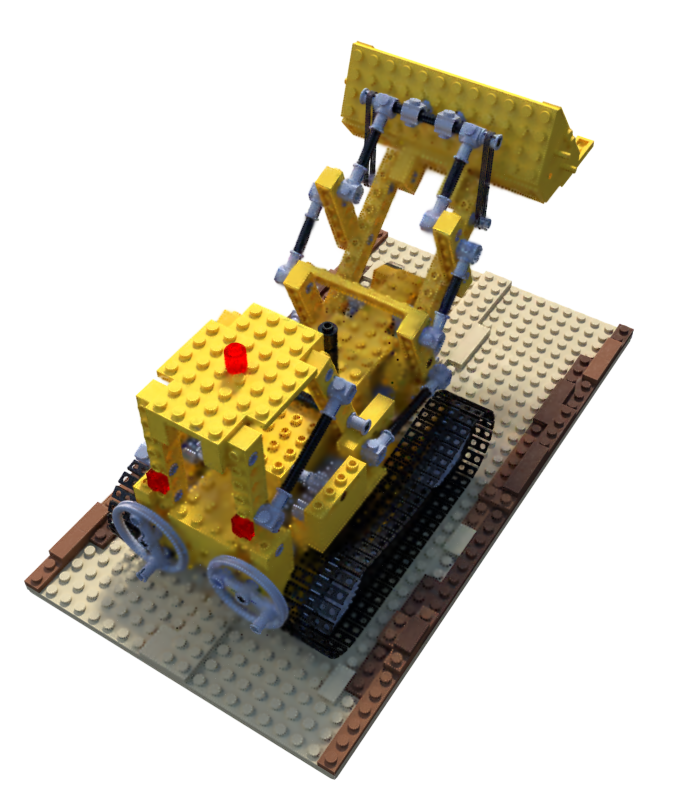}
    &   \includegraphics[width=\hsize,valign=m]{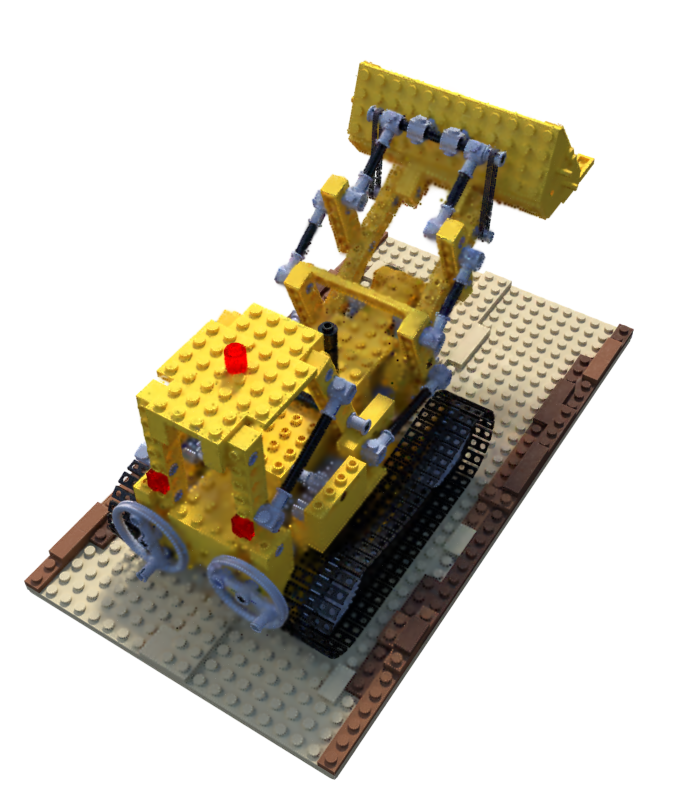}
     &   \includegraphics[width=\hsize,valign=m]{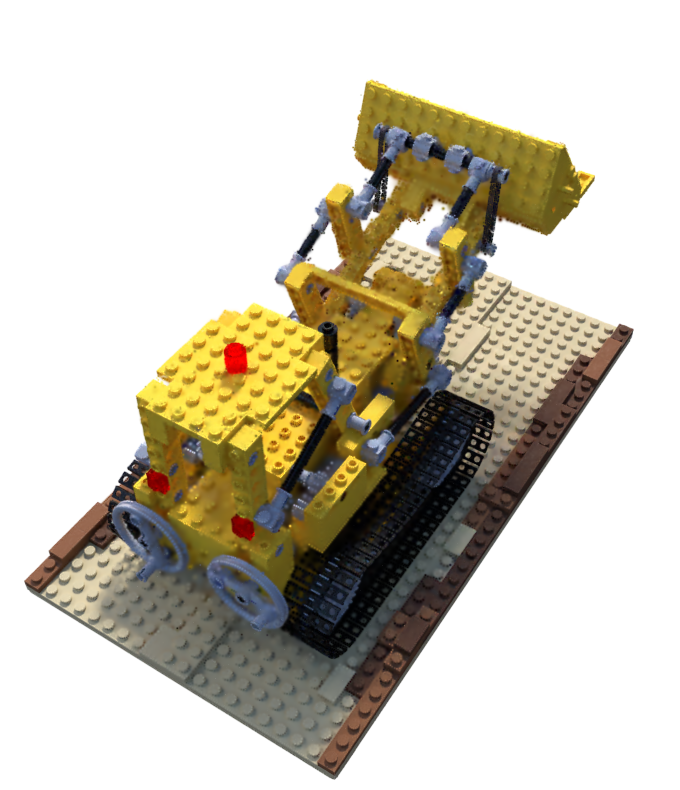}
    &   \includegraphics[width=\hsize,valign=m]{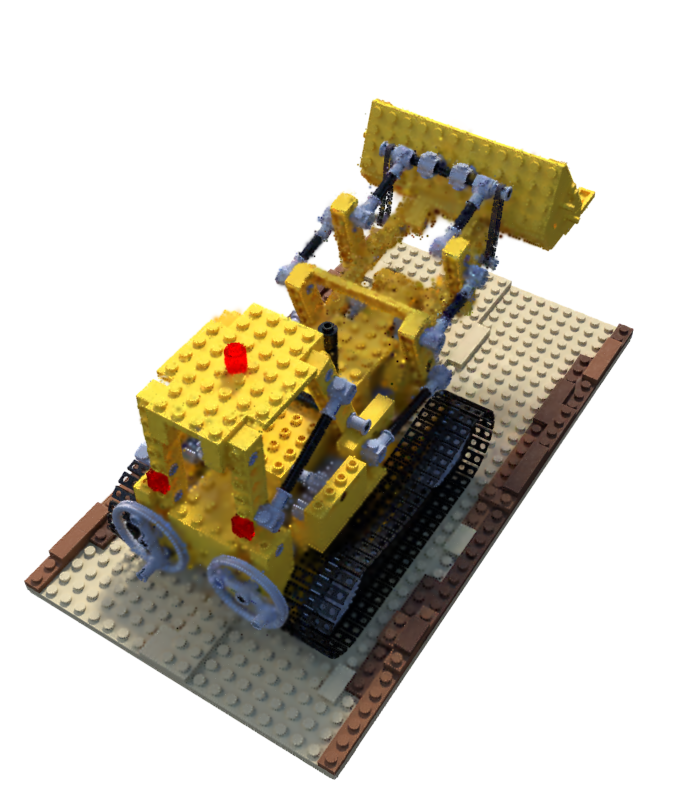}
    &   \includegraphics[width=\hsize,valign=m]{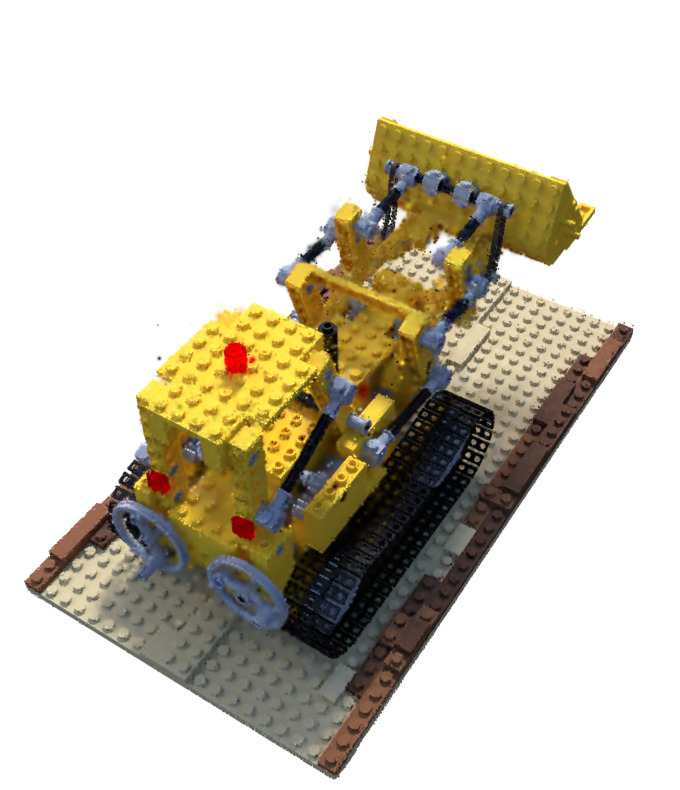}
    &   \includegraphics[width=\hsize,valign=m]{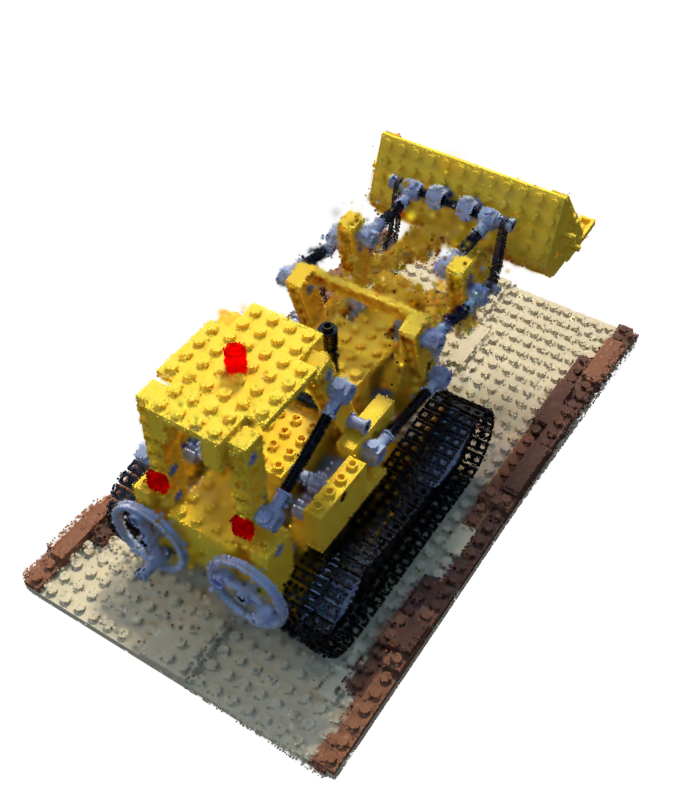}
    &   \includegraphics[width=\hsize,valign=m]{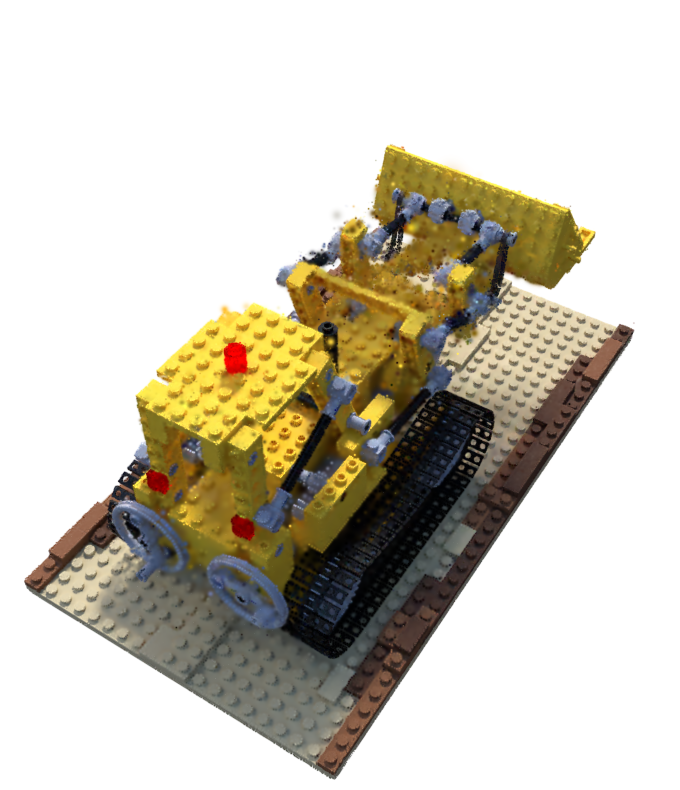}
    \\

    \midrule

    \includegraphics[width=\hsize,valign=m]{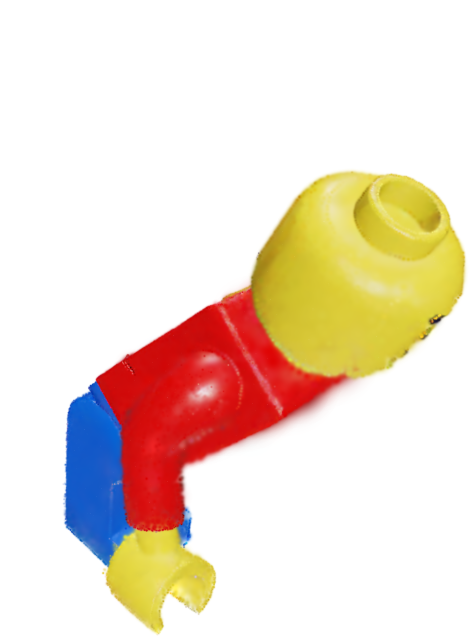}
    &   \includegraphics[width=\hsize,valign=m]{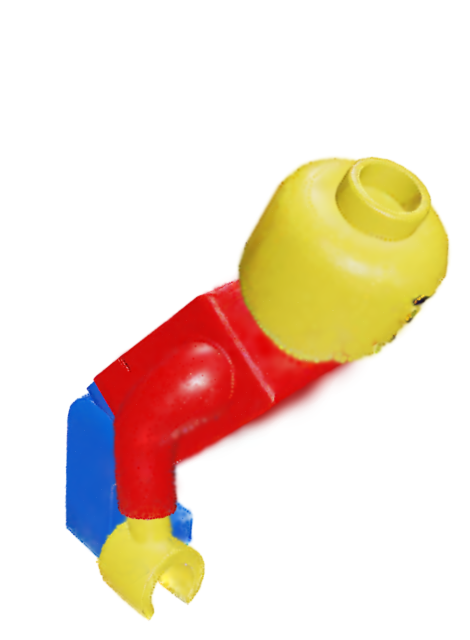}
    &   \includegraphics[width=\hsize,valign=m]{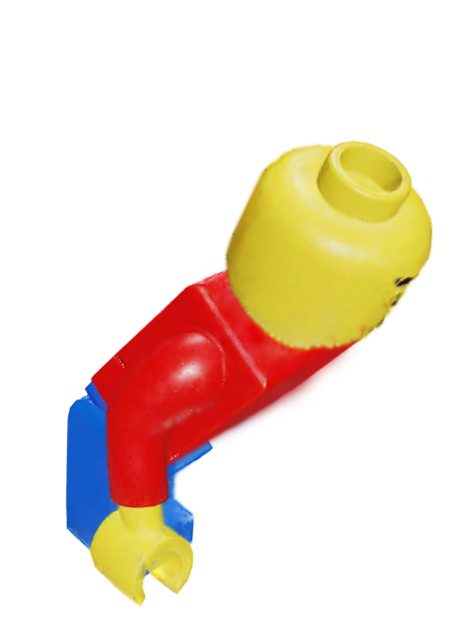}
    &   \includegraphics[width=\hsize,valign=m]{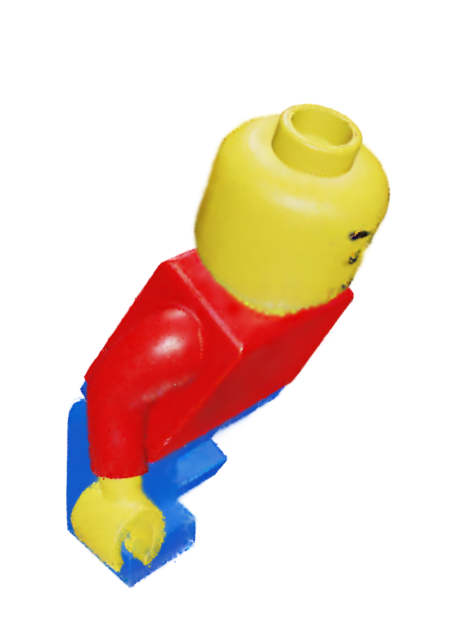}
     &   \includegraphics[width=\hsize,valign=m]{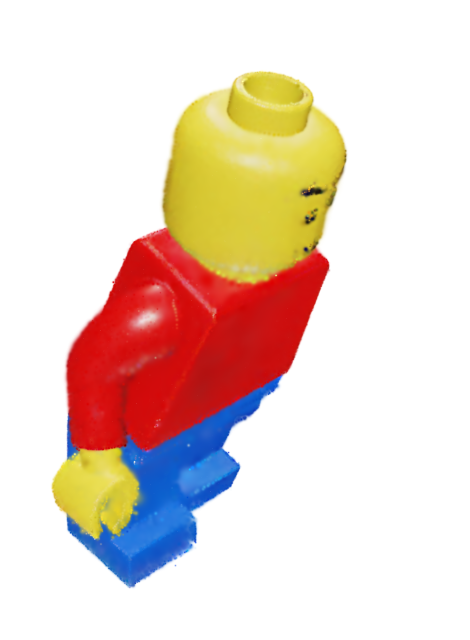}
    &   \includegraphics[width=\hsize,valign=m]{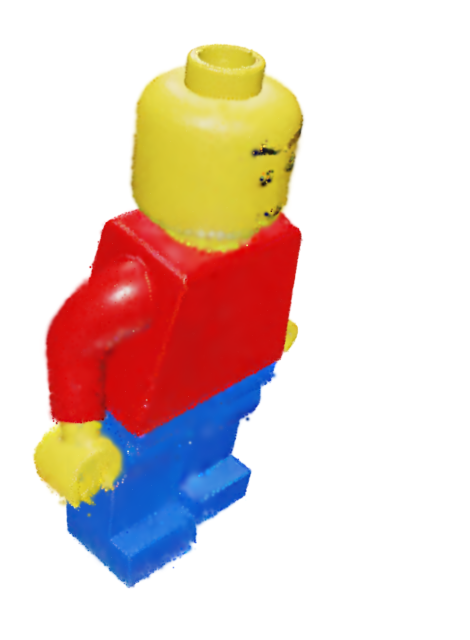}
    &   \includegraphics[width=\hsize,valign=m]{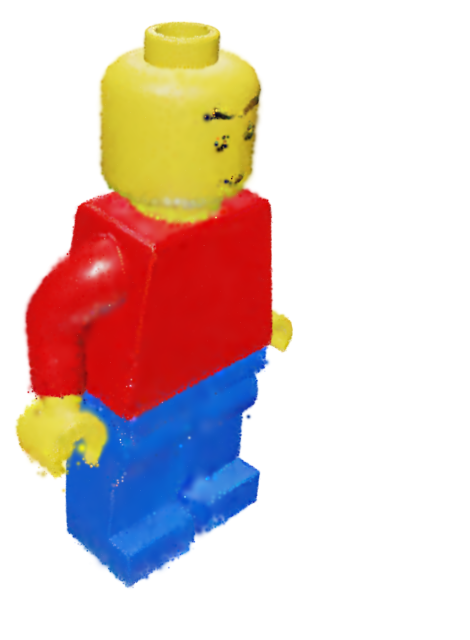}
    &   \includegraphics[width=\hsize,valign=m]{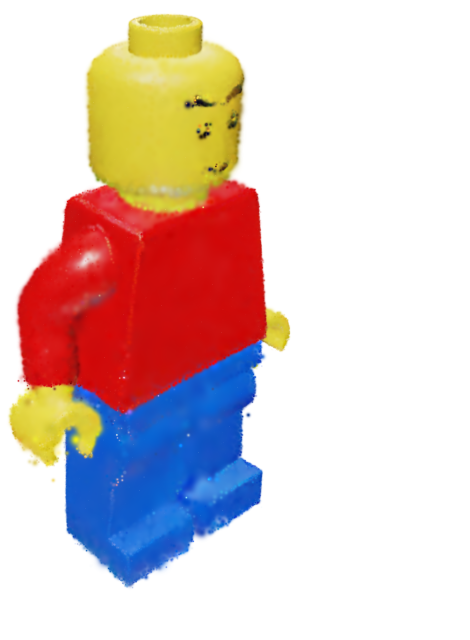}
    &   \includegraphics[width=\hsize,valign=m]{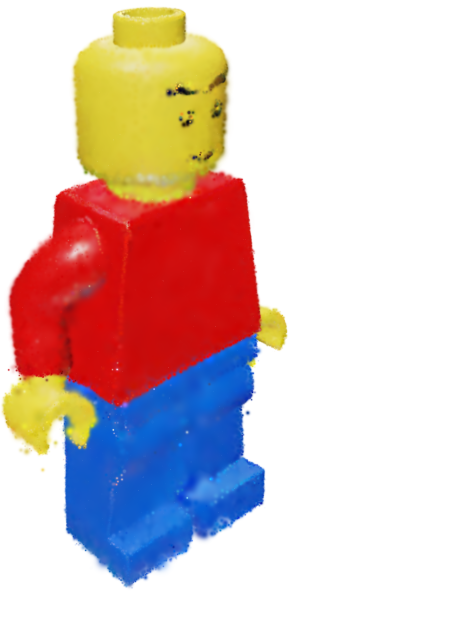}
    \\

    \midrule

    \includegraphics[width=\hsize,valign=m]{images/inter_extra/lamp/00134_-0.40.png}
    &   \includegraphics[width=\hsize,valign=m]{images/inter_extra/lamp/00134_-0.20.png}
    &   \includegraphics[width=\hsize,valign=m]{images/inter_extra/lamp/00134_0.00.png}
    &   \includegraphics[width=\hsize,valign=m]{images/inter_extra/lamp/00134_0.25.png}
     &   \includegraphics[width=\hsize,valign=m]{images/inter_extra/lamp/00134_0.50.png}
    &   \includegraphics[width=\hsize,valign=m]{images/inter_extra/lamp/00134_0.75.png}
    &   \includegraphics[width=\hsize,valign=m]{images/inter_extra/lamp/00134_1.00.png}
    &   \includegraphics[width=\hsize,valign=m]{images/inter_extra/lamp/00134_1.20.png}
    &   \includegraphics[width=\hsize,valign=m]{images/inter_extra/lamp/00134_1.40.png}
    \\  

    \midrule

    \includegraphics[width=\hsize,valign=m]{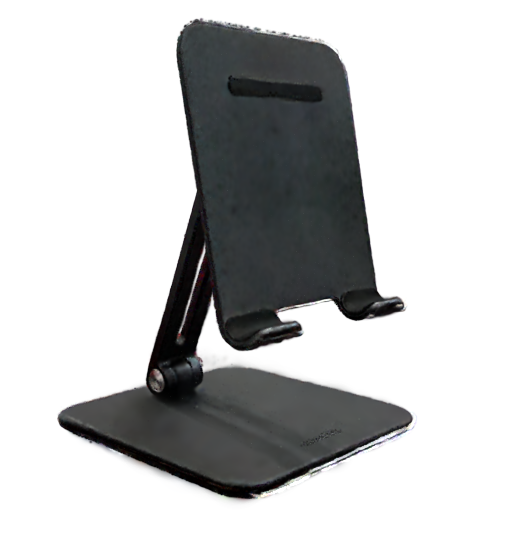}
    &   \includegraphics[width=\hsize,valign=m]{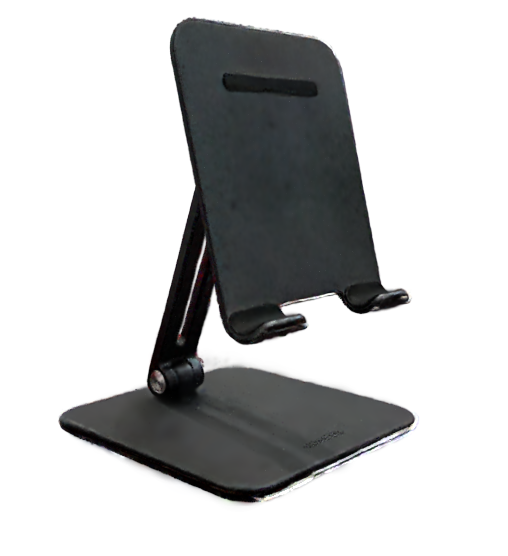}
    &   \includegraphics[width=\hsize,valign=m]{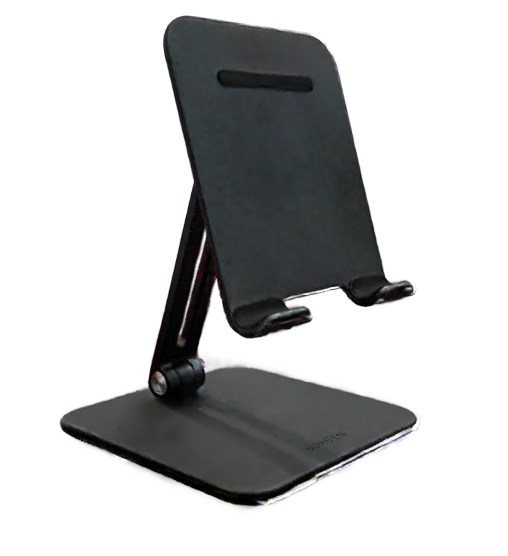}
    &   \includegraphics[width=\hsize,valign=m]{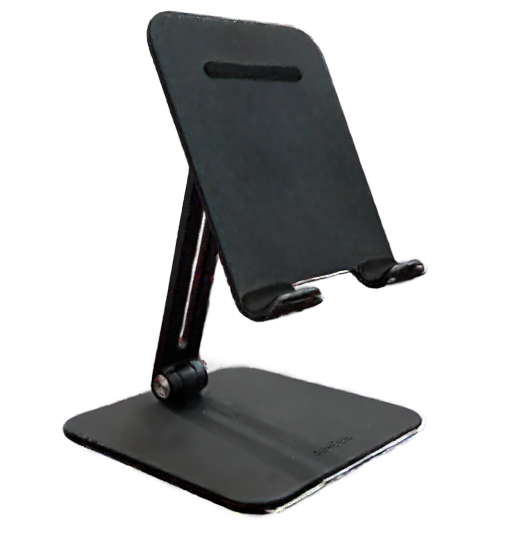}
     &   \includegraphics[width=\hsize,valign=m]{images/suppl_inter_extra_papr/stand/00069_0.25.png}
    &   \includegraphics[width=\hsize,valign=m]{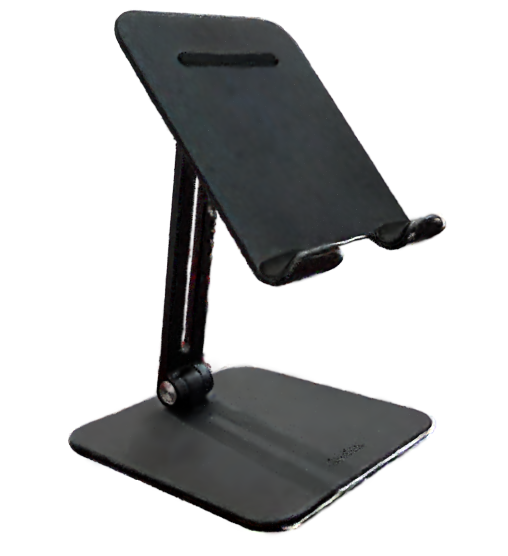}
    &   \includegraphics[width=\hsize,valign=m]{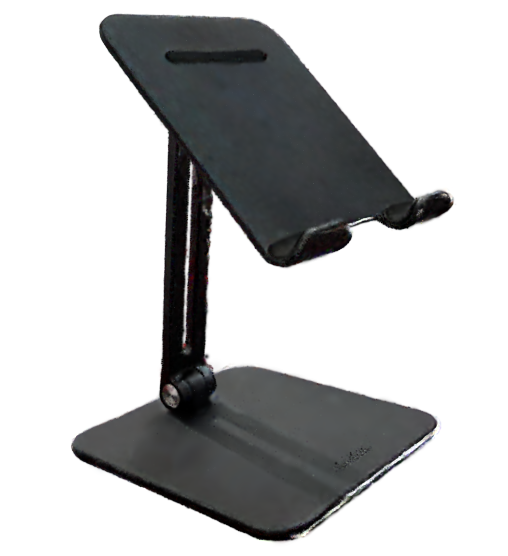}
    &   \includegraphics[width=\hsize,valign=m]{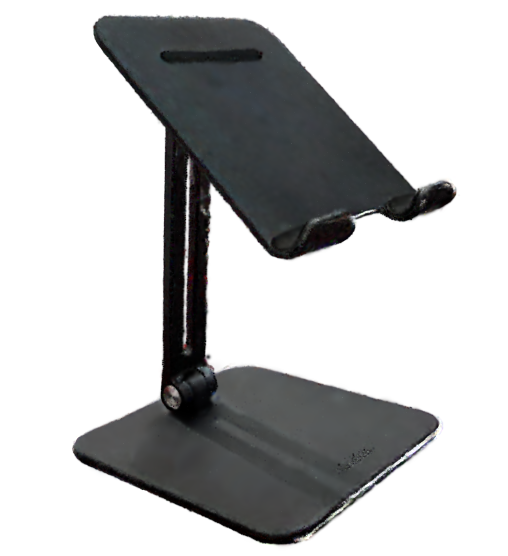}
    &   \includegraphics[width=\hsize,valign=m]{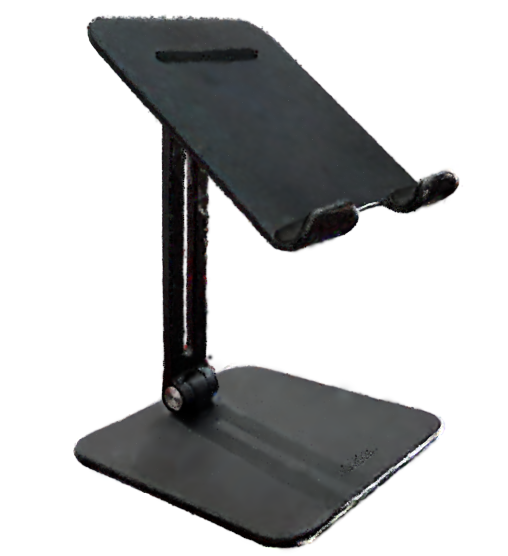}
    \\

\end{tabularx}
\caption{\textbf{Additional Interpolation and Extrapolation Results.} The figure shows interpolation and extrapolation novel-view synthesis results using our method on the PAPR in Motion dataset~\cite{peng2024papr}. From top to bottom, the scenes displayed are \texttt{Dolphin}, \texttt{Butterfly}, \texttt{Giraffe}, \texttt{Crab}, \texttt{Lego Bulldozer}, \texttt{Lego Man}, \texttt{Lamp}, and \texttt{Stand}.}
\label{fig:suppl_inter_extra_local}
\end{figure*}